\documentclass[prl,twocolumn,amsmath,amssymb,superscriptaddress]{revtex4}

\usepackage{epsfig,amsmath}
\usepackage{subfigure}
\usepackage{graphicx}
\usepackage{dcolumn}
\usepackage{stmaryrd}
\usepackage{mathrsfs}
\usepackage{pifont}
\usepackage{amsthm}
\usepackage{amssymb}
\usepackage{bm}
\usepackage{latexsym}
\usepackage{color}
\usepackage{epstopdf}
\usepackage{amsfonts}
\usepackage{units}
\usepackage{times}
\usepackage{upgreek}
\usepackage{tabularx}
\usepackage{booktabs}
\usepackage[colorlinks=true,linkcolor=blue,citecolor=blue,urlcolor=blue]{hyperref}
 \usepackage{float}

\newcommand{\tr}{{\rm Tr}}

\newcommand{\ket}[1]{|#1\rangle}
\newcommand{\bra}[1]{\langle #1|}
\newcommand{\proj}[1]{\ket{#1}\bra{#1}}

\begin{document}

\title{Probe of Generic Quantum Contextuality and Nonlocal Resources for Qubits}

\author{Wei Li}
\thanks{These authors contributed equally to this work.}
\affiliation{College of Physics and Hebei Key Laboratory of Photophysics Research and Application, Hebei Normal University, Shijiazhuang, Hebei 050024, China}

\author{Min-Xuan Zhou}
\thanks{These authors contributed equally to this work.}
\affiliation{Institute of Physics, Chinese Academy of Sciences, Beijing 100190, China}
\affiliation{School of Physical Sciences, University of Chinese Academy of Sciences, Beijing 100049, China}

\author{Yun-Hao Shi}
\affiliation{Institute of Physics, Chinese Academy of Sciences, Beijing 100190, China}

\author{Z. D. Wang}
\email{zwang@hku.hk}
\affiliation{Frontier Research Institute for Physics, South China Normal University, Guangzhou 510006, China}
\affiliation{HK Institute of Quantum Science \& Technology and Department of Physics, The University of Hong Kong, Pokfulam Road, Hong Kong, China}
\affiliation{Hong Kong Branch for Quantum Science Center of Guangdong-Hong Kong-Macau Greater Bay Area, Shenzhen 518045, China}

\author{Heng Fan}
\email{hfan@iphy.ac.cn}
\affiliation{Institute of Physics, Chinese Academy of Sciences, Beijing 100190, China}
\affiliation{School of Physical Sciences, University of Chinese Academy of Sciences, Beijing 100049, China}
\affiliation{Beijing Academy of Quantum Information Sciences, Beijing 100193, China}

\author{Yan-Kui Bai}
\email{ykbai@semi.ac.cn}
\affiliation{College of Physics and Hebei Key Laboratory of Photophysics Research and Application, Hebei Normal University, Shijiazhuang, Hebei 050024, China}
\affiliation{HK Institute of Quantum Science \& Technology and Department of Physics, The University of Hong Kong, Pokfulam Road, Hong Kong, China}


\begin{abstract}
We reveal that the entropic uncertainty relation with a quantum memory is able to intrinsically connect local generic contextuality addressed in the pioneering work by Spekkens \cite{rw05pra} and nonlocal quantum resources such as entanglement and Bell nonlocality. Based on the constructed optimal set for any given single-qubit state, we prove rigorously a faithful criterion to witness the generic contextuality in the scenario of local quantum state preparation. Furthermore, within the framework of quantum resource distribution, it is proved that there exist quantitative trade-off relations between local preparation contextuality and bipartite entanglement or Bell nonlocality in a shared quantum system, which are captured by two inequalities where the local and nonlocal quantum resources can coexist. The faithful criterion and quantitative inequalities are all experimentally testable, which are verified through two independent well-designed experiments on the Quafu quantum cloud platform.
\end{abstract}

\maketitle


\textit{Introduction.}---In quantum theory, contextuality  \cite{rw05pra,sr07pra,eps60p,sk67jmm,cab96pla,kly08prl,lap11nat,md16nc,xu20prl,cab21prl,bc22rmp,che25prl}, entanglement \cite{e35pr,kv89pra,per93book,ter00pra,rh09rmp}, Bell nonlocality \cite{js64p,jf69prl,nie10book,bru14rmp}, and uncertainty relations (URs) \cite{r29pr,dd83prl,kk86prd,hm88prl,ls15pra,pjc17rmp} are characteristic properties and have intimate relationships, which have now become the cornerstones of quantum science and technology. The Kochen-Specker (KS) theorem \cite{eps60p,sk67jmm} states that quantum mechanics is in conflict with classical models, and the KS contextuality \cite{bc22rmp} is a typical nonclassical conception in various theories which can serve as an important physical resource in universal quantum computation \cite{mh14nat} and quantum communication complexity \cite{gup23prl}. Apart from the well-known entanglement monogamous inequalities \cite{c00pra,kw04pra,o06prl,fan07pra,b09pra,b14prl}, Kurzy\'{n}ski \textit{et al} showed that a fundamental monogamy relation limits the local KS contextuality and Bell nonlocality in a bipartite system \cite{pk14prl}, and experimental verification was performed on the photon systems showing the mutually repulsive property of the two nonclassicalities \cite{xz16prl}. Recently, Xue \textit{et al} further showed that local KS contextuality and nonlocality in a generalized Bell scenario can coexist \cite{px23prl}. The similar trade-off relations between local KS contextuality and entanglement or nonlocality were also derived in various theories \cite{jia16pra,saha17pra,cs17pra}.

Besides the KS contextuality, the generalized contextuality introduced by Spekkens \cite{rw05pra} focuses on some operational procedures in physical theories, where the preparation contextuality can indicate the nonclassicality in the scenario of quantum state preparation (QSP) and exhibit quantum advantages in certain tasks such as parity-oblivious multiplexing \cite{rws09prl} and quantum minimum-error state discrimination \cite{ds18prx}. Moreover, preparation contextuality in a specific set of single-party states can be used to detect the nonclassicality of the URs \cite{lc22prl}. It is also noted that entanglement and Bell nonlocality are crucial quantum resources in quantum information processing \cite{ekt91prl,ben96prl,rau01prl,bri09np,aci07prl,nad22nat}. However, within the framework of quantum resource distribution (QRD), it is still an open problem whether there exist the quantitative trade-off relations between this generic contextuality and entanglement or nonlocality, although some qualitative positive correlations have been identified for various theories in the framework of resource conversion via remote state preparation (RSP) \cite{ban15pra,ham17prl,sah19pra,wri23prl,mp24prl}. The key challenge in solving this critical problem is twofold: First, the absence of a faithful criterion for witnessing the contextuality in the local QSP procedure even for the simplest four-state set associated with any given quantum state; Second, the lack of effective methods to establish the intrinsic relationships between local preparation contextuality and nonlocal quantum resources.

In this Letter, we analytically prove that there exist the trade-off relations between local preparation contextuality and bipartite entanglement or Bell nonlocality evaluated by the violation of Clauser-Horne-Shimony-Holt (CHSH) inequality \cite{jf69prl} in qubit systems. By virtue of the optimal four-state set for any given single-qubit state $\rho_A$ constructed by $B_2$-orbit realizability condition, we present a faithful criterion to witness the local contextuality via the entropic uncertainty relation (EUR). Then we establish the trade-off relations between local preparation contextuality and nonlocal quantum resources in a shared bipartite quantum state $\rho_{AB}$, which are captured by two quantitative inequalities. Finally,  we verify the theoretical results on solid-state superconducting systems via two independent experiments on the Quafu quantum cloud computing cluster \cite{qbac}, where the local contextuality and nonlocal quantum resources are detected synchronously.


\textit{Preliminaries.}---Preparation contextuality \cite{rw05pra,sr07pra} establishes the impossibility of a noncontextual ontological model with preparation equivalence classes in an operational theory. In this case, the classical ontological model reproduces the predictions of operational quantum theory if and only if \cite{rw05pra,lc22prl} $\mathbb{P}(y|M,P)=\sum_{\lambda\in \Gamma} \xi(y|M,\lambda)\mu(\lambda|P)$, in which the probability on the left hand specifies an outcome $y$ that results from a measurement $M$ given a preparation procedure $P$ and $\lambda$ on the right hand is an ontic state in the state space $\Gamma$ with $\mu(\lambda|P)$ being a probability distribution over the ontic states for the preparation procedure and $\xi(y|M,\lambda)$ being a conditional probability distribution of the outcome $y$ resulted from the measurement $M$ on this ontic state. The operational equivalence of two preparation procedures is defined as $P\simeq P'\Leftrightarrow \mathbb{P}(y|M,P)=\mathbb{P}(y|M,P')$ \cite{rw05pra} for all measurements $M$ and outcomes $y$. Then the preparation noncontextuality in quantum theory requires that two equivalent preparations, which are related to the same density operator, must have the same distribution over the ontic state space
\begin{equation}\label{01}
	P\simeq P'\Rightarrow \mu(\lambda|P)=\mu(\lambda|P').
\end{equation}
When the condition in Eq. \eqref{01} is violated, the quantum state preparations $P$ and $P'$ exhibit the nonclassicality which is referred to as preparation contextuality.

Entanglement is a kind of nonlocal resource in composite quantum systems, and the entangled state can not be written as convex sum of the tensor product of subsystem states \cite{kv89pra,rh09rmp}. In bipartite systems, Bell state $\ket{\Psi}_{AB}=(\ket{00}+\ket{11})/\sqrt{2}$ is a typical entangled state \cite{b92prl}, and the negative conditional von Neumann entropy $S(A|B)=S(AB)-S(B)$ is an effective signature for entanglement detection in which $S(\rho)=-\tr[\rho\mbox{log}\rho]$ \cite{di05prsa}. Bell nonlocality is a stronger nonclassical correlation \cite{js64p}, and the nonlocality in two-qubit systems can be probed by violation of the CHSH inequality \cite{jf69prl}
\begin{equation}\label{02}
	\langle B\rangle=|\langle A_{0}B_{0}\rangle+\langle A_{0}B_{1}\rangle+\langle A_{1}B_{0}\rangle-\langle A_{1}B_{1}\rangle|\leq 2,
\end{equation}
where $A_i$ and $B_i$ with $i=0,1$ are local measurement operators, and the correlation function $\langle B\rangle>2$ indicates the existence of Bell nonlocality.

The EUR in quantum information theory has the form \cite{hm88prl}
\begin{equation}\label{03}
	H(X)+H(Z)\geq \mbox{log}_2\frac{1}{c},
\end{equation}
where $H(X)$ is the Shannon entropy of the probability distribution of the outcomes when $X$ is measured (the meaning of $H(Z)$ is similar), and the constant $1/c$ quantifies the complementarity of the observables. Moreover, by introducing a quantum memory, the EUR can predict the outcomes of the measurements more precisely and can be expressed as $H(X|B)+H(Z|B)\geq \mbox{log}_2(1/c)+S(A|B)$ in which $H(X|B)$ being conditional von Neumann entropy of the post-measurement state given the memory $B$ \cite{mb10np,cfl11np, dr11np}.


\begin{figure}
	\epsfig{figure=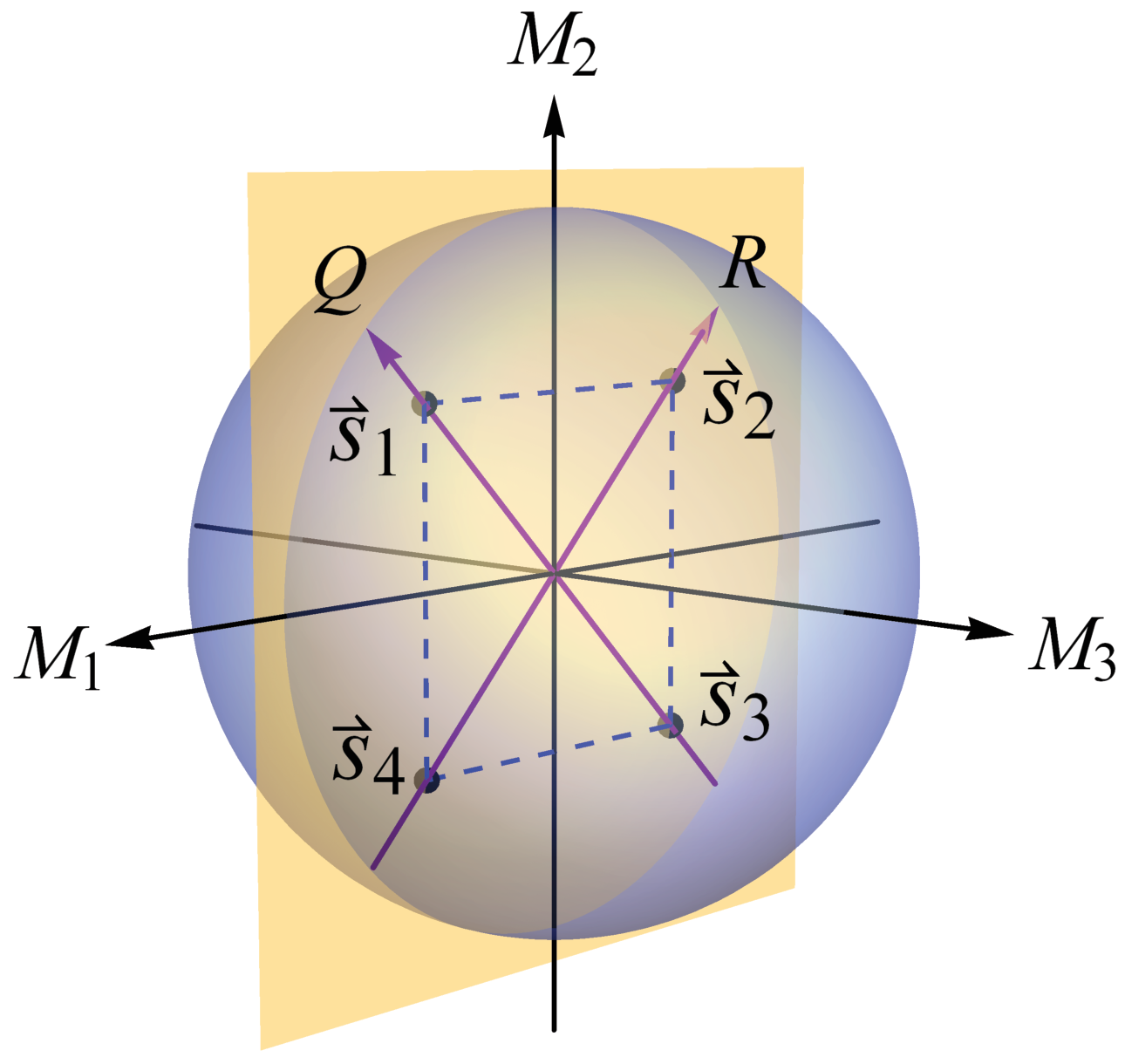,width=0.3\textwidth}
	\caption{The optimal four-state set $\Lambda_{\vec{s}_{1}}$ of a single-qubit state $\vec{s}_1$  generated by the $B_2$-orbit realizability with two complementary measurements $M_1$ and $M_2$ in a Bloch sphere, where the pair of measurements $Q$ and $R$ are optimal observables for the EUR in Eq. \eqref{07} to faithfully witness the preparation contextuality.}
	\label{Fig1}
\end{figure}

\textit{A faithful criterion witnessing preparation contextuality.}--- It is still an unsolved and challenging problem, even for the simplest four-state set in qubit theory, to faithfully witness the preparation contextuality in the set associated with any given quantum state. Here, we solve this problem by decomposing it into two stages, where we first construct the optimal four-state set of any given single-qubit state and then prove a faithful criterion to witness the contextuality in this optimal set.

Now we consider an arbitrary single-qubit state which is represented by the polarization vector $\vec{s}_{1}$ in a Bloch sphere. Associated with this quantum state, we have a pair of complementary measurements
\begin{equation}\label{04}
M_1=(Q-R)/\sqrt{2},~~M_2=(Q+R)/\sqrt{2},
\end{equation}
where $Q$ is the measurement along the direction of polarization vector $\vec{s}_1$ and $R$ is a complementary one to $Q$ as shown in Fig. \ref{Fig1}. Then we can find three other states $\vec{s}_2$, $\vec{s}_3$ and $\vec{s}_4$ under the action of the symmetry group of a square in the $\hat{M}_1\hat{M}_2$ plane with reflections (the Coxeter group $B_2$ \cite{hum90book}), which satisfy the equal predictability property
\begin{eqnarray}\label{05}
	&&\langle M_1\rangle_{\vec{s}_{1}}=\langle M_2\rangle_{\vec{s}_{1}},\nonumber\\
	&&\langle M_1\rangle_{\vec{s}_{1}}=-\langle M_1\rangle_{\vec{s}_{2}}=-\langle M_1\rangle_{\vec{s}_{3}}=\langle M_1\rangle_{\vec{s}_{4}},\nonumber\\
	&&\langle M_2\rangle_{\vec{s}_{1}}=\langle M_2 \rangle_{\vec{s}_{2}}=-\langle M_2\rangle_{\vec{s}_{3}}=-\langle M_2\rangle_{\vec{s}_{4}}.	
\end{eqnarray}
Moreover, the four-state set $\Lambda_{\vec{s}_{1}}$ satisfies the operational equivalent relation
\begin{equation}\label{06}
	\frac{1}{2}\vec{s}_{1}+\frac{1}{2}\vec{s}_{3}=\frac{1}{2}\vec{s}_{2}+\frac{1}{2}\vec{s}_{4},
\end{equation}
where the preparation procedure of $\{\vec{s}_{1}, \vec{s}_{3}\}$ is denoted by $P$ and the one of $\{\vec{s}_{2}, \vec{s}_{4}\}$ is characterized by $P'$. We refer to the relations in Eqs. \eqref{05} and \eqref{06} as \textit{$B_2$-orbit realizability condition}, for which the generated set $\Lambda_{\vec{s}_1}$ in Fig. \ref{Fig1} satisfies the $B_2$ symmetry and we have the following lemma.

\emph{Lemma 1}.---For an arbitrary single-qubit state $\vec{s}_1$, the four-state set $\Lambda_{\vec{s}_1}$ generated by the $B_2$-orbit realizability condition in the $\hat{M}_1\hat{M}_2$ plane of Bloch sphere is optimal.

We note that the authors of Ref. \cite{lc22prl} constructed a kind of four-state sets with the $A_1^2$ symmetry, which are specific ones for the task of characterizing the nonclassical property of URs and depend on the expectation values of the given observables. Here, our constructed set $\Lambda_{\vec{s}_1}$ has higher symmetry due to $A_1^2$ being a subgroup of $B_2$ \cite{hum90book}, and its optimality means that any other four-state set of $\vec{s}_1$ with the $A_1^2$ symmetry admits of a classical explanation provided that the optimal set is preparation noncontextual (see the analytical proof of Lemma 1 in Sec. I of the Supplemental Material (SM) \cite{sup25bai}).

Next, we analyze the preparation contextuality in the QSP of the optimal four-state set $\Lambda_{\vec{s}_1}$ associated with any given single-qubit state $\vec{s}_1$, and can obtain the following result.

\emph{Theorem 1}.---For a single-qubit state $\vec{s}_{1}$, the four-state set $\Lambda_{\vec{s}_{1}}$ exhibits the preparation contextuality if and only if the EUR resulting from a pair of optimal complementary observables $Q$ and $R$ on $\vec{s}_1$ satisfies
\begin{equation}\label{07}
	H(Q)+H(R)< 1+C,
\end{equation}
where $H(Q)$ and $H(R)$ are the Shannon entropies of the post-measurement states, and the constant is a binary entropy function $C=h[(2-\sqrt{2})/4]\simeq 0.6009$. Moreover, for any pair of complementary measurements $Q'$ and $R'$, the optimal set $\Lambda_{\vec{s}_{1}}$ exhibits preparation contextuality when the sufficient condition $H(Q')<C$ or $H(R')<C$ for $\vec{s}_{1}$ is satisfied.

\begin{figure}
	\epsfig{figure=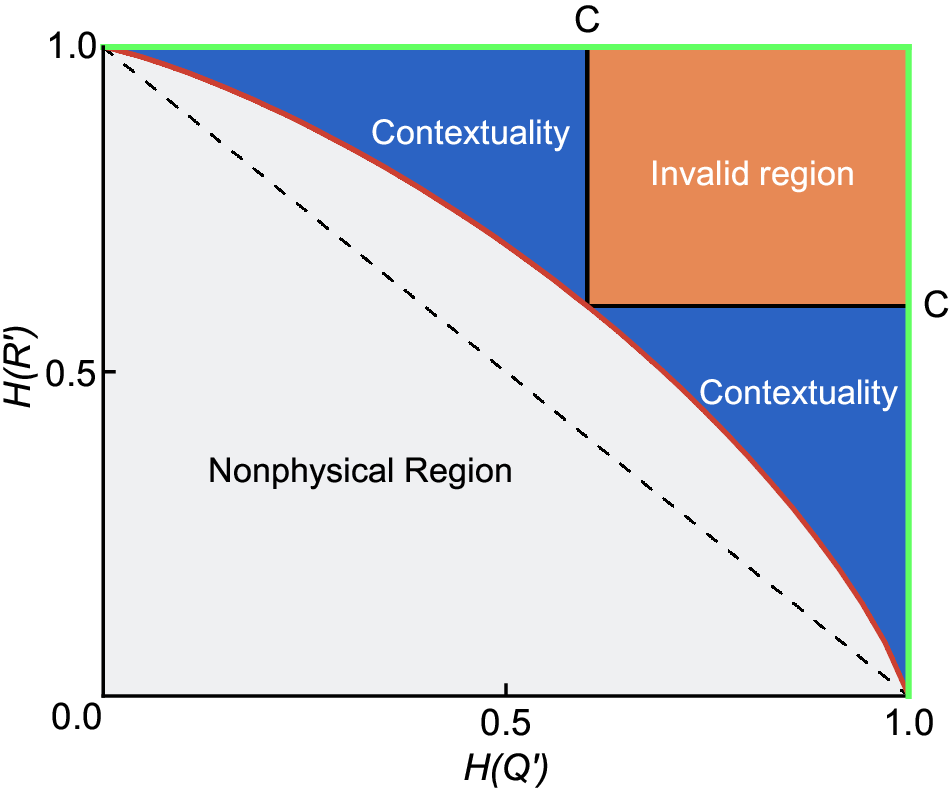, width=0.36\textwidth}
	\caption{The regional diagram for witnessing the preparation contextuality in the optimal four-state set $\Lambda_{\vec{s}_{1}}$, which is divided by the EUR of observables $\{Q',R'\}$. The Blue areas indicate regions of the contextuality, the orange area represents the invalid region being inconclusive, the gray area signifies the nonphysical one, and the upper and right-hand boundaries (green solid lines) correspond to the optimal observables $\{Q,R\}$ providing the faithful witness.}
	\label{Fig2}
\end{figure}

The analytical proof of Theorem 1 is given in Sec. II of the SM \cite{sup25bai}. According to Eq. \eqref{07} in this theorem, the preparation contextuality in the optimal four-state set $\Lambda_{\vec{s}_{1}}$ can be faithfully witnessed (see Sec. IIA of the SM \cite{sup25bai}), where the optimal observables $Q$ and $R$ related to the single-qubit state $\vec{s}_1$ are shown in Fig. \ref{Fig1}. It is noted that this pair of optimal observables $\{Q, R\}$ is not unique, and we can obtain the same results when $Q$ and $R$ are interchanged. In Fig. \ref{Fig2}, we present the regional diagram of quantum contextuality (blue areas), which is divided by the EUR of an arbitrary pair of complementary observables $\{Q',R'\}$. As shown in the figure, the nonphysical region is determined by the relation $H(Q')+H(R')< 1$ (the black dashed line indicating the boundary) and the complementary property of two observables (the red solid line), the invalid region (orange area) means that the detection of preparation contextuality based on $Q'$ and $R'$ is inconclusive (see Sec. IIB of the SM \cite{sup25bai}), and the upper and the right-hand boundaries (the green solid lines) correspond to the optimal observables $\{Q,R\}$ in the faithful criterion.


\emph{The trade-off relations between local preparation contextuality and nonlocal quantum resources}.---We consider a nonlocal scenario where the two parties, Alice and Bob, share a bipartite quantum state $\rho_{AB}$. Given the reduced state $\rho_A$ with the polarization vector $\vec{s}_1$, there exists the optimal four-state set $\Lambda(\rho_A)$ in Alice's party, for which the contextuality in the local QSP can be witnessed by the faithful criterion in Eq. \eqref{07}. We first study the relationship between local preparation contextuality and bipartite entanglement within the framework of QRD, and can derive the following relations.

\emph{Theorem 2}.---Let $\rho_{AB}$ be a two-qubit state. Then the preparation contextuality in local QSP of the optimal four-state set $\Lambda(\rho_A)$ and bipartite entanglement in the shared state $\rho_{AB}$ are restricted by the quantitative trade-off relation
\begin{equation}\label{08}
	1\leq H_{QR}(A)+S(A|B)\leq 3,
\end{equation}
where $H_{QR}(A)=H_{A}(Q)+H_{A}(R)$ is the EUR for subsystem $\rho_A$ to witness the contextuality, and the negative conditional entropy $S(A|B)$ is the lower bound of distillable entanglement. When the shared quantum state is pure, we have the equality $H_{QR}(A)+S(A|B)=1$.

The analytical proof of Theorem 2 is presented in Sec. III of the SM \cite{sup25bai}. According to Eq. \eqref{08}, we can obtain that the local preparation contextuality and bipartite entanglement can coexist while they are not freely distributable. The upper bound is attained for the maximally mixed state $\rho_{AB}=I/4$, and the lower bound corresponds to a pure state $\ket{\psi_{AB}}$. In particular, we find that when the value of two-qubit entanglement is large enough ($-S(A|B)\geq C$), the optimal four-state set $\Lambda(\rho_A)$ is necessarily noncontextual due to $H_{QR}(A)\geq 1+C$ according to our derived trade-off relation.

Moreover, in the shared bipartite system, when nonlocal quantum resource is characterized by Bell nonlocality, we have the following result.

\emph{Theorem 3}.---For a two-qubit state $\rho_{AB}$, there exists a quantitative trade-off relation between local preparation contextuality of $\Lambda(\rho_A)$ and the maximal Bell nonlocality in the shared bipartite state, which is expressed as
\begin{equation}\label{09}
	1\leq H_{QR}(A)+[2-\langle B\rangle_{\mathrm{max}}] \leq 4,
\end{equation}
where $H_{QR}$ is the EUR for subsystem A with the optimal measurements, and the term in the brackets is used to evaluate the magnitude of the deviation from classical bound with $\langle B\rangle_{\mathrm{max}}$ being the maximum of the CHSH correlation in the two-qubit state. When $\rho_{AB}$ is a pure state, the upper bound of the above inequality is $4-2\sqrt{2}$.

In Eq. \eqref{09}, the upper bound is attained for the mixed state $\rho_{AB}=I_{AB}/4$, and the lower bound is achieved by the product state $\ket{\phi}_{A}\otimes \ket{\varphi}_{B}$ with $\ket{\phi}_A$ being an eigenstate of $Q$. Furthermore, according to the lower bound, the local preparation contextuality in the optimal set $\Lambda(\rho_A)$ will disappear when Bell nonlocality is strong enough. The proof of Theorem 3 can be found in Sec. IV of the SM \cite{sup25bai}.


\begin{figure*}
\epsfig{figure=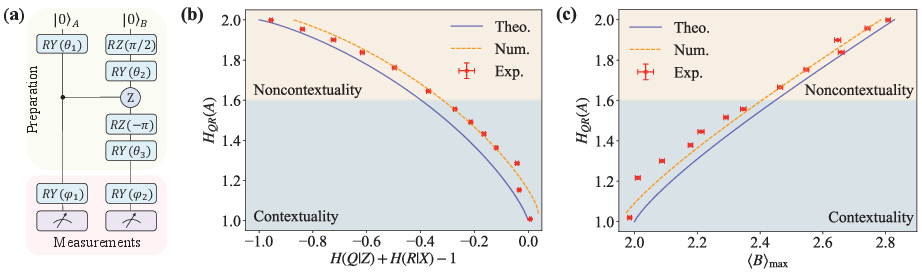,width=0.95\textwidth}
\caption{Demonstration for the trade-off relations between local preparation contextuality and nonlocal quantum resources. (a) Quantum circuit for state preparation and measurements. The upper block prepares the two-qubit state $\ket{\psi(\beta_i)}$ via modulation angles ($\theta_1,\theta_2,\theta_3$), while the lower block performs local measurements with rotation parameters ($\varphi_1,\varphi_2$) and a default measurement being $Z$. (b) Synchronous detection of local preparation contextuality and bipartite entanglement. The blue solid line indicates the theoretical prediction, the orange dashed line represents a numerical simulation based on two-qubit gate fidelity, and the 13 red data points show experimental results corresponding to states $\ket{\psi(\beta_i)}$ with the value of parameter $\beta_i$ increasing from bottom to top. The error bars represent the standard error obtained from 30 replicates of the experiment, and each set of experiments consists of 3,000 single-shot measurements for the prepared state. The light blue and light beige shaded areas correspond to the contextuality and noncontextuality regions, respectively. (c) Synchronous detection of local preparation contextuality and Bell nonlocality. Here the blue solid line is theoretical prediction, the orange dashed line shows noisy numerical simulation, and the 13 red data points represent the experimental outcomes. The function $\langle B\rangle_{\mathrm{max}}$ is the maximal Bell correlation, for which the optimal measurements on subsystem $A$ are $\{A_0=Q, A_1=R\}$ and the optimal ones $\{B_0, B_1\}$ on subsystem $B$ can be derived via the state parameter $\beta_i$ and the Horodecki parameter $\mathcal{M}$.}\label{Fig3}
\end{figure*}

\emph{Synchronous detection of local preparation contextuality and nonlocal quantum resources.}---In the experiments, we prepare a kind of two-qubit pure states, which has the form
\begin{equation}\label{10}	
\ket{\psi(\beta)}=\sqrt{\alpha}\ket{\phi_{1}}_A\ket{0}_B+\sqrt{\beta}\ket{\phi_{2}}_A\ket{1}_B,
\end{equation}
where two orthogonal local states are $\ket{\phi_{1}}_A=\cos(\pi/8)\ket{0}+\sin(\pi/8)\ket{1}$ and  $\ket{\phi_{2}}_A=\sin(\pi/8)\ket{0}-\cos(\pi/8)\ket{1}$, and the parameters satisfy $\alpha=1-\beta$ with $\beta\in[0,0.5]$. According to Theorem 1, the preparation contextuality of the set $\Lambda(\rho_A)$ can be witnessed faithfully via the EUR
\begin{equation}\label{11}
 H_{QR}(A)=H(Q)+H(R)=1+h(\beta),
\end{equation}
where the optimal measurements for the reduced state $\rho_A$ are $Q=(X+Z)/\sqrt{2}$ and $R=(-X+Z)/\sqrt{2}$ with $X$ and $Z$ being Pauli matrices, and the state-dependent binary entropy function is $h(\beta)=-\beta \mbox{log}_2\beta-(1-\beta)\mbox{log}_2(1-\beta)$.

Based on the faithful criterion in Eq. \eqref{07} of Theorem 1, the set $\Lambda(\rho_A)$ is preparation contextual when the parameter $\beta<(2-\sqrt{2})/4$. Moreover, the detection of bipartite entanglement is based on the EUR with a quantum memory \cite{mb10np}, where the conditional entropy as an effective entanglement indicator can be expressed as $S(A|B)\leq H(Q|B)+H(R|B)-1$. Since the conditional entropy is not decreasing under the measurement on subsystem $B$, we can obtain the measurable upper bound for this entanglement indicator
\begin{equation}\label{12}
	H(Q|Z)+H(R|X)-1\geq S(A|B),
\end{equation}
where the numerical optimization for the minimal bound results in the measurements $Z$ and $X$ on qubit $B$ (see Sec. VI of the SM \cite{sup25bai}), and the two-qubit state must be entangled when the left hand of the inequality is negative.

The experimental verification between local contextuality and bipartite entanglement is performed on $Baihua$ chip \cite{qbac}, for which the hardware information is given in Sec. V of the SM \cite{sup25bai}. A series of two-qubit states $\ket{\psi(\beta_i)}$ with the state parameter $\beta_i$ changing from $0$ to $0.5$ and $i=1,...,13$ are prepared. The product state is the case of $\beta_1=0$, and the maximal entangled state corresponds to $\beta_{13}=0.5$. The quantum circuits for state preparation and local measurements are shown in Fig. \ref{Fig3}(a). The preparation circuit is composed of one two-qubit controlled-$Z$ gate and five local rotation gates with parameters $\theta_1$, $\theta_2$, and $\theta_3$. The local measurements $Q$, $R$, $X$, and $Z$ in Eqs. \eqref{11} and \eqref{12} are made by modulating the rotation angles $\varphi_1$ and $\varphi_2$ in the measurement circuit. In Fig. \ref{Fig3}(b), the experimental results are plotted (from bottom to top, 13 red data points correspond to state parameters from $\beta_1=0$ to $\beta_{13}=0.5$), where the $x$-axis is the entanglement indicator $H(Q|Z)+H(R|X)-1$ and $y$-axis is the contextuality witness $H_{QR}(A)$ (see the detailed experimental data in Sec. VI of the SM \cite{sup25bai}). For each $\beta_i$, we repeat 30 sets of experiments where each set consists of 3000 single-shot measurements. The blue solid line is the result of theoretical prediction, and the orange dashed line represents the noisy numerical simulation according to the fidelity of two-qubit controlled-$Z$ gate \cite{xu24arXiv}. The two regions of contextuality (the light blue area) and noncontextuality (the light beige area) are divided by the criterion in Eq. \eqref{07}. The experimental results further verify our theoretical predictions in Theorem 2, where the six experimental data in light blue shaded region (except the bottom one) probe synchronously the local preparation contextuality and bipartite entanglement, and the six data in the light beige shaded region show that the local contextuality will disappear when the two-qubit entanglement is strong enough.

Another experimental verification is the trade-off relation between the local contextuality and Bell nonlocality as shown in Theorem 3. For the prepared state $\ket{\psi(\beta)}$, we choose the optimal measurements on subsystem $A$ to be $A_0=Q$ and $A_1=R$ given in Eq. \eqref{11}, and the optimal measurements on subsystem $B$ have generic forms $B_{0}=a_{0}X+b_{0}Y+c_{0}Z$ and $B_{1}=a_{1}X+b_{1}Y+c_{1}Z$ in which the coefficients $a_i$ and $b_i$ are determined by an optimization procedure in terms of the relation $\langle B\rangle_{\mathrm{max}}=2\sqrt{\mathcal{M}}$ with $\mathcal{M}$ being the Horodecki parameter \cite{hr95pla,cs17prl}. By adjusting the rotation angles $\varphi_1$ and $\varphi_2$ in the measurement circuit of Fig. \ref{Fig3}(a), the above local measurements can be performed. The details of measurements $B_0$ and $B_1$ are listed in Sec. VII of the SM \cite{sup25bai}. When the correlation $\langle B\rangle_{\mathrm{max}}>2$, the two-qubit state exhibits Bell nonlocality. In Fig. \ref{Fig3}(c), the experimental results are displayed, where the functions $H_{QR}(A)$ and $\langle B\rangle_{\mathrm{max}}$ are detected for 13 prepared states $\ket{\psi(\beta_i)}$ with $i=1, 2,\dots, 13$ from bottom to top. The experimental data (13 red data points) agree with the noisy numerical simulation (the orange dashed line), and the error bars represent the standard error based on 30 sets of experiments. In the light blue shaded region, local preparation contextuality and Bell nonlocality are probed synchronously (except for the bottom data corresponding to a product state). The trade-off relation in Theorem 3 is also demonstrated in the light beige shaded region, where the set $\Lambda(\rho_A)$ is necessarily noncontextual when the two-qubit Bell nonlocality is stronger than a certain value \cite{note1}.


\emph{Conclusion and outlook.}---By virtue of a faithful criterion to witness preparation contextuality in the optimal four-state set $\Lambda(\rho_A)$ \cite{note2}, we have solved the critical problem that there exist quantitative trade-off relations between the local contextuality and bipartite entanglement or Bell nonlocality in a shared quantum system $\rho_{AB}$. Theorem 1 solves a non-trivial problem that is dual to the one addressed in Ref. \cite{lc22prl}. Moreover, unlike the mutually repulsive relation between local KS contextuality and Bell nonlocality \cite{pk14prl,xz16prl}, the local preparation contextuality and nonlocal quantum resources can coexist with their trade-off relations captured by the two inequalities in Theorems 2 and 3. The operational characteristics of preparation contextuality depend on both the single-qubit state and the specific preparation procedure. For the optimal eight-state set $\widetilde{\Lambda}(\vec{s}_1)$ with the $B_3$ symmetry, the faithful criterion for witnessing the contextuality can be expressed as
\begin{equation}\label{13}
	H(Q)+H(R)< 1+C',
\end{equation}
where the parameter is $C'=h[(3-\sqrt{3})/6]\simeq 0.7440$. In comparison with Eq. \eqref{07} in Theorem 1 for the four-state set, the upper bound in Eq. \eqref{13} is further improved due to the higher symmetry of the eight-state set. In this case, the trade-off relations in Theorems 2 and 3 are still satisfied. The proof of Eq. \eqref{13} (presented as Theorem 4) and corresponding analysis on the trade-offs are provided in Secs. IX and X of the SM \cite{sup25bai}. In addition, the exploration on generic Bell scenarios is an important direction for future study.

Our theoretical results are experimentally validated using the solid-state superconducting systems, which reveal the intrinsic relationships between local preparation contextuality and nonlocal quantum resources. Furthermore, the trade-off relations we established provide the fundamental constraints for designing various future applications harnessing multiple quantum resources jointly.

W.L. and Y.K.B. would like to thank Zhen-Peng Xu for helpful discussions. This work was supported by NSFC (Grants No. 11575051, No. 12347165, No. 12404330, and No. 12404405), Hebei NSF (Grant No. A2021205020), Hebei 333 Talent Project (No. B20231005), the Guangdong Provincial Quantum Science Strategic Initiative (Grant No. GDZX2404001), the NSFC/RGC JRS grant (RGC Grant No. N-HKU774/21), the GRF (Grant No. 17303023) of Hong Kong, and the fund of Hebei Normal University (Grant No. L2026J02). Y.H.S. acknowledges support from NSFC (Grant No. 12504593) and China Postdoctoral Science Foundation (Grant No. GZB20240815). H.F. acknowledges support from NSFC (Grants No. 92265207, No. T2121001, No. U25A6009, and No. 92365301). M.X.Z. was sponsored by CPS-Yangtze Delta Region Industrial Innovation Center of Quantum and Information Technology-MindSpore Quantum Open Fund. We use the MindSpore Quantum framework \cite{xu24arXiv} to perform numerical simulations of quantum circuits. The simulation codes used in this study are publicly available at https://github.com/ucas-zmx/QCNR.

\onecolumngrid
\clearpage

\renewcommand{\theequation}{S\arabic{equation}}
\renewcommand{\thefigure}{S\arabic{figure}}
\setcounter{equation}{0}
\setcounter{figure}{0}
\setcounter{page}{1}
\setcounter{secnumdepth}{5}

\begin{center}
\textbf{\large SUPPLEMENTAL MATERIAL:\\ Probe of Generic Quantum Contextuality and Nonlocal Resources for Qubits}
\end{center}

\begin{center}
Wei Li,$^{1,*}$ Min-Xuan Zhou,$^{2,3,*}$ Yun-Hao Shi,$^{2}$  Z. D. Wang,$^{4,5,6,\dag}$ Heng Fan,$^{2,3,7,\ddag}$ and Yan-Kui Bai$^{1,5,\S}$\\
\vspace{0.3em}

\small {
\textit{$^1$College of Physics and Hebei Key Laboratory of Photophysics Research and Application,\\
Hebei Normal University, Shijiazhuang, Hebei 050024, China\\
$^2$Institute of Physics, Chinese Academy of Sciences, Beijing 100190, China\\
$^3$School of Physical Sciences, University of Chinese Academy of Sciences, Beijing 100049, China\\
$^4$Frontier Research Institute for Physics, South China Normal University, Guangzhou 510006, China\\
$^5$HK Institute of Quantum Science \& Technology and Department of Physics,\\
The University of Hong Kong, Pokfulam Road, Hong Kong, China\\
$^6$Hong Kong Branch for Quantum Science Center of Guangdong-Hong Kong-Macau Greater Bay Area, Shenzhen 518045, China\\
$^7$Beijing Academy of Quantum Information Sciences, Beijing 100193, China\\}
}
\end{center}

\onecolumngrid
\begin{center}
\normalsize
\textbf{\normalsize Contents}
\end{center}

\begin{tabbing}
\hspace{50.2em}\=\hspace{15em}\=\kill
{\color{blue}\textbf{I}.~~}\hyperref[LEMMA 1]{\textbf{Proof of Lemma 1}} \>{2} \\[2ex]

{\color{blue}\textbf{II}.~~}\hyperref[THEOREM 1]{\textbf{Proof of Theorem 1}} \>{5} \\
\quad~~{\color{blue}A.~~}\hyperref[THEOREM 1A]{Proof of the faithful criterion in Theorem 1} \>{5}\\
\quad~~{\color{blue}B.~~}\hyperref[THEOREM 1B]{Proof of the sufficient condition in Theorem 1 and the
explanation of invalid region} \>{7}\\
\quad~~{\color{blue}C.~~}\hyperref[THEOREM 1C]{The difference between Theorem 1 and the nonclassical
property of the UR.} \>{10}\\[2ex]

{\color{blue}\textbf{III}.~~}\hyperref[THEOREM 2]{\textbf{Proof of Theorem 2}} \>{12} \\[2ex]

{\color{blue}\textbf{IV}.~~}\hyperref[THEOREM 3]{\textbf{Proof of Theorem 3}} \>{13} \\[2ex]

{\color{blue}\textbf{V}.~~}\hyperref[DEVICE]{\textbf{Device information and readout correction in the experiments}} \>{15} \\[2ex]

{\color{blue}\textbf{VI}.~~}\hyperref[EXPERIMENTAL 1]{\textbf{Experimental detection of local preparation contextuality and bipartite entanglement}} \>{15} \\[2ex]

{\color{blue}\textbf{VII}.~~}\hyperref[EXPERIMENTAL 2]{\textbf{Experimental detection of local preparation contextuality and Bell nonlocality}} \>{16} \\[2ex]

{\color{blue}\textbf{VIII}.~~}\hyperref[SINGLE]{\textbf{Probe of quantum contextuality via a single optimal measurement}} \>{18}\\[2ex]

{\color{blue}\textbf{IX}.~~}\hyperref[IX]{\textbf{The faithful criterion for witnessing preparation contextuality in the optimal eight-state set}} \>{19} \\
\quad~~{\color{blue}A.~~}\hyperref[IXA]{The optimal eight-state set with $B_3$ symmetry} \>{19}\\
\quad~~{\color{blue}B.~~}\hyperref[IXB]{The faithful criterion for the optimal eight-state set} \>{21}\\[2ex]

{\color{blue}\textbf{X}.~~}\hyperref[X]{\textbf{Trade-off relations for preparation contextuality in the optimal eight-state set}} \>{22} \\[2ex]

\quad{\color{blue}~~~}\hyperref[ref-sup]{\textbf{References}} \>{23}\\
\end{tabbing}

\clearpage
\twocolumngrid

\section{Proof of Lemma 1}
\label{LEMMA 1}

In Lemma 1 of the main text, we construct an optimal four-state set $\Lambda(\vec{s}_1)$ for any given single-qubit state $\vec{s}_1$ by the $B_2$-orbit realizability condition. The optimality property of  $\Lambda(\vec{s}_1)$ lies in that any other four-state set of $\vec{s}_1$ with the $A_1^2$ symmetry admits of a classical explanation, provided that the optimal set is preparation noncontextual. Next, we give the Lemma and its analytical proof.

\emph{Lemma 1}.---For an arbitrary single-qubit state $\vec{s}_1$, the four-state set $\Lambda_{\vec{s}_1}$ generated by the $B_2$-orbit realizability condition in the $\hat{M}_1\hat{M}_2$ plane of Bloch sphere is optimal.

\emph{Proof.} The proof procedure is composed of two stages. We first derive the criterion for witnessing the preparation contextuality in the optimal four-state set generated by the $B_2$-orbit realizability condition, and then prove the optimality property of the optimal set in contrast to any other four-state set with the $A_1^2$ symmetry.

We consider an arbitrary single-qubit state which is represented by the polarization vector $\vec{s}_{1}$ in a Bloch sphere. Associated with this state $\vec{s}_{1}$, we can find a pair of complementary measurements
\begin{eqnarray}\label{s01}
M_1=\frac{Q-R}{\sqrt{2}},~~M_2=\frac{Q+R}{\sqrt{2}},
\end{eqnarray}
where $Q$ is the measurement along the direction of this polarization vector and the measurement $R$ is complementary to $Q$ as shown in Fig. 1 of the main text. By introducing an additional measurement $M_3$ which is orthogonal to the plane of $\hat{M}_1\hat{M}_2$, the single-qubit state in a Bloch sphere can be expressed as
\begin{equation}\label{s02}
	\rho(\vec{s}_1)=\frac{I+\vec s_1\cdot \vec{M}}{2},
\end{equation}
where $\vec{s}_1$ is the polarization vector and $\vec M=\{M_1,M_3,M_2\}$ is the operator basis with the operator $M_i$ being a function of Pauli matrices.

According to the relation between this single-qubit state and related measurements, the polarization vector $\vec{s}_1$ is coplanar with $\{M_1, M_2\}$, and the angle between $\vec{s}_1$ and $M_1$ is $\pi/4$ which is the same as the angle between $\vec{s}_1$ and $M_2$. Then the expectation values of the complementary measurements $\{M_1, M_2\}$ on the single-qubit state $\rho(\vec{s}_1)$ can be written as
\begin{equation}\label{s03}
	\begin{split}
		\langle M_1 \rangle_{\vec s_1}&=\tr(M_1\rho)=\frac{\sqrt{2}}{2}s_1,\\
		\langle M_2 \rangle_{\vec s_1}&=\tr(M_2\rho)=\frac{\sqrt{2}}{2}s_1,
	\end{split}
\end{equation}
where $s_1=|\vec{s}_1|$ is the magnitude of the polarization vector for the quantum state.

In the main text, we define the $B_2$-orbit realizability condition, which includes the equal predictability
\begin{eqnarray}\label{s04}
    &&\langle M_1\rangle_{\vec{s}_{1}}=\langle M_2\rangle_{\vec{s}_{1}}, \nonumber\\	
    &&\langle M_1\rangle_{\vec{s}_{1}}=-\langle M_1\rangle_{\vec{s}_{2}}=-\langle M_1\rangle_{\vec{s}_{3}}=\langle M_1\rangle_{\vec{s}_{4}},\nonumber\\
	&&\langle M_2\rangle_{\vec{s}_{1}}=\langle M_2 \rangle_{\vec{s}_{2}}=-\langle M_2\rangle_{\vec{s}_{3}}=-\langle M_2\rangle_{\vec{s}_{4}},	
\end{eqnarray}
and the operational equivalent relation in the four-state set preparation
\begin{equation}\label{s05}
	\frac{1}{2}\vec{s}_{1}+\frac{1}{2}\vec{s}_{3}=\frac{1}{2}\vec{s}_{2}+\frac{1}{2}\vec{s}_{4}.
\end{equation}
Then, we can construct the four-state set $\Lambda_{\vec{s}_1}$ via the $B_2$-orbit realizability as illustrated in Fig. 1 of the main text. The preparation procedure of $\{\vec{s}_{1}, \vec{s}_{3}\}$ is denoted by $P$ and the one of $\{\vec{s}_{2}, \vec{s}_{4}\}$ is characterized by $P'$. In contrast to other four-state sets with the $A_1^2$ symmetry \cite{lc22prl-s} which serves for a different task for identifying the nonclassical property of the uncertainty relation, our constructed set $\Lambda_{\vec{s}_1}$ has higher symmetry due to the Coxeter $A_1^2$ group being a subgroup of the Coxeter $B_2$ group \cite{hum90book-s}.

Next, we analyze the preparation contextuality in the four-state set $\Lambda_{\vec{s}_1}$ with the $B_2$ symmetry, and derive the criterion that the procedure of quantum state preparation (QSP) admits of a noncontextual ontological model. Here, without loss of generality, we consider an ontic state space composed of four ontic states \cite{rw05pra-s,srw07pra-s}. Then, associated with the quantum state $\vec{s}_1$, there exists a probability distribution of ontic states, which has a parameterized expression as
\begin{equation}\label{s06}
	\mu _{1}=\begin{pmatrix}a & b & c & d\end{pmatrix},
\end{equation}
where the elements $a,b,c,d\in [0,1]$ are the probabilities for the corresponding ontic states, and the normalization condition of the distribution means
\begin{equation}\label{s07}
	a + b + c + d = 1.
\end{equation}
For a pair of complementary measurements $M_1$ and $M_2$ with the binary outcomes, the conditional probability distribution of the outcomes based on the ontological model \cite{srw07pra-s} can be expressed as
\begin{eqnarray}\label{s08}
	\begin{split}
		&\xi_{+1|M_1}=\begin{pmatrix}0 & 1 & 0 & 1\end{pmatrix}^\top,\\
		&\xi_{-1|M_1}=\begin{pmatrix}1 & 0 & 1 & 0\end{pmatrix}^\top,\\
		&\xi_{+1|M_2}=\begin{pmatrix}1 & 1 & 0 & 0\end{pmatrix}^\top,\\
		&\xi_{-1|M_2}=\begin{pmatrix}0 & 0 & 1 & 1\end{pmatrix}^\top.
	\end{split}
\end{eqnarray}
Then, according to Eqs. \eqref{s06} and \eqref{s08}, the expectation values for the pair of measurements $\{M_1,~M_2\}$ on state $\vec{s}_1$ in the ontic state space have the forms
\begin{equation}\label{s09}
	\begin{split}
		&\left \langle M_1 \right \rangle_{\vec{s}_{1}} =(\xi_{+1|M_1}-\xi_{-1|M_1}) \cdot \mu_{1},\\
		&\left \langle M_2 \right \rangle_{\vec{s}_{1}} =(\xi_{+1|M_2}-\xi_{-1|M_2}) \cdot \mu_{1}.
	\end{split}
\end{equation}
Combining the ontological model explanation in Eq. \eqref{s09} and the normalization condition of the ontic state probability distribution in Eq. \eqref{s07}, we can obtain
\begin{eqnarray}\label{s10}
	a&=&1-c-\frac{1+\left \langle M_1 \right \rangle_{\vec{s}_{1}} }{2},\nonumber\\
	b&=&c+\left \langle M_1 \right \rangle_{\vec{s}_{1}},\nonumber\\
	d&=&1-c-\frac{1+\left \langle M_1 \right \rangle_{\vec{s}_{1}} }{2},
\end{eqnarray}
where we use the $B_2$ symmetry $\left\langle M_1 \right \rangle_{\vec{s}_{1}}=\left\langle M_2 \right \rangle_{\vec{s}_{1}}$. Moreover, in terms of the equal predictability condition in Eq. \eqref{s04}, the ontic state probability distributions corresponding to other three states $\vec{s}_i$ with $i=2, 3, 4$ can be written as
\begin{equation}\label{s11}
	\begin{split}
		&\mu_{2}=\begin{pmatrix}b+\kappa   & a-\kappa   & d-\kappa  & c+\kappa  \end{pmatrix},\\
		&\mu_{3}=\begin{pmatrix}d-\nu   & c+\nu  &b+\nu  & a-\nu  \end{pmatrix},\\
		&\mu_{4}=\begin{pmatrix}c+\tau  & d-\tau   &a-\tau & b+\tau\end{pmatrix},
	\end{split}
\end{equation}
where the parameters $\kappa$, $\nu$, and $\tau$ range in $[-c,1-c]$.

Furthermore, according to the operational equivalence relation of the QSP procedures $P$ and $P'$ in Eq. \eqref{s05}, the assumption of the noncontextual ontological model requires
\begin{equation}\label{s12}
	\frac{1}{2}\mu _{1}+\frac{1}{2}\mu _{3}=\frac{1}{2}\mu _{2}+\frac{1}{2}\mu _{4},
\end{equation}
where the left hand is the probability distribution $\mu(\lambda|P)$ and the right hand represents the probability distribution $\mu(\lambda|P')$. Substituting Eqs. \eqref{s06}, \eqref{s10} and \eqref{s11} into Eq. \eqref{s12}, we can derive the condition
\begin{equation}\label{s13}
	1-2c-\langle M_1\rangle_{\vec{s}_1}-\nu =2c+\langle M_1\rangle_{\vec{s}_1}+\kappa+\tau.
\end{equation}
Therefore, combining the $B_2$ symmetry of our constructed four state set, the preparation noncontextual property implies
\begin{equation}\label{s14}
\begin{split}
	\langle M_1 \rangle_{\vec{s}_1}= \langle M_2 \rangle_{\vec{s}_1}&=\frac{1-4c -(\kappa+\nu+\tau)}{2}\\
&\leq \frac{1-c}{2}\\
&\leq \frac{1}{2},
\end{split}
\end{equation}
where the condition in Eq. \eqref{s13} is used in the first equality, the second inequality is derived by the ranges of parameters $\kappa$, $\nu$, and $\tau$ in Eq. \eqref{s11}, and the third inequality holds when we choose the probability of the ontic state $c=0$. In summary, the preparation procedure of the four-state set $\Lambda_{\vec{s}_1}(\vec{s}_1,\vec{s}_2,\vec{s}_3,\vec{s}_4)$ with the $B_2$ symmetry in quantum theory admits of the noncontextual ontological explanation, when the expectation values of measurements $M_1$ and $M_2$ on $\vec{s}_1$ satisfy the third inequality in Eq. \eqref{s14}.

On the other hand, the four-state set $\Lambda_{\vec{s}_1}(\vec{s}_1,\vec{s}_2,\vec{s}_3,\vec{s}_4)$  exhibits the preparation contextuality when the expectation values obey the following inequality
\begin{equation}\label{s15}
	\langle M_1 \rangle_{\vec{s}_1}= \langle M_2 \rangle_{\vec{s}_1} >\frac{1}{2},
\end{equation}
which indicates that the optimal four-state set related to $\vec{s}_1$ in quantum theory is in conflict with the explanation of noncontextual ontological model given in Eqs. \eqref{s13} and \eqref{s14}.

\begin{figure}
	\epsfig{figure=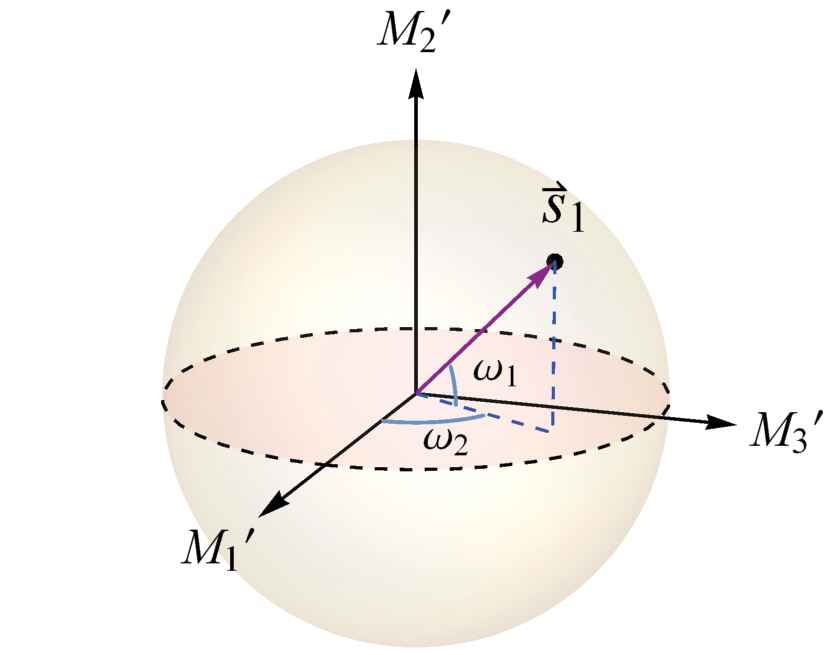,width=0.3\textwidth}
	\caption{The Bloch representation of a single-qubit state in the operator basis $\{M_1',M_3',M_2'\}$, where $\vec{s}_1$ is the polarization vector with the latitude and longitude are $\omega_1$ and $\omega_2$, respectively.}\label{figs1}
\end{figure}

\begin{figure*}
	\epsfig{figure=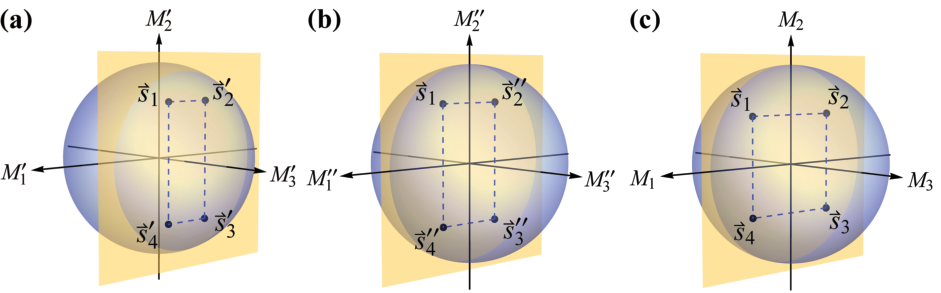,width=0.95\textwidth}
	\caption{Three typical four-state sets with the different symmetries for a given single-qubit state $\vec{s}_1$ in the Bloch sphere. (a) The four-state set $\Lambda'_{\vec{s}_1}(\omega_1\neq 0, \omega_2\neq 0)$ with the $A_1^2$ symmetry. (b) The four-state set $\Lambda''_{\vec{s}_1}(\omega_1\neq \pi/4, \omega_2 =0)$ with the $A_1^2$ symmetry. (c) The optimal four-state set $\Lambda_{\vec{s}_1}(\omega_1=\pi/4, \omega_2= 0)$ with the $B_2$ symmetry.}\label{figs2add}
\end{figure*}

Now, we prove the optimality property of the four-state set $\Lambda_{\vec{s}_1}$ with the $B_2$ symmetry, and show that any other four-state set $\Lambda'_{\vec{s}_1}$ with the $A_1^2$ symmetry admits of the noncontextual ontological model provided that the optimal set is preparation noncontextual. For a given single-qubit state $\vec{s}_1$, the four-state set with the $A_1^2$ symmetry is generated as follows. Take a pair of projection measurements $\{M_1', M_2'\}$, where $M_1'$ is aligned in an arbitrary direction and $M_2'$ is complementary to $M_1'$ along the orthogonal direction. Moreover, by introducing an additional measurement $M_3'$ that is orthogonal to the plane $\hat{M}_1'\hat{M}_2'$, the single-qubit state $\vec{s_1}$ can be expressed as $\rho=\frac{1}{2}(I+\vec s_1\cdot \vec{M}')$, where $\vec M'=\{M_1',M_3',M_2'\}$ is the operator basis of the Bloch sphere. As illustrated in Fig. \ref{figs1}, a schematic diagram is provided where the vector $\vec{s}_1$ can be determined by the vector magnitude $s_1=|\vec{s}_1|$ and the azimuthal angle (latitude $\omega_1$ and longitude $\omega_2$). The expectation values of the pair of complementary measurements $\{M_1',M_2'\}$ on the single-qubit state $\vec{s}_1$ have the forms
\begin{equation}\label{s16}
	\begin{split}
		\langle M_1' \rangle_{\vec s_1}&=\tr(M_1'\rho)=s_1\cos\omega_1\cos\omega_2,\\
		\langle M_2' \rangle_{\vec s_1}&=\tr(M_2'\rho)=s_1\sin\omega_1.
	\end{split}
\end{equation}
Then, according to the reflection actions under the Coxeter group $A_1^2$ \cite{hum90book-s}, the other three qubit states $\vec{s_i}'$ with $i=2, 3, 4$ can be determined which satisfy the $A_1^2$ equal predictability
\begin{eqnarray}\label{s17}
	&&\langle M_1'\rangle_{\vec{s}_{1}}=-\langle M_1'\rangle_{\vec{s_2}'}=-\langle M_1'\rangle_{\vec{s_3}'}=\langle M_1'\rangle_{\vec{s_4}'},\nonumber\\
	&&\langle M_2'\rangle_{\vec{s}_{1}}=\langle M_2' \rangle_{\vec{s_2}'}=-\langle M_2'\rangle_{\vec{s_3}'}=-\langle M_2'\rangle_{\vec{s_4}'},
\end{eqnarray}
and the preparation equivalent relation $\frac{1}{2}\vec{s}_{1}+\frac{1}{2}\vec{s_3}'=\frac{1}{2}\vec{s_2}'+\frac{1}{2}\vec{s_4}'$ similar to Eq. \eqref{s05}. Therefore, based on the $A_1^2$-orbit realizability \cite{lc22prl-s}, a four-state set $\Lambda'_{\vec{s}_1}(\vec{s}_1, \vec{s_2}', \vec{s_3}',\vec{s_4}')$ with the $A_1^2$ symmetry can be constructed, where the preparation procedure of $\{\vec{s}_{1}, \vec{s_{3}}'\}$ is denoted by $P$ and the one of $\{\vec{s_2}', \vec{s_4}'\}$ is characterized by $P'$.

Next, we analyze the preparation contextuality in the four-state set $\Lambda'_{\vec{s}_1}$ with the noncontextual ontological model. For the four-state set $\Lambda'_{\vec{s}_1}$, the preparation equivalence in the ontological model requires $P\simeq P'\Rightarrow \frac{1}{2}\mu_1(\lambda|P)+\frac{1}{2}\mu_3'(\lambda|P)=\frac{1}{2}\mu_2'(\lambda|P')+\frac{1}{2}\mu_4'(\lambda|P')$ in which $\mu_1$ and $\mu'_i$ with $i=2,3,4$ are the corresponding probability distributions of ontic states associated with the four single-qubit states. After a similar analysis as those in Eqs. \eqref{s06}-\eqref{s12}, we can obtain that the four-state set $\Lambda'_{\vec{s}_1}$ with the $A_1^2$ symmetry is preparation noncontextual when the joint predictability of the pair of measurements $\{M_1', M_2'\}$ on $\vec{s}_1$ satisfies
\begin{eqnarray}\label{s18}
	\langle M_1' \rangle_{\vec{s}_1}+\langle M_2' \rangle_{\vec{s}_1}&=&s_1\cos\omega_1\cos\omega_2+s_1\sin\omega_1\nonumber\\
	&\leq& 1.
\end{eqnarray}
Correspondingly, when the joint predictability for the pair of complementary measurements $M_1'$ and $M_2'$ on $\vec{s}_1$ violates the above inequality, i.e.,
\begin{equation}\label{j18}
		\langle M_1' \rangle_{\vec{s}_1}+\langle M_2' \rangle_{\vec{s}_1} > 1,
\end{equation}
the preparation procedure of the four-state set  $\Lambda'_{\vec{s}_1}$ is in conflict with the classical ontological model with preparation equivalence and exhibits the operational contextuality.

In this stage, we study the optimality of our constructed four-state set $\Lambda_{\vec{s}_1}$ with the $B_2$ symmetry by analyzing the maximum of the joint predication of $M'_1$ and $M_2'$ which can be written as
\begin{eqnarray}\label{s19}
	\langle M_1' \rangle_{\vec s_1}+\langle M_2'  \rangle_{\vec s_1}
	&=&s_1\cos\omega_1\cos \omega_2+s_1\sin\omega_1\nonumber\\
    &\leq& s_1\cos\omega_1+s_1\sin\omega_1\nonumber\\
    &\leq& s_1\cos\frac{\pi}{4}+s_1\sin\frac{\pi}{4}\nonumber\\
	&=&\sqrt{2}s_1\nonumber\\
    &=&2\langle M_1 \rangle_{\vec{s}_1},
\end{eqnarray}
where in the first inequality we choose the longitude $\omega_2=0$, the second inequality holds for the latitude $\omega_1=\pi/4$, and in the last equality we use the property in Eq. \eqref{s03}. In Fig. \ref{figs2add}, we present the three typical four-state sets associated with the given single-qubit state $\vec{s}_1$, where the two sets with the $A_1^2$ symmetry are $\Lambda'_{\vec{s}_1}(\omega_1\neq 0, \omega_2\neq 0)$, $\Lambda_{\vec{s}_1}''(\omega_1\neq \pi/4, \omega_2=0)$ and the optimal set with the $B_2$ symmetry is $\Lambda_{\vec{s}_1}(\omega_1=\pi/4, \omega_2= 0)$. Then, combining Eq. \eqref{s19} with the noncontextual criteria in Eqs. \eqref{s14} and \eqref{s18} for the four-state sets with the $B_2$ and $A_1^2$ symmetries respectively, we have the following relation
\begin{equation}\label{z19}
\langle M_1 \rangle_{\vec s_1}=\langle M_2  \rangle_{\vec s_1}\leq 1/2\Rightarrow \langle M_1' \rangle_{\vec s_1}+\langle M_2'  \rangle_{\vec s_1}\leq 1,
\end{equation}
which indicates the optimality of the four-state set $\Lambda_{\vec{s}_1}$ with the $B_2$ symmetry, i.e., any other four-state set $\Lambda'_{\vec{s}_1}$ with the $A_1^2$ symmetry admits of the noncontextual ontological model provided that the optimal set $\Lambda_{\vec{s}_1}$ is preparation noncontextual. The proof of Lemma 1 in the main text is completed. \hfill$\blacksquare$

The above analytical proof for Lemma 1 in the main text shows the optimality of the four-state set $\Lambda(\vec{s}_1)$ associated with any given single-qubit state. In addition, as a byproduct of this Lemma, we can obtain the following Corollary.

\emph{Corollary 1.}---For any given single-qubit state $\rho(\vec{s}_1)$, the optimal four-state set $\Lambda_{\vec{s}_1}$ with the $B_2$ symmetry is noncontextual and admits of a generalized ontological model with preparation equivalence when the measurement expectations satisfy
\begin{equation}\label{c21}
	\langle M_1 \rangle_{\vec{s}_1}= \langle M_2 \rangle_{\vec{s}_1} \leq\frac{1}{2},
\end{equation}
and the optimal four-state set is in conflict with the classical ontological model and exhibit the quantum contextuality in the preparation procedure when the measurement expectations obey
\begin{equation}\label{c22}
	\langle M_1 \rangle_{\vec{s}_1}= \langle M_2 \rangle_{\vec{s}_1} >\frac{1}{2},
\end{equation}
where the expressions of the complementary measurements $M_1$ and $M_2$ have the form shown in Eq. (1) of the main text. Moreover, for any four-state set $\Lambda'_{\vec{s}_1}$ with the $A_1^2$ symmetry associated with an arbitrary pair of complementary measurements $M_1'$ and $M_2'$, it is noncontextual and has a classical explanation of the ontological model with preparation equivalence when $\langle M_1' \rangle_{\vec{s}_1}+ \langle M_2' \rangle_{\vec{s}_1} \leq 1$, and the set $\Lambda'_{\vec{s}_1}$ exhibits preparation contextuality when $\langle M_1' \rangle_{\vec{s}_1}+ \langle M_2' \rangle_{\vec{s}_1} >1$.

\emph{Proof.} This corollary follows from Eqs. \eqref{s14}-\eqref{s15} and \eqref{s18}-\eqref{j18} in the proof of Lemma 1. \hfill$\blacksquare$

\begin{figure}
	\epsfig{figure=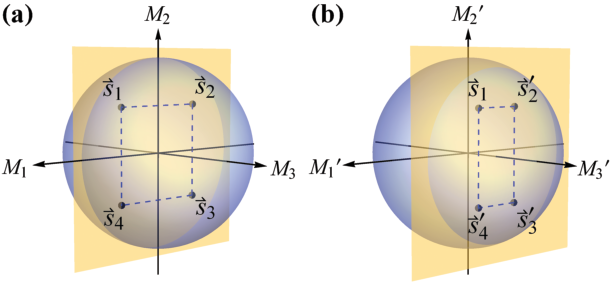,width=0.48\textwidth}
	\caption{Two four-state sets for the given quantum state in Eq. \eqref{z20}. (a) The optimal four-state set $\Lambda_{\vec{s}_1}(\vec{s}_1, \vec{s_2}, \vec{s_3},\vec{s_4})$ generated by the $B_2$-orbit realizability. (b) The four-state set $\Lambda_{\vec{s}_1}'(\vec{s}_1, \vec{s_2}', \vec{s_3}',\vec{s_4}')$ generated by the $A_1^2$-orbit realizability.}\label{figs2}
\end{figure}

Next, we investigate a concrete example for the optimality of the four-state set with the $B_2$ symmetry. We consider a single-qubit state $\rho(\vec{s}_1)$ with the given polarization vector $\vec{s_1}$, which has the following form in the Bloch representation
 \begin{equation}\label{z20}
 	\rho(\vec{s}_1)=\frac{1}{2}(I+\frac{1}{4} X+\frac{\sqrt{3}}{4} Y+\frac{1}{2} Z ),
 \end{equation}
where $X$, $Y$ and $Z$ denote the Pauli matrices of corresponding directions, respectively. After some calculation, we have the norm of the polarization is
\begin{equation}\label{z22}
	s_1=|\vec{s}_1|=\frac{\sqrt{2}}{2}.
\end{equation}
For this quantum state, two measurements $M_1$ and $M_2$ can be constructed according to the method given in Eq. (4) of the main text, then combining the measurement $M_3$ orthogonal to the $\hat{M}_1\hat{M}_2$ plane we obtain a set of operator basis which can be written as
\begin{eqnarray}\label{z23}
	M_1&=&\frac{1}{2}X+\frac{\sqrt{3}}{2}Y, \nonumber\\
	M_2&=&Z, \nonumber\\
	M_3&=&-\frac{\sqrt{3}}{2}X+\frac{1}{2}Y.
\end{eqnarray}
In the new operator basis $\{M_1, M_3, M_2\}$, the single-qubit state can be written as
\begin{equation}\label{s21}
	\rho(\vec{s}_1)=\frac{I+\frac{1}{2} M_1+\frac{1}{2} M_2}{2},
\end{equation}
and using the $B_2$-orbit realizability condition we can generate the optimal four-state set $\Lambda_{\vec{s}_1}(\vec{s}_1, \vec{s_2}, \vec{s_3},\vec{s_4})$ as illustrated in Fig. \ref{figs2}(a). The preparation contextuality in the optimal four-state set can be witnessed by the expectation of measurements $M_1$ and $M_2$. After some calculation, we can obtain
\begin{eqnarray}\label{s22}
	\langle M_1 \rangle_{\vec s_1}=\langle M_2 \rangle_{\vec s_1}=\frac{\sqrt{2}}{2}s_1=\frac{1}{2}.
\end{eqnarray}
According to the criterion in Eq. \eqref{c21} in Corollary 1, we have the result that the optimal four-state set $\Lambda_{\vec{s}_1}(\vec{s}_1, \vec{s_2}, \vec{s_3},\vec{s_4})$ is noncontextual in the QSP and admits of a classical ontological model with preparation equivalence.

Moreover, we investigate another four-state set of the same quantum state in Eq. \eqref{z20}, which has the $A_1^2$ symmetry. Without loss of generality, we can choose the operator basis $\{M_1'=X, M_3'=Y, M_2'=Z\}$. As illustrated in Fig. \ref{figs2}(b), the four-state set $\Lambda_{\vec{s}_1}'(\vec{s}_1, \vec{s_2}', \vec{s_3}',\vec{s_4}')$ of $\vec{s}_1$ with the $A^2_1$ symmetry can be generated by the $A_1^2$-orbit realizability \cite{lc22prl-s}. The preparation contextuality in the four state set  $\Lambda_{\vec{s}_1}'$ can be witnessed by the joint expectations of measurements $M_1'$ and $M_2'$, and we can derive the following result
\begin{eqnarray}\label{s20}
	\langle M_1' \rangle_{\vec s_1}+\langle M_2' \rangle_{\vec s_1}&=&s_1\cos\frac{\pi}{4}\cos\frac{\pi}{3}+s_1\sin\frac{\pi}{4}\nonumber\\
	&=&\frac{3}{4}.
\end{eqnarray}
Thus, in terms of the criterion in Eq. \eqref{s18}, the four-state set with the $A_1^2$ symmetry $\Lambda_{\vec{s}_1}'(\vec{s}_1, \vec{s_2}', \vec{s_3}',\vec{s_4}')$ is preparation noncontextual and admits a classical ontological model.

In summary, for the given single-qubit state $\rho(\vec{s}_1)$ in Eq. \eqref{z20}, we construct two four-state sets with the $B_2$ symmetry and the $A_1^2$ symmetry as shown in Fig. \ref{figs2}. The results in Eqs. \eqref{s22} and \eqref{s20} further verify Lemma 1 in the main text: i.e., any other four-state set with the $A_1^2$ symmetry admits of a classical ontological explanation provided that the optimal set generated by $B_2$-orbit realizability is preparation noncontextual.


\section{Proof of Theorem 1}
\label{THEOREM 1}
In the main text, Theorem 1 describes the criteria for witnessing the preparation contextuality in the optimal four-state set associated with any given single-qubit via the EUR of two complementary measurements. This section is composed of three parts. In Sec. IIA, we provide an analytical proof for the faithful criterion in Theorem 1, the proof of the sufficient criterion in Theorem 1 is given in Sec. IIB, and in Sec. IIC we discuss the difference of our faithful criterion from the classical bound for UR derived by Catani \textit{et al} in Ref. \cite{lc22prl-s} for a different task.

\subsection{Proof of the faithful criterion in Theorem 1}
\label{THEOREM 1A}
In Theorem 1 of the main text, the faithful criterion for witnessing the preparation contextuality in the optimal four-state set $\Lambda_{\vec{s}_1}$ is stated as follows.

\emph{The faithful criterion in Theorem 1}.---For a single-qubit state $\vec{s}_{1}$, the four-state set $\Lambda_{\vec{s}_{1}}$ exhibits the preparation contextuality if and only if the EUR resulting from a pair of optimal complementary observables $Q$ and $R$ on $\vec{s}_1$ satisfies
\begin{equation}\label{s23}
	H(Q)+H(R)< 1+C,
\end{equation}
where $H(Q)$ and $H(R)$ are the Shannon entropies of the post-measurement states, and the constant is a binary entropy function $C=h[(2-\sqrt{2})/4]\simeq 0.6009$.

\textit{Proof.} We consider a single-qubit state represented in the Bloch sphere as
\begin{equation}\label{j23}
	\rho(\vec{s}_1)=\frac{I+\vec{s}_1\cdot \vec{M}}{2},
\end{equation}
where $\vec{s}_1$ is the polarization vector and the operator basis is $\{M_1, M_3, M_2\}$ as shown in Fig. 1 of the main text. According to the $B_2$-orbit realizability condition given in Eqs. (5) and (6) of the main text, we can generate the optimal four-state set $\Lambda_{\vec{s}_1}$ with the $B_2$ symmetry. Moreover, in terms of the Lemma 1 in the main text, the optimality property of the four-state set $\Lambda_{\vec{s}_1}$ lies that any other four-state set $\Lambda'_{\vec{s}_1}$ with the $A_1^2$ symmetry admits of a classical noncontextual explanation provided that the optimal one $\Lambda_{\vec{s}_1}$ with the $B_2$ symmetry is preparation noncontextual. In particular, the quantitative expression for this optimality is given by Eq. \eqref{z19} in Sec. I of this Supplemental Material.

Now, we analyze the two complementary measurements $Q$ and $R$ of the entropic uncertainty relation (EUR) in Eq. \eqref{s23}. Based on the relation in Eq. (4) of the main text, we can derive
\begin{eqnarray}\label{s26}
	Q&=&\frac{\sqrt{2}}{2}\left(M_1+M_2\right),\nonumber\\
	R&=&\frac{\sqrt{2}}{2}\left(-M_1+M_2\right),
\end{eqnarray}
where $M_1$ and $M_2$ compose the operator basis of the Bloch sphere in Fig. 1 of the main text. For the two-outcome measurement $Q$, we can express it in the diagonal representation
\begin{eqnarray}\label{j33}
	Q=Q_+-Q_-,	
\end{eqnarray}
where $Q_+$ and $Q_-$ are the eigenvectors with the eigenvalues $+1$ and $-1$ respectively, and have the forms
\begin{eqnarray}\label{j34}
	Q_+&=&\frac{I+\frac{\sqrt{2}}{2}M_1+\frac{\sqrt{2}}{2}M_2}{2},\nonumber\\
	Q_-&=&\frac{I-\frac{\sqrt{2}}{2}M_1-\frac{\sqrt{2}}{2}M_2}{2}.
\end{eqnarray}
For the single-qubit state $\rho(\vec{s}_1)$ with the given polarization vector $\vec{s}_1$ in Eq. \eqref{j23}, the Shannon function of the measurement $Q$ can be expressed as
\begin{eqnarray}\label{s27}
	H(Q)=h(q)=-q\mbox{log}_2q-(1-q)\mbox{log}_2(1-q),
\end{eqnarray}
where, without loss of generality, we set $q\leq (1-q)$ and let $q$ be the probability of the outcome for eigenvector $Q_-$. Then, according to Eqs. \eqref{j23} and \eqref{j34}, we have
\begin{eqnarray}\label{s28} 
	q=\tr\left[\rho(\vec{s}_1)Q_-\right]=\frac{1}{2}-\frac{\sqrt{2}}{2}\langle M_1\rangle_{\vec{s}_1},
\end{eqnarray}
where $\langle M_1\rangle_{\vec{s}_1}$ is the expectation value of the measurement $M_1$ on the single-qubit state $\rho(\vec{s}_1)$, and we use the equal predictability $\langle M_1\rangle_{\vec{s}_1}=\langle M_2\rangle_{\vec{s}_1}$ of the Coxeter group $B_2$ given in Eq. (5) of the main text. Substituting Eq. \eqref{s28} into Eq. \eqref{s27}, we have the Shannon entropy
\begin{equation}\label{j37}
	H(Q)=h\left(\frac{1}{2}-\frac{\sqrt{2}}{2}\langle M_1\rangle_{\vec{s}_1}\right),
\end{equation}
where $h(\cdot)$ is the binary entropy function.

Similarly, the two-outcome measurement $R$ in its diagonal representation can be written as
\begin{equation}\label{j38}
	R=R_+-R_-,
\end{equation}
where two eigenvectors with eigenvalues $\pm 1$ have the forms
\begin{eqnarray}\label{j39}
	R_+&=&\frac{I+\frac{-\sqrt{2}}{2}M_1+\frac{\sqrt{2}}{2}M_2}{2},\nonumber\\
	R_-&=&\frac{I+\frac{\sqrt{2}}{2}M_1-\frac{\sqrt{2}}{2}M_2}{2}.
\end{eqnarray}
For the single-qubit state $\rho(\vec{s}_1)$, the Shannon function of the measurement $R$ can be expressed as
\begin{eqnarray}\label{s29}
	H(R)=h(r)=-r\mbox{log}_2r-(1-r)\mbox{log}_2(1-r),
\end{eqnarray}
where we let $r$ be the probability of the outcome for eigenvector $R_-$. Combining Eq. \eqref{j23} and Eq. \eqref{j39}, we can derive
\begin{eqnarray}\label{s30}
    r=\tr\left[\rho(\vec{s}_1)R_-\right]=\tr\left(\frac{I}{4}\right)=\frac{1}{2},
\end{eqnarray}
where we use the equal predication property $\langle M_1\rangle_{\vec{s}_1}-\langle M_2\rangle_{\vec{s}_1}=0$. Substituting Eq. \eqref{s30} to Eq. \eqref{s29}, we can obtain the Shannon entropy
\begin{equation}\label{s31}
	H(R)=h(r)=h\left(\frac{1}{2}\right)=1.	
\end{equation}
Therefore, based on Eqs. \eqref{j37} and \eqref{s31}, the EUR of the pair of measurements $Q$ and $R$ can be expressed as
\begin{equation}\label{s32}
	H(Q)+H(R)=1+h\left(\frac{1}{2}-\frac{\sqrt{2}}{2}\langle M_1\rangle_{\vec{s}_1}\right).
\end{equation}

\begin{figure}
	\epsfig{figure=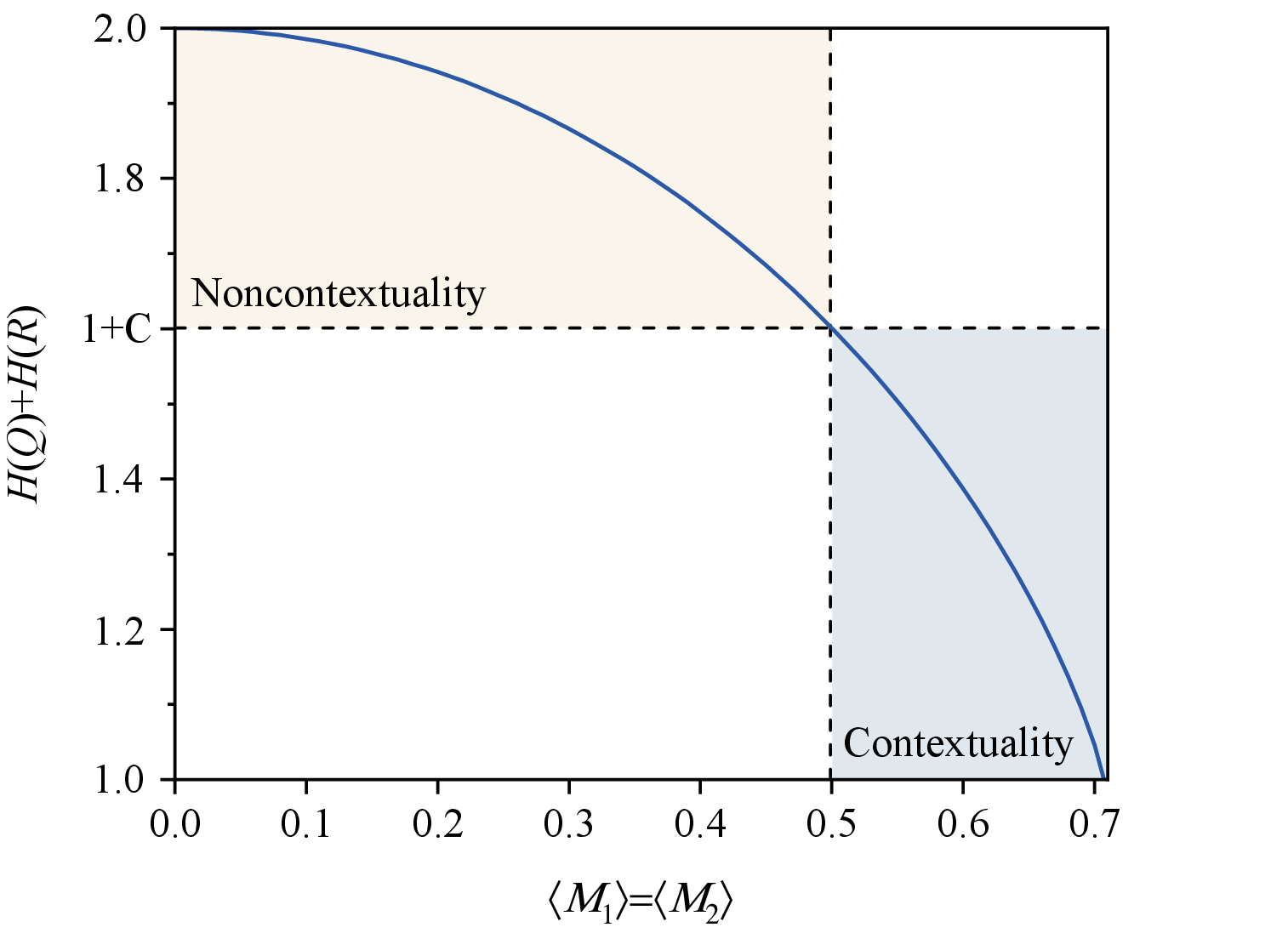,width=0.45\textwidth}
	\caption{The quantitative illustration of the faithful criterion in Theorem 1, where the EUR $H(Q)+H(R)$ (the blue solid line) in the upper left region (the light beige shaded area) indicates the noncontextuality in the quantum state preparation (QSP) of $\Lambda_{\vec{s}_1}$ and the EUR in lower right region (the light blue shaded area) denotes the preparation contextuality of the set $\Lambda_{\vec{s}_1}$.}
	\label{figs3add}
\end{figure}

In the optimal four-state set $\Lambda_{\vec{s}_1}$, the expectation values of two measurements have the property $\langle M_1\rangle_{\vec{s}_{1}}=\langle M_2\rangle_{\vec{s}_{1}}$ due to the $B_2$ symmetry, and both values range in $[0, \sqrt{2}/2]$. After some analysis on the property of binary entropy function, we can obtain that the EUR in Eq. \eqref{s32} is a monotonically decreasing function of the expectation value $\langle M_1\rangle_{\vec{s}_1}$. In Fig. \ref{figs3add}, the EUR $H(Q)+H(R)$ is plotted as a function of the expectation value of the measurement $M_1$ on the single-qubit state (the blue solid line), which further illustrates the monotonically property.

Next, based on Eq. \eqref{s32} and its monotonic property, we prove the faithful criterion in Theorem 1 of the main text. As shown in Eq. \eqref{s23}, when the criterion is satisfied, we have
\begin{eqnarray}\label{j44}
	&&H(Q)+H(R)<1+C\nonumber\\	
	&&\Rightarrow \frac{1}{2}-\frac{\sqrt{2}}{2}\langle M_1\rangle_{\vec{s}_{1}}<\frac{1}{2}-\frac{\sqrt{2}}{4}\nonumber\\
	&&\Rightarrow \langle M_1\rangle_{\vec{s}_{1}} >\frac{1}{2}\nonumber\\
	&&\Rightarrow \langle M_1\rangle_{\vec{s}_{1}} =\langle M_2\rangle_{\vec{s}_{1}}>\frac{1}{2},
\end{eqnarray}
where in the second inequality we use Eq. \eqref{s32} and the monotonic property of binary function $h(\cdot)$ as well as the constant $C=h[(2-\sqrt{2})/4]$, and in the last equation we use the $B_2$ symmetry $\langle M_1\rangle_{\vec{s}_{1}}=\langle M_2\rangle_{\vec{s}_{1}}$. According to the Eq. \eqref{c22} in Corollary 1, we can obtain that, when the EUR $H(Q)+H(R)<1+C$, the optimal four-state set $\Lambda_{\vec{s}_1}$ is preparation contextual and cannot be explained by the generalized noncontextual ontological model.

On the other hand, when the criterion in Eq. \eqref{s23} is violated, we can derive
\begin{eqnarray}\label{j45}
&&H(Q)+H(R)\geq 1+C\nonumber\\	
&&\Rightarrow \frac{1}{2}-\frac{\sqrt{2}}{2}\langle M_1\rangle_{\vec{s}_{1}}\geq\frac{1}{2}-\frac{\sqrt{2}}{4}\nonumber\\
&&\Rightarrow \langle M_1\rangle_{\vec{s}_{1}}\leq\frac{1}{2}\nonumber\\
&&\Rightarrow \langle M_1\rangle_{\vec{s}_{1}} =\langle M_2\rangle_{\vec{s}_{1}}\leq\frac{1}{2},
\end{eqnarray}
where we use the expression in Eq. \eqref{s32}, the property of binary entropy function, and the equal prediction of the $B_2$ symmetry. Thus, according to the Eq. \eqref{c21} in Corollary 1, we have the conclusion that when the EUR in Eq. \eqref{s23} is violated (\textit{i.e}., $H(Q)+H(R)\geq1+C$), the optimal four-state set $\Lambda_{\vec{s}_1}$ is noncontextual and admits of a classical ontological model with preparation equivalence.

In Fig. \ref{figs3add}, the corresponding relation captured by Eqs. \eqref{j44}-\eqref{j45} between the faithful criterion via the EUR and the expectation values $\langle M_1\rangle_{\vec{s}_{1}}=\langle M_2\rangle_{\vec{s}_{1}}$ in Corollary 1 is plotted. The EUR $H(Q)+H(R)$ is plotted as a monotonically decreasing function of the expectation value $\langle M_1\rangle_{\vec{s}_{1}}$ (the blue solid line), where the expectation values $\langle M_1\rangle_{\vec{s}_{1}}=\langle M_2\rangle_{\vec{s}_{1}}$ ranges in $[0, \sqrt{2}/2]$ due to the $B_2$ symmetry. As illustrated in the figure, when $H(Q)+H(R)< 1+C$, it follows that $\langle M_1\rangle_{\vec{s}_{1}}=\langle M_2\rangle_{\vec{s}_{1}}>1/2$ which corresponds to the light blue shaded region indicating the preparation contextuality of the optimal set $\Lambda_{\vec{s}_1}$; when $H(Q)+H(R)\geq 1+C$, it follows that $\langle M_1\rangle_{\vec{s}_{1}}=\langle M_2\rangle_{\vec{s}_{1}}\leq 1/2$ which corresponds to the light beige shaded region denoting the noncontextuality in the preparation of the optimal set $\Lambda_{\vec{s}_1}$. Moreover, we refer to the pair of complementary measurements $Q$ and $R$ as the optimal ones, since the criterion $H(Q)+H(R)< 1+C$ is faithful to witness the preparation contextuality in the optimal four-state set $\Lambda_{\vec{s}_1}$ generated by the $B_2$-orbit realizability.

In summary, by comparing Eqs. \eqref{j44}-\eqref{j45} with Corollary 1 in Sec. I of the SM, we prove that the criterion in Eq. \eqref{s23} is a faithful one to witness the preparation contextuality in the optimal four-state set $\Lambda_{\vec{s}_1}$ for any given single-qubit state $\vec{s}_1$, which is quantitatively illustrated in Fig. \ref{figs3add}. The proof of the faithful criterion in Theorem 1 is completed.
 \hfill$\blacksquare$

\subsection{Proof of the sufficient condition in Theorem 1 and the explanation of  invalid region}
\label{THEOREM 1B}

In Theorem 1 of the main text, besides the faithful criterion via the EUR, we also present a sufficient criterion to witness the preparation contextuality in the optimal four-state set $\Lambda_{\vec{s}_1}$. In this subsection, we first provide an analytical  proof for the sufficient condition, and then give an explanation on the invalid region (the orange area) in Fig. 2 of the main text. The sufficient criterion in Theorem 1 is stated as follows.

\emph{The sufficient criterion in Theorem 1.}--- For any pair of complementary measurements $Q'$ and $R'$, the optimal set $\Lambda_{{s}_1}$ exhibits preparation contextuality when the sufficient condition
\begin{equation}\label{s35} 
	H(Q')<C~~\mbox{or}~~ H(R')<C
\end{equation}
for $\vec{s}_{1}$ is satisfied.

\begin{figure}
	\epsfig{figure=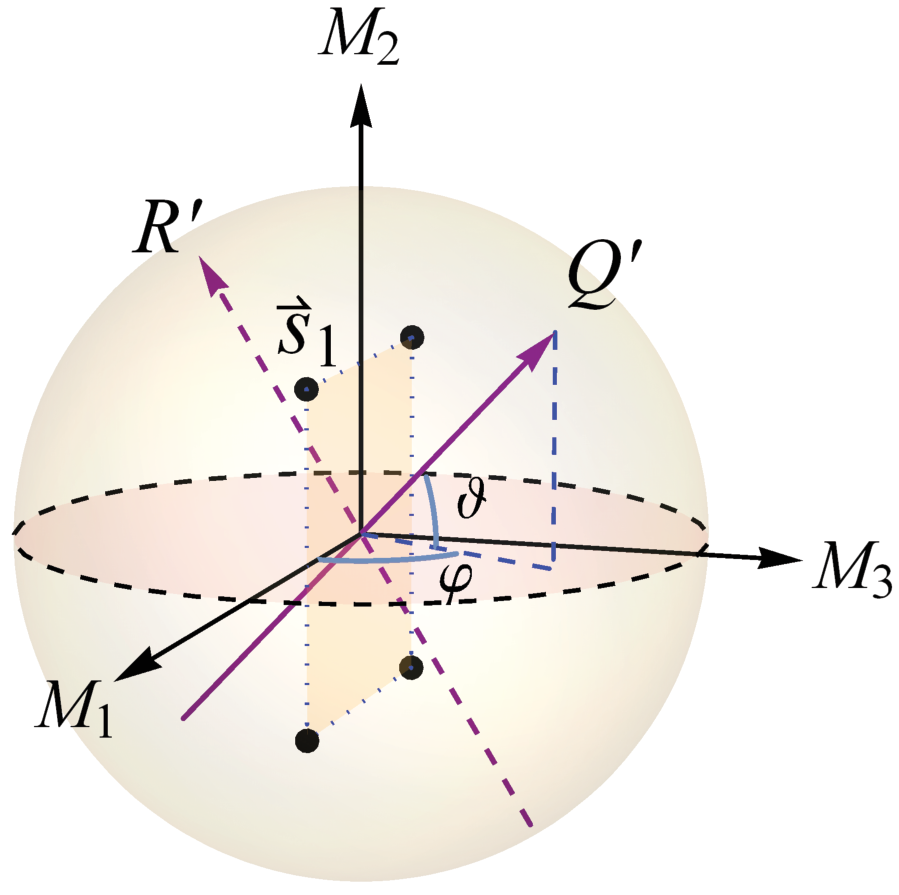,width=0.3\textwidth}
	\caption{The optimal four-state set $\Lambda_{\vec{s}_{1}}$ with the $B_2$ symmetry in the Bloch sphere with the operator basis $\{M_1, M_3, M_2\}$, and $Q'$ and $R'$ are a pair of arbitrary complementary measurements.}\label{figs3}
\end{figure}

\textit{Proof.} For an arbitrary pair of complementary measurements $Q'$ and $R'$ in the Bloch representation, they can be expressed as
\begin{equation}\label{s36} 
	\begin{split}
		Q'&=\cos\vartheta (\cos\varphi M_1 +\sin\varphi M_3)+\sin\vartheta M_2,\\
		R'&=\cos\vartheta_1 (\cos\varphi_1 M_1 +\sin\varphi_1 M_3)+\sin\vartheta_1 M_2,
	\end{split}
\end{equation}
where $(\vartheta, \varphi)$ is the azimuthal angle (latitude $\vartheta$ and longitude $\varphi$) for the measurement $Q'$ as shown in Fig. \ref{figs3}, and $(\vartheta_1, \varphi_1)$ is the azimuthal angle for the measurement $R'$. Moreover, due to the complementary property of measurements $Q'$ and $R'$, these angle parameters in Eq. \eqref{s36} should satisfy the following relation
\begin{eqnarray}\label{s37} 
	&&\cos\vartheta \cos\varphi\cos\vartheta_1 \cos\varphi_1+\cos\vartheta\sin\varphi\cos\vartheta_1\sin\varphi_1\nonumber\\
	&&+\sin\vartheta\sin\vartheta_1=0.
\end{eqnarray}

The Shannon function for the measurement $Q'$ on $\vec{s}_1$ can be written as
\begin{eqnarray}\label{s38} 
	H(Q')&=&h(q')\nonumber\\
         &=&-q'\mbox{log}_2q'-(1-q')\mbox{log}_2(1-q'),
\end{eqnarray}
where, without loss of generality, we set $q'\leq (1-q')$ and let $q'$ be the probability for the outcome of eigenvector $Q_+'$ with eigenvalue being $+1$. After some derivation, we can obtain the probability
\begin{eqnarray}\label{s39}
	q'&=&\tr\left[\rho(\vec{s}_1) Q_+'\right]\nonumber\\
	&=&\frac{1}{2}[1-|(\cos\vartheta \cos\varphi+\sin\vartheta)\langle M_1\rangle_{\vec{s}_{1}}|],
\end{eqnarray}
where we use the expression of single-qubit state $\rho(\vec{s}_1)$ in Eq. \eqref{j23}, the eigenvector $Q_+'=[I+\cos\vartheta\cos\varphi M_1+\sin\vartheta M_2+\cos\vartheta\sin\varphi M_3]/2$, and the $B_2$ symmetry $\langle M_1\rangle_{\vec{s}_{1}}=\langle M_2\rangle_{\vec{s}_{1}}$ in Eq. (5) of the main text. Substituting Eq. \eqref{s39} into Eq. \eqref{s38}, we have
\begin{eqnarray}\label{j51} 
	H(Q')&=&h\left(\frac{1-|(\cos\vartheta \cos\varphi+\sin\vartheta)\langle M_1\rangle_{\vec{s}_{1}}|}{2}\right),
\end{eqnarray}
where $h(\cdot)$ is the binary entropy function. In this case, when the sufficient condition for the measurement $Q'$ is satisfied, we can obtain
\begin{eqnarray}\label{j52} 
	&&H(Q')<C\nonumber\\
	&\Rightarrow& h\left(\frac{1-|(\cos\vartheta \cos\varphi+\sin\vartheta)\langle M_1\rangle_{\vec{s}_{1}}|}{2}\right)\nonumber\\
	&&        <h\left(\frac{1}{2}-\frac{\sqrt{2}}{4}\right)\nonumber\\
	&\Rightarrow& |(\cos\vartheta \cos\varphi+\sin\vartheta)\langle M_1\rangle_{\vec{s}_{1}}|> \frac{\sqrt{2}}{2}\nonumber\\
	&\Rightarrow& |(\cos\vartheta+\sin\vartheta) \langle M_1\rangle_{\vec{s}_{1}}| > \frac{\sqrt{2}}{2}\nonumber\\
	&\Rightarrow& \langle M_1\rangle_{\vec{s}_{1}}=\langle M_2\rangle_{\vec{s}_{1}} > \frac{1}{2},
\end{eqnarray}
where in the second inequality we use the expression of $H(Q')$ in Eq. \eqref{j51} and the constant $C=h[(1-\sqrt{2}/2)/2]$, in the third inequality we use the monotonically decreasing property of the binary function, in the fourth inequality the maximum of $\mbox{cos}\varphi$ for $\varphi=0$ is chosen, and the last inequality holds for the maximum of $\cos\vartheta+\sin\vartheta$ with $\vartheta=\pi/4$ and the $B_2$ symmetry  $\langle M_1\rangle_{\vec{s}_{1}}=\langle M_2\rangle_{\vec{s}_{1}}$. According to the Corollary 1 in Sec. I of this SM, we have the result via Eq. \eqref{j52} that, when the sufficient criterion $H(Q')<C$ is satisfied, the optimal four-state set $\Lambda_{\vec{s}_1}$ exhibits the preparation contextuality.

In a similar way, for the measurement $R'$, we have
\begin{eqnarray}\label{s40} 
	H(R')=h(r')=-r'\mbox{log}_2r'-(1-r')\mbox{log}_2(1-r'),
\end{eqnarray}
where we set $r'\leq (1-r')$ being the probability of eigenvalue $+1$ for eigenvector $R_+'=[I+\cos\vartheta_1\cos\varphi_1 M_1+\sin\vartheta_1 M_2+\cos\vartheta_1\sin\varphi_1 M_3]/2$. After some derivation, we can obtain
\begin{eqnarray}\label{s41} 
	r'&=&\tr\left[\rho(\vec{s}_1) R_+'\right]\nonumber\\
	&=&\frac{1}{2}[1-|(\cos\vartheta_1 \cos\varphi_1+\sin\vartheta_1)\langle M_1\rangle_{\vec{s}_1}|],
\end{eqnarray}
where we use the Eq. \eqref{j23} and the $B_2$ symmetry $\langle M_1\rangle_{\vec{s}_{1}}=\langle M_2\rangle_{\vec{s}_{1}}$ in Eq. (5) in the main text. Substituting Eq. \eqref{s41} into Eq. \eqref{s40}, we have
\begin{eqnarray}\label{j55}
	H(R')=h\left(\frac{1-|(\cos\vartheta_1\cos\varphi_1+\sin\vartheta_1)\langle M_1\rangle_{\vec{s}_1}|}{2}\right).
\end{eqnarray}
Therefore, when the sufficient criterion for the measurement $R'$ is satisfied, we can obtain
\begin{eqnarray}\label{j56}
	&&H(R')<C\nonumber\\
	&\Rightarrow& h\left(\frac{1-|(\cos\vartheta_1 \cos\varphi_1+\sin\vartheta_1)\langle M_1\rangle_{\vec{s}_1}|}{2}\right)\nonumber\\
	&&<h\left(\frac{1}{2}-\frac{\sqrt{2}}{4}\right)\nonumber\\
	&\Rightarrow& |(\cos\vartheta_1 \cos\varphi_1+\sin\vartheta_1)\langle M_1\rangle_{\vec{s}_{1}}|> \frac{\sqrt{2}}{2}\nonumber\\
	&\Rightarrow& |(\cos\vartheta_1+\sin\vartheta_1) \langle M_1\rangle_{\vec{s}_{1}}| > \frac{\sqrt{2}}{2}\nonumber\\
	&\Rightarrow& \langle M_1\rangle_{\vec{s}_{1}}=\langle M_2\rangle_{\vec{s}_{1}} > \frac{1}{2},
\end{eqnarray}
where in the second inequality we employ the expression for $H(R')$ in Eq. \eqref{j55}, in the third inequality we leverage the monotonically decreasing property of the binary entropy function, in the fourth inequality we select the maximum value of
$\cos\varphi_1$ with $\varphi_1=0$, and in the last inequality we use the $B_2$ symmetry condition $\langle M_1\rangle_{\vec{s}_{1}}=\langle M_2\rangle_{\vec{s}_{1}}$  and the property that $\cos\vartheta_1+\sin\vartheta_1$ attains its maximum at $\vartheta_1=\pi/4$. According to Corollary 1 in Sec. I of this SM, we conclude that when the sufficient criterion $H(R')<C$ is satisfied, the optimal four-state set $\Lambda(\vec{s}_{1})$ is preparation contextual.

In conclusion, combining Eq. \eqref{j52}, Eq. \eqref{j56} and the Corollary 1 in Sec. I of the SM, we can obtain that $H(Q')<C$ or $H(R')<C$ for any pair of complementary measurements $Q'$ and $R'$ is a sufficient criterion for witnessing the preparation contextuality in the optimal four-state set $\Lambda_{\vec{s}_1}$. Then, the proof of the sufficient criterion in Theorem 1 is completed.
\hfill$\blacksquare$

In Fig. 2 of the main text, the orange area denotes \textit{``invalid region"} which corresponds to the Shannon entropies violating of the sufficient criterion in Theorem 1 of the main text, \textit{i.e.}, $H(Q')\geq C~\mbox{and}~H(R')\geq C$.

In the following analysis, we will give the explanation on invalid region of Shannon entropies $H(Q')$ and $(R')$, where the detection of preparation contextuality in the optimal four-state set $\Lambda_{\vec{s}_1}$ is inconclusive according to a pair of arbitrary complementary measurements $Q'$ and $R'$.

Based on Eq. \eqref{j51}, we can derive
\begin{eqnarray}\label{s45} 
	H(Q')&=&h\left(\frac{1-|(\cos\vartheta \cos\varphi+\sin\vartheta)\langle M_1\rangle_{\vec{s}_{1}}|}{2}\right)\nonumber\\
	&\geq& h(\frac{1-\sqrt{2}\langle M_1\rangle_{\vec{s}_{1}}}{2})\nonumber\\
	&=&  H(Q),
\end{eqnarray}
where in the first inequality we use the monotonically decreasing property of the binary entropy as a function of the absolute value $|(\cos\vartheta \cos\varphi+\sin\vartheta)\langle M_1\rangle_{\vec{s}_{1}}|$ with the parameters chosen to be $\varphi=0$ and $\vartheta=\pi/4$, and the last equality holds due to Eq. \eqref{j37} with $Q$ being the optimal measurement. In addition, when the sufficient condition in Eq. \eqref{s35} is violated, the entropy for the measurement $R'$ has the property
\begin{equation}\label{d58} 
    C\leq H(R')\leq 1=H(R),
\end{equation}
where $R$ is the optimal measurement and we use the relation in Eq. \eqref{s31}. Furthermore, combining Eq. \eqref{s45} and Eq. \eqref{d58}, we can derive
\begin{eqnarray}\label{s46}
   H(Q')\geq C &\mbox{and}& H(R')\geq C \nonumber\\
	&\nRightarrow& H(Q)+H(R)\geq 1+C,
\end{eqnarray}
which means that when the sufficient criterion in Eq. \eqref{s35} is violated, the optimal four-state set $\Lambda_{\vec{s}_1}$ is not necessarily noncontextual in terms of the faithful criterion in Eq. \eqref{s23}.

\begin{figure}
	\epsfig{figure=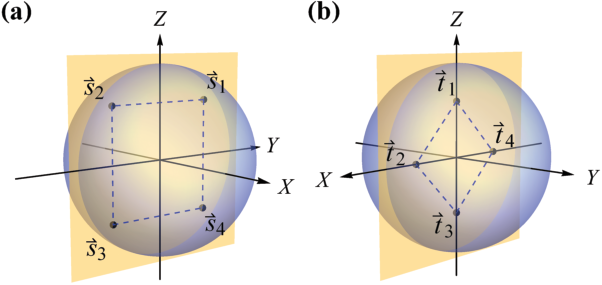,width=0.48\textwidth}
	\caption{Two optimal four-state sets $\Lambda_{\vec{s}_1}$ in panel (a) and $\Lambda_{\vec{t}_1}$ in panel (b), where both the two sets does not satisfy the sufficient criterion in Eq. \eqref{s35}. The set $\Lambda_{\vec{s}_1}$ is preparation contextual but the set $\Lambda_{\vec{t}_1}$ is noncontextual, which exemplifies that the detection based non-optimal measurement $Q'$ and $R'$ fails and is inconclusive when the sufficient criterion loses its efficacy. }\label{figs4add}
\end{figure}

Next, as a concrete example, we consider a single-qubit state with the polarization vector $\vec{s}_1=(0,3/5,3/5)$, and analyze a pair of non-optimal measurements $Q'$ and $R'$ which can be written as
\begin{eqnarray}\label{j60} 
   Q'=(X+Z)/\sqrt2,~~R'=(X-Z)/\sqrt2,
\end{eqnarray}
where $X$ and $Z$ are Pauli matrices serving as the operator basis in Bloch representation. For the given qubit state $\rho(\vec{s}_1)$, we can generate the optimal four-state set $\Lambda_{\vec{s}_1}$ by the $B_2$-orbit realizability presented in the main text. The four quantum states in the optimal set $\Lambda_{\vec{s}_1}=\{\vec{s}_1, \vec{s}_2, \vec{s}_3, \vec{s}_4\}$ can be expressed as
\begin{equation}\label{j61} 
	\begin{aligned}
		&\rho(\vec{s}_1) = \begin{pmatrix}
			\cfrac{4}{5} & -\cfrac{3i}{10} \\[6pt]
			\cfrac{3i}{10} & \cfrac{1}{5}
		\end{pmatrix},\quad
		\rho(\vec{s}_2) = \begin{pmatrix}
			\cfrac{4}{5} & \cfrac{3i}{10} \\[6pt]
			-\cfrac{3i}{10} & \cfrac{1}{5}
		\end{pmatrix}, \\[4pt]
		&\rho(\vec{s}_3)=\begin{pmatrix}
			\cfrac{1}{5} & \cfrac{3i}{10} \\[6pt]
			-\cfrac{3i}{10} & \cfrac{4}{5}
		\end{pmatrix},\quad
		\rho(\vec{s}_4)=\begin{pmatrix}
			\cfrac{1}{5} & -\cfrac{3i}{10} \\[6pt]
			\cfrac{3i}{10} & \cfrac{4}{5}
		\end{pmatrix}.
	\end{aligned}
\end{equation}
After some calculation, we have the Shannon entropies
\begin{eqnarray}\label{j62}
	&&H(Q')=h\left[(10-3\sqrt2)/20\right]\approx 0.8660 >C,\nonumber\\
	&&H(R')=h\left[(10-3\sqrt2)/20\right]\approx 0.8660>C,
\end{eqnarray}
which indicates that the sufficient criterion in Eq. \eqref{s35} is violated. On the other hand, after some analysis, we can obtain that the optimal measurements for the faithful criterion are
\begin{eqnarray}\label{j63}
	Q=(Y+Z)/\sqrt2,~~R=(Y-Z)/\sqrt2.
\end{eqnarray}
Thus, according to the faithful criterion in Eq. \eqref{s23}, we have the EUR
\begin{eqnarray}\label{j64}
	H(Q)+H(R)&=&1+h\left[(5-3\sqrt2)/10\right]\nonumber\\
	&\approx& 1.3870 \nonumber\\
	&<&1+C,
\end{eqnarray}
which identifies the optimal four-state set $\Lambda_{\vec{s}_1}$ shown in Fig. \ref{figs4add}(a) being preparation contextual.

In order to further illustrate the \textit{``invalid region"} in Fig. 2 of the main text, we consider another example where the single-qubit state has the polarization vector $\vec{t}_1=(0,0,3/5)$. Using the $B_2$-orbit realizability method, we can generate the optimal four-state set $\Lambda_{\vec{t}_1}=\{\vec{t}_1, \vec{t}_2, \vec{t}_3, \vec{t}_4\}$ as shown in Fig. \ref{figs4add}(b), in which the four quantum states have the forms
\begin{equation}\label{j65}
	\begin{aligned}
		&\rho(\vec{t}_1)=\begin{pmatrix}
			\cfrac{4}{5} & 0 \\[6pt]
			0 & \cfrac{1}{5}
		\end{pmatrix},\quad
		\rho(\vec{t}_2)=\begin{pmatrix}
			\cfrac{1}{2} & \cfrac{3}{10} \\[6pt]
			\cfrac{3}{10} & \cfrac{1}{2}
		\end{pmatrix}, \\[4pt]
		&\rho(\vec{t}_3)=\begin{pmatrix}
			\cfrac{1}{5} & 0 \\[6pt]
			0 & \cfrac{4}{5}
		\end{pmatrix},\quad
		\rho(\vec{t}_4)=\begin{pmatrix}
			\cfrac{1}{2} & -\cfrac{3}{10} \\[6pt]
			-\cfrac{3}{10} & \cfrac{1}{2}
		\end{pmatrix}.
	\end{aligned}
\end{equation}
After some calculation for the non-optimal measurements $Q'$ and $R'$ in Eq. \eqref{j60} on the single-qubit state $\vec{t}_1$, we can obtain
\begin{eqnarray}\label{j66}
	&&H(Q')\approx 0.8660 >C,\nonumber\\
	&&H(R')\approx 0.8660>C,
\end{eqnarray}
which indicates the violation of the sufficient criterion. On the other hand, we can derive that the optimal measurements for $\vec{t}_1$ are $Q=Z$ and $R=X$, respectively. Then, we check the preparation contextuality in the optimal four-state set $\Lambda_{\vec{t}_1}=\{\vec{t}_1, \vec{t}_2, \vec{t}_3, \vec{t}_4\}$ in terms of the faithful criterion in Eq. \eqref{s23}, and have the result
\begin{eqnarray}\label{j67}
	H(Q)+H(R)&=&1+h[1/5]\nonumber\\
	&\approx& 1.7219\nonumber\\
	&>&1+C,
\end{eqnarray}
which indicates that the four-state set $\Lambda_{\vec{t}_1}$ is preparation noncontextual.

To sum up, the \textit{``invalid region"} in Fig. 2 of the main text refers to the fact that, when the sufficient criterion in Theorem 1 is not satisfied, the detection of preparation contextuality via the non-optimal measurements $Q'$ and $R'$ fails and is inconclusive. As exemplified by the two concrete four-state sets $\Lambda_{\vec{s}_1}$ and $\Lambda_{\vec{t}_1}$ shown in Fig. \ref{figs4add}, where the sufficient criterion loses its efficacy and the two sets are contextual and noncontextual, respectively.

\subsection{The difference between Theorem 1 and the nonclassical property of the UR.}
\label{THEOREM 1C}
In Theorem 1 of the main text, we present a faithful criterion via the entropic uncertainty relation (EUR) to witness the preparation contextuality in the optimal four-state set $\Lambda_{\vec{s}_1}$ for any given single-qubit state $\rho(\vec{s}_1)$, which can be expressed as
\begin{equation}\label{d01} 
	H(Q)+H(R) < 1+C\simeq 1.6009.
\end{equation}
Recently, Catani \textit{et al} focused on a different problem as stated in the title \textit{"What is Nonclassical about Uncertainty Relations"} \cite{lc22prl-s}, where they explored that, to what extent, the joint predictability of uncertainty relation (UR), such as  $\Delta X^2+\Delta Z^2 \geq 1$, can be interpreted by a generalized noncontextual model. For qubit theory, they obtain that the classical boundary for the UR of the pair of measurements (Pauli matrices on x and z directions) is
\begin{equation}\label{s49} 
	|\langle X \rangle_{\vec s_1}|+|\langle Z  \rangle_{\vec s_1}|\leq 1,
\end{equation}
which establishes the relation between the classical property of the UR and the preparation noncontextuality of a specific four-state set $\Lambda'_{\vec{s}_1}$ of single-qubit state $\rho(\vec{s}_1)$ with the $A_1^2$ symmetry.

Here, we would like to point out that the motivation of Theorem 1 in our work is to establish a faithful criterion via the EUR to witness preparation contextuality in the optimal four-state set $\Lambda_{\vec{s}_1}$ of any given single-qubit state. However, the specific four-state set $\Lambda'_{\vec{s}_1}$ in the work of Catani et al \cite{lc22prl-s} serves for a different task and is used to establish a classical explanation on the joint predictability of the UR. \textit{Therefore, the two criteria given in Eqs. \eqref{d01} and \eqref{s49} serve for distinct tasks, and then they are essentially different.}

\begin{figure}
	\epsfig{figure=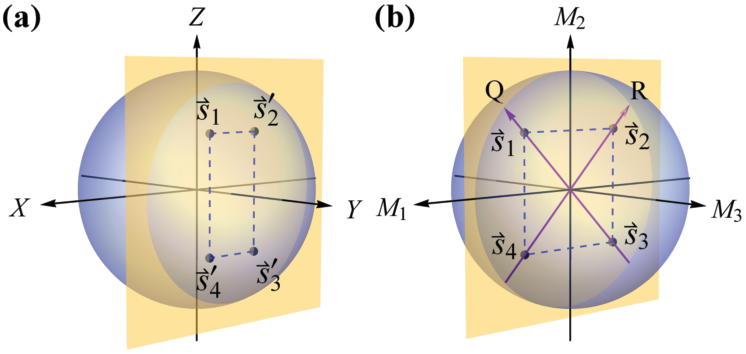,width=0.48\textwidth}
	\caption{Two four-state sets for detecting the nonclassical property of the UR and witnessing the preparation contextuality in the QSP, respectively. (a) The specific four-state set $\Lambda'_{\vec{s}_1}(\vec{s}_1, \vec{s}\hspace{1pt}'_2, \vec{s}\hspace{1pt}'_3, \vec{s}\hspace{1pt}'_4)$ with the $A_1^2$ symmetry related to measurements $X$ and $Z$. (b) The optimal four-state set $\Lambda_{\vec{s}_1}(\vec{s}_1,\vec{s}_2,\vec{s}_3,\vec{s}_4)$ with the $B_2$ symmetry with $M_1=(Q-R)/\sqrt{2}$ and $M_2=(Q+R)/\sqrt{2}$ with the optimal measurements $Q$ and $R$ given in Eq. \eqref{s53}.}\label{figs5add}
\end{figure}

Next, we will analyze two sets of examples and clarify the difference and relation between the two criteria. First, we consider a single-qubit state which can be written as
\begin{equation}\label{z50}
	\rho(\vec{s}_1)=\frac{1}{2}(I+\frac{3\sqrt{2}}{16} X+\frac{3\sqrt{6}}{16} Y+\frac{3\sqrt{2}}{8} Z ),
\end{equation}
where $X$, $Y$, $Z$ are Pauli matrices and consist of the operator basis in the Bloch representation. In order to judge the classical or nonclassical property of the UR ($\Delta X^2+\Delta Z^2 \geq 1$), we calculate the expectation values of the related measurements on the single-qubit state and can obtain
\begin{equation}\label{s50}
	|\langle X \rangle_{\vec s_1}|+|\langle Z  \rangle_{\vec s_1}|=\frac{9\sqrt{2}}{16}\simeq 0.7955.
\end{equation}
According the criterion in Eq. \eqref{s49}, we can get that the joint prediction of the UR is classical and admits of a generalized noncontextual ontological explanation. Based on the $A_1^2$ realizability associated with $X$ and $Z$ \cite{lc22prl-s}, the four-state set $\Lambda'_{\vec{s}_1}(\vec{s}_1, \vec{s}\hspace{1pt}'_2, \vec{s}\hspace{1pt}'_3, \vec{s}\hspace{1pt}'_4)$ can be generated as illustrated in Fig. \ref{figs5add}(a), which is preparation noncontextual and has the form
\begin{equation}\label{s51}
	\begin{split}
		\Lambda'_{\vec{s}_1}=& \{\frac{1}{2}(I+\frac{3\sqrt{2}}{16} X+\frac{3\sqrt{6}}{16} Y+\frac{3\sqrt{2}}{8} Z ),\\
		&   \frac{1}{2}(I-\frac{3\sqrt{2}}{16} X+\frac{3\sqrt{6}}{16} Y+\frac{3\sqrt{2}}{8} Z ),\\
		&   \frac{1}{2}(I-\frac{3\sqrt{2}}{16} X+\frac{3\sqrt{6}}{16} Y-\frac{3\sqrt{2}}{8} Z ),\\
		&   \frac{1}{2}(I+\frac{3\sqrt{2}}{16} X+\frac{3\sqrt{6}}{16} Y-\frac{3\sqrt{2}}{8} Z )\}.
	\end{split}
\end{equation}
The specificity of the four-state set $\Lambda'_{\vec{s}_1}$ lies that it is $A_1^2$ symmetric and depends on the expectation values of the observables on $\vec{s}_1$ in the UR. On the other hand, for the same single-qubit state given in Eq. \eqref{z50}, we can generate the optimal four-state set $\Lambda_{\vec{s}_1}(\vec{s}_1, \vec{s}_2, \vec{s}_3, \vec{s}_4)$ in terms of the $B_2$-orbit realizability presented in the main text of our work, which is illustrated in Fig. \ref{figs5add}(b). Thus, we can obtain the four-state set with the $B_2$ symmetry which can be expressed as
\begin{equation}\label{s52}
	\begin{split}
		\Lambda_{\vec{s}_1}=& \{\frac{1}{2}(I+\frac{3\sqrt{2}}{16} X+\frac{3\sqrt{6}}{16} Y+\frac{3\sqrt{2}}{8} Z ),\\
		&   \frac{1}{2}(I-\frac{3\sqrt{2}}{16} X-\frac{3\sqrt{6}}{16} Y+\frac{3\sqrt{2}}{8} Z ),\\
		&   \frac{1}{2}(I-\frac{3\sqrt{2}}{16} X-\frac{3\sqrt{6}}{16} Y-\frac{3\sqrt{2}}{8} Z ),\\
		&   \frac{1}{2}(I+\frac{3\sqrt{2}}{16} X+\frac{3\sqrt{6}}{16} Y-\frac{3\sqrt{2}}{8} Z )\}.
	\end{split}
\end{equation}
Moreover, the optimal measurements $Q$ and $R$ have the forms
\begin{equation}\label{s53}
	\begin{split}
		Q&=\frac{\sqrt{2}}{4} X+\frac{\sqrt{6}}{4} Y+\frac{\sqrt{2}}{2} Z, \\
		R&=-\frac{\sqrt{2}}{4} X-\frac{\sqrt{6}}{4} Y+\frac{\sqrt{2}}{2} Z.
	\end{split}
\end{equation}
Based on the optimal observables $Q$ and $R$, we can calculate the EUR and have the result
\begin{eqnarray}\label{s54}
	H(Q)+H(R)&=&1+h\left(\frac{1}{8}\right) \simeq 1.5436.
\end{eqnarray}
According to the faithful criterion in Eq. \eqref{d01}, we can get the conclusion that the procedure of quantum state preparation (QSP) for the optimal set $\Lambda_{\vec{s}_1}$ is contextual. Therefore, for a given single-qubit state $\rho(\vec{s}_1)$, even if the UR is classical and admits of a generalized ontological explanation described by the specific four-state set $\Lambda'_{\vec{s}_1}$ shown in Fig. \ref{figs5add}(a), the optimal four-state set $\Lambda_{\vec{s}_1}$ shown in Fig. \ref{figs5add}(b) can still exhibit the preparation contextuality.

\begin{figure}
	\epsfig{figure=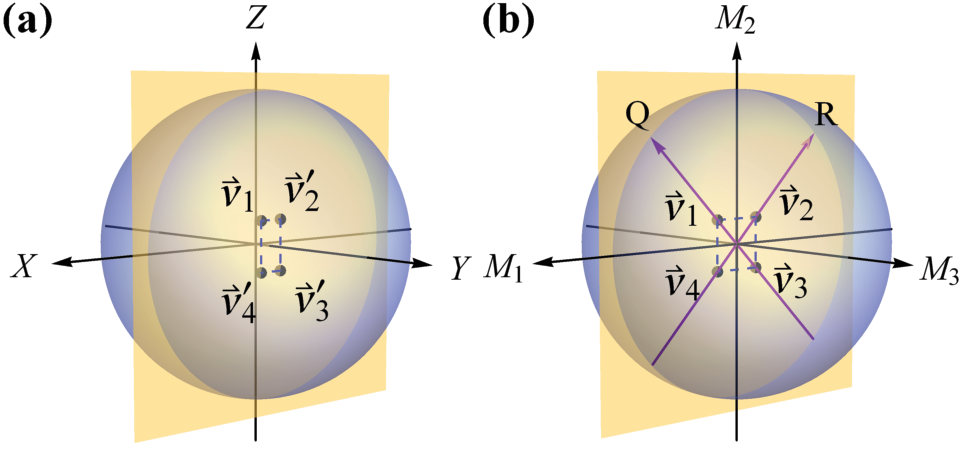,width=0.48\textwidth}
	\caption{Two four-state sets for detecting the classical interpretation of the UR and witnessing the preparation noncontextuality in the QSP. (a) The specific four-state set $\Lambda'_{\vec{v}_1}(\vec{v}_1, \vec{v}\hspace{1pt}'_2, \vec{v}\hspace{1pt}'_3, \vec{v}\hspace{1pt}'_4)$ with the $A_1^2$ symmetry related to measurements $X$ and $Z$. (b) The optimal four-state set $\Lambda_{\vec{v}_1}(\vec{v}_1,\vec{v}_2,\vec{v}_3,\vec{v}_4)$ with the $B_2$ symmetry with $M_1=(Q-R)/\sqrt{2}$ and $M_2=(Q+R)/\sqrt{2}$ with the optimal measurements $Q$ and $R$ given in Eq. \eqref{s53}.}\label{figs6add}
\end{figure}

We now consider another case that, for a given single-qubit state $\rho(\vec{v}_1)$, its optimal four-state set $\Lambda_{\vec{v}_1}$ with the $B_2$ symmetry is preparation noncontextual according to the faithful criterion in Eq. \eqref{d01}. In this case, using Lemma 1 in the main text, we can get that any other four-state set $\Lambda'_{\vec{v}_1}$ with the $A_1^2$ symmetry associated with a pair of corresponding complementary measurements $\{M_1', M_2'\}$ is necessarily noncontextual. Therefore, the joint predictability of the uncertainty relation ($\Delta (M_1')^2+\Delta (M_2')^2\geq 1$) admits of a generalized noncontextual explanation. Since the pair of measurements $M_1'$ and $M_2'$ are chosen arbitrarily, we can get the criterion in Eq. \eqref{s49} when setting $M_1'=X$ and $M_2'=Z$. As a result, we can obtain
\begin{equation}\label{j76} 
	H(Q)+H(R)\geq 1+C \Rightarrow |\langle X \rangle_{\vec v_1}|+|\langle Z  \rangle_{\vec v_1}|\leq 1,
\end{equation}
which gives the relation that \textit{``When the optimal set $\Lambda_{\vec{v}_1}$ is preparation noncontextual, the UR of $X$ and $Z$ on $\vec{v}_1$ necessarily admits of a classical explanation"}.

As a concrete example, we consider a single-qubit state
\begin{eqnarray}\label{d74}
	\rho(\vec{v}_1)=\frac{1}{2}(I+\frac{\sqrt{2}}{16} X+\frac{\sqrt{6}}{16} Y+\frac{\sqrt{2}}{8} Z),
\end{eqnarray}
where $X, Y, Z$ are Pauli matrices serving as the operator basis of Bloch representation. As shown in Fig. \ref{figs6add}(b), its optimal four-state set
$\Lambda_{\vec{v}_1}(\vec{v}_1, \vec{v}_2, \vec{v}_3, \vec{v}_4)$ generated by the $B_2$-orbit realizability has the form
\begin{equation}\label{s55}
	\begin{split}
		\Lambda_{\vec{v}_1}=& \{\frac{1}{2}(I+\frac{\sqrt{2}}{16} X+\frac{\sqrt{6}}{16} Y+\frac{\sqrt{2}}{8} Z),\\
		&   \frac{1}{2}(I-\frac{\sqrt{2}}{16} X-\frac{\sqrt{6}}{16} Y+\frac{\sqrt{2}}{8} Z ),\\
		&   \frac{1}{2}(I-\frac{\sqrt{2}}{16} X-\frac{\sqrt{6}}{16} Y-\frac{\sqrt{2}}{8} Z ),\\
		&   \frac{1}{2}(I+\frac{\sqrt{2}}{16} X+\frac{\sqrt{6}}{16} Y-\frac{\sqrt{2}}{8} Z )\}.
	\end{split}
\end{equation}
According to the faithful criterion in Eq. \eqref{d01}, we have
\begin{eqnarray}\label{s56}
	H(Q)+H(R)&=&1+h\left(\frac{3}{8}\right)\approx 1.9544\nonumber\\
	&\geq& 1+C,
\end{eqnarray}
which indicates the optimal set $\Lambda_{\vec{v}_1}$ is preparation noncontextual.
For the same quantum state given in Eq. \eqref{d74}, we have the expectation values of the observables $X$ and $Z$ in the UR
\begin{equation}\label{s58} 
	\langle X \rangle_{\vec{v}_1}+\langle Z  \rangle_{\vec{v}_1}=\frac{3\sqrt{2}}{16}\approx 0.2652\leq 1,
\end{equation}
which indicates that the UR is classical according to the criterion in Eq. \eqref{s49}. As shown in Fig. \ref{figs6add}(a), the noncontextual four-state set $\Lambda'_{\vec{v}_1}(\vec{v}_1, \vec{v}\hspace{1pt}'_2, \vec{v}\hspace{1pt}'_3, \vec{v}\hspace{1pt}'_4)$ with the $A_1^2$ symmetry is
\begin{equation}\label{s57}
	\begin{split}
		\Lambda'_{\vec{v}_1}=& \{\frac{1}{2}(I+\frac{\sqrt{2}}{16} X+\frac{\sqrt{6}}{16} Y+\frac{\sqrt{2}}{8} Z ),\\
		& \frac{1}{2}(I-\frac{\sqrt{2}}{16} X+\frac{\sqrt{6}}{16} Y+\frac{\sqrt{2}}{8} Z ),\\
		& \frac{1}{2}(I-\frac{\sqrt{2}}{16} X+\frac{\sqrt{6}}{16} Y-\frac{\sqrt{2}}{8} Z ),\\
		& \frac{1}{2}(I+\frac{\sqrt{2}}{16} X+\frac{\sqrt{6}}{16} Y-\frac{\sqrt{2}}{8} Z )\}.
	\end{split}
\end{equation}

In summary, the two criteria in Eqs. \eqref{d01} and \eqref{s49} serve for distinct tasks, and there exists the optimal set of the given single-qubit state is preparation contextual even if the UR of the same single-qubit state is classical. Moreover, due to the optimality of the optimal set indicated by Lemma 1 of the main text, we have the relation that the UR is necessarily classical when the optimal set is preparation noncontextual.

\section{Proof of Theorem 2}
\label{THEOREM 2}
In Theorem 2 of the main text, we consider a shared two-qubit state $\rho_{AB}$, and present a trade-off relation between the local preparation contextuality of the optimal four-state set $\Lambda(\rho_A)$ and bipartite entanglement within the framework of quantum resource distribution, which has the form
\begin{equation}\label{s59} 
	1\leq H_{QR}(A)+S(A|B)\leq 3,	
\end{equation}
where $H_{QR}(A)=H_A(Q)+H_A(R)$ is the EUR of the optimal measurements $\{Q, R\}$ on the subsystem $A$, and $S(A|B)=S(AB)-S(B)$ is quantum conditional entropy serving as entanglement indicator. When the two-qubit system is in a pure state, the trade-off relation becomes an equality and can be expressed as
\begin{equation}\label{s60} 
	H_{QR}(A)+S(A|B)=1.
\end{equation}
We now provide the analytical proof of Theorem 2.

\textit{Proof.}---We first analyze the upper bound of the inequality in Eq. \eqref{s59}, which corresponds to the maximally mixed state $\rho_{AB}=I/4$. For this bipartite state, both the EUR and the conditional entropy can achieve their maximal values
\begin{eqnarray}\label{s61} 
	&&H_{QR}(A)=H_A(Q)+H_A(R)=2,\nonumber\\
	&&S(A|B)=S(AB)-S(B)=1,
\end{eqnarray}
where we use the reduced state of subsystem being $\rho_A=I/2$ and the subadditive property of von Neumann entropy \cite{nie10book-s}
\begin{equation}\label{s62} 
	S(AB)\leq S(A)+S(B).
\end{equation}
In this case, the four-state set $\Lambda(\rho_A)$ is preparation noncontextual according to the faithful criterion in Theorem 1 of the main text, and the two-qubit state is separable due to the positive value of the conditional entropy.

In order to derive the lower bound of the inequality given in Eq. \eqref{s59}, we analyze the Shannon entropy $H_A(Q)$, where the optimal measurement $Q$ is along the direction of the polarization vector of the reduced quantum state $\rho_A=\tr_B[\rho_{AB}]$. In the eigenvector space of the measurement $Q$, the single-qubit state can be expressed as
\begin{equation}\label{s63} 
	\rho_A=\frac{1}{2}(1+s_1)Q_++\frac{1}{2}(1-s_1)Q_-,
\end{equation}
where $s_1=|\vec{s}_1|$ is the modulus of the polarization vector, and $Q_\pm$ are the eigenvectors corresponding to the eigenvalues $\pm 1$ which satisfy the property $Q=Q_+-Q_-$ and take the forms
\begin{eqnarray}\label{s64} 
	Q_+&=&\frac{1}{2}\left(I+\frac{\vec{s}_1}{s_1}\cdot \vec{M}\right), \nonumber\\
	Q_-&=&\frac{1}{2}\left(I-\frac{\vec{s}_1}{s_1}\cdot \vec{M}\right).
\end{eqnarray}
Therefore, the post-measurement state can be written as
\begin{eqnarray}\label{s65} 
	\rho_Q^A&=&Q_+(\rho_A)Q_+ +Q_-(\rho_A)Q_-\nonumber\\
	&=&\frac{1}{2}(1+s_1)Q_++\frac{1}{2}(1-s_1)Q_-\nonumber\\
	&=&\rho_A,
\end{eqnarray}
where the expression of $\rho_A$ in Eq. \eqref{s63} is used in the last equality. Then, we have the property that the Shannon entropy of the post-measurement state $\rho_Q^A$ is equal to the corresponding von Neumann entropy of the single-qubit state $\rho_A$, i.e.,
\begin{equation}\label{s66} 
	H(Q)=h\left[\frac{1}{2}(1+s_1)\right]=S(A),
\end{equation}
where $h(\cdot)$ is the binary entropy function. Next, we analyze the Shannon entropy $H(R)$, for which the post-measurement state is
\begin{eqnarray}\label{s67} 
	\rho_R^A&=&R_+(\rho_A)R_+ +R_-(\rho_A)R_-\nonumber\\
	&=&R_+\left[\frac{1}{2}(1+s_1)Q_++\frac{1}{2}(1-s_1)Q_-\right]R_+ \nonumber\\
	&+&R_-\left[\frac{1}{2}(1+s_1)Q_++\frac{1}{2}(1-s_1)Q_-\right]R_- \nonumber\\
	&=&\frac{1}{2}R_+ +\frac{1}{2}R_-\nonumber\\
	&=&\frac{I}{2},
\end{eqnarray}
where $R_\pm$ with the corresponding eigenvalues $\pm 1$ are the eigenvectors of measurement $R$, and we use the expression of $\rho_A$ in Eq. \eqref{s63}, the complementary property between the measurements $Q$ and $R$, and the completeness $R_++R_-=I$. Based on Eq. \eqref{s67}, the Shannon entropy for the measurement $R$ on $\rho_A$ can be written as $H(R)=h(1/2)=1$, and then we have the EUR
\begin{eqnarray}\label{s68} 
	H_{QR}(A)&=&H_A(Q)+H_A(R)\nonumber\\
	&=&1+S(A).
\end{eqnarray}
Combining Eq. \eqref{s68} with the quantum conditional entropy for $\rho_{AB}$, we can derive the lower bound of the quantitative inequality in Eq. \eqref{s59}
\begin{eqnarray}\label{s69} 
	&&H_{QR}(A)+S(A|B) \nonumber\\
	&=&H_A(Q)+H_A(R)+S(AB)-S(B) \nonumber\\
	&=&1+[S(AB)+S(A)-S(B)]\nonumber\\
	&\geq& 1+[S(AB)-|S(A)-S(B)|]\nonumber\\
	&\geq& 1,
\end{eqnarray}
where, in the first inequality, we use the relation $S(A)-S(B)\geq -|S(A)-S(B)|$, and, in the second inequality, we apply the subadditivity of von Neumann entropy $S(AB)\geq |S(A)-S(B)|$ \cite{nie10book-s}. Thus, by combining the upper and lower bounds given in Eqs. \eqref{s61} and \eqref{s69}, we can obtain the trade-off relation between local preparation contextuality and bipartite quantum entanglement as given in Eq. \eqref{s59}, which proves the trade-off relation for the mixed state case in Theorem 2.

When the shared two-qubit state is pure, the quantitative trade-off relation in Eq. \eqref{s59} becomes
\begin{eqnarray}\label{s70}
    &&H_{QR}(A)+S(A|B)\nonumber\\
    &&=H_A(Q)+H_A(R)+S(A|B)\nonumber\\
	&&=1+S(A)+S(AB)-S(B)\nonumber\\
	&&=1,
\end{eqnarray}
where we used the relation in Eq. \eqref{s68} and the properties $S(AB)=0$ as well as $S(A)=S(B)$ for a pure two-qubit state. Therefore, we prove the trade-off relation in Eq. \eqref{s60} for the pure state case, which completes the proof of Theorem 2 in the main text.
\hfill$\blacksquare$


\section{Proof of Theorem 3}
\label{THEOREM 3}
In Theorem 3 of the main text, within the framework of quantum resource distribution, we investigate a shared two-qubit state $\rho_{AB}$ and show that there exists a quantitative trade-off relation between local preparation contextuality and Bell nonlocality which has the form
\begin{equation}\label{s71}  
	1\leq H_{QR}(A)+[2-\langle B\rangle_{\mathrm{max}}] \leq 4,
\end{equation}
where $H_{QR}(A)$ witnesses the preparation contextuality in the optimal four-state set $\Lambda(\rho_A)$ and the term $2-\langle B\rangle_{\mathrm{max}}$ detects the deviation magnitude of the nonlocality from boundary of maximal classical correlation with $\langle B\rangle_{\mathrm{max}}$ being the maximal CHSH correlation for $\rho_{AB}$. Moreover, when the two-qubit state is pure, the trade-off relation can be expressed as
\begin{equation}\label{s72} 
	1\leq H_{QR}(A)+[2-\langle B\rangle_{\mathrm{max}}]\leq 4-2\sqrt{2}.
\end{equation}
The proof of Theorem 3 in the main text is as follows.

\textit{Proof.}---In the Bloch representation, the two-qubit state can be expressed as \cite{hr95pla-s}
\begin{eqnarray}\label{s73} 
	\rho_{AB}&=&\frac{1}{4}(I_A\otimes I_B+\vec{a}\cdot\vec{\sigma}\otimes I_B+I_A\otimes\vec{b}\cdot\vec{\sigma}\nonumber\\
	&&+\sum_{j,k=1}^{3}T_{jk}\sigma_j\otimes\sigma_k),
\end{eqnarray}
where $\vec{\sigma}=(\sigma_x, \sigma_y, \sigma_z)$ are Pauli matrices, $\vec{a}=(a_x,a_y,a_z)^\top$ is the polarization vector of $\rho_A$ with $a_i=\tr[\rho_A\sigma_i]$ (similarly, $\vec{b}$ is the polarization vector of $\rho_B$), and $T_{ij}=\tr[\rho_{AB}\sigma_i\otimes \sigma_j]$ is the element of a $3\times 3$ correlation matrix $T$.

We first consider the case of a two-qubit pure state which, after local unitary transformations, can be expressed as its Schmidt decomposition form \cite{pa93-s,tbm00pra-s}
\begin{equation}\label{s74}
	\ket{\psi}_{AB}=\sqrt{\lambda_1}\ket{00}+\sqrt{\lambda_2}\ket{11},
\end{equation}
where the nonnegative $\lambda_i$s are the Schmidt coefficients satisfying the relation $\lambda_1=1-\lambda_2$. After some derivation, we obtain the correlation matrix $T$ for this pure state which can be expressed as
\begin{equation}\label{s75} 
	T=\begin{pmatrix}
		2\sqrt{\lambda_1\lambda_2} & 0  & 0 \\
		0 & -2\sqrt{\lambda_1\lambda_2} & 0 \\
		0  & 0 & 1
	\end{pmatrix},
\end{equation}
where we use the property $\lambda_1+\lambda_2=1$. The maximal Bell nonlocality of the two-qubit quantum state can be obtained via the Horodecki parameter \cite{hr95pla-s}
\begin{eqnarray}\label{s76} 
	\langle B\rangle_{\mathrm{max}}&=&\sqrt{4\mathcal{M}(\ket{\psi}_{AB})}\nonumber\\
	&=&2\sqrt{m_1+m_2}\nonumber\\
	&=&2\sqrt{1+4\lambda_1\lambda_2},
\end{eqnarray}
where the parameter is $\mathcal{M}(\ket{\psi}_{AB})=m_1+m_2=1+4\lambda_1\lambda_2$ with $m_1=1$ and $m_2=4\lambda_1\lambda_2$ being the two largest eigenvalues of the matrix $S=T^\top T$ in which $T$ is the correlation matrix given in Eq. \eqref{s75}. Then, the deviation magnitude of the nonlocality from the boundary of maximal classical correlation can be written as
\begin{equation}\label{s77} 
	2-\langle B\rangle_{\mathrm{max}}=2-2\sqrt{1+4\lambda_1\lambda_2}.
\end{equation}
Moreover, for the shared two-qubit pure state $\ket{\psi}_{AB}$, its quantum conditional entropy is
\begin{eqnarray}\label{s78}
	S(A|B)&=&S(AB)-S(B)\nonumber\\
	&=&-S(B)\nonumber\\
	&=&\lambda_1\mbox{log}_2 \lambda_1 +(1-\lambda_1)\mbox{log}_2 (1-\lambda_1),
\end{eqnarray}
where we use the property $S(AB)=0$ for the pure state and the diagonal form of the reduced density matrix $\rho_B^d=\mbox{diag}\{\lambda_1,1-\lambda_1\}$.

\begin{figure}
	\epsfig{figure=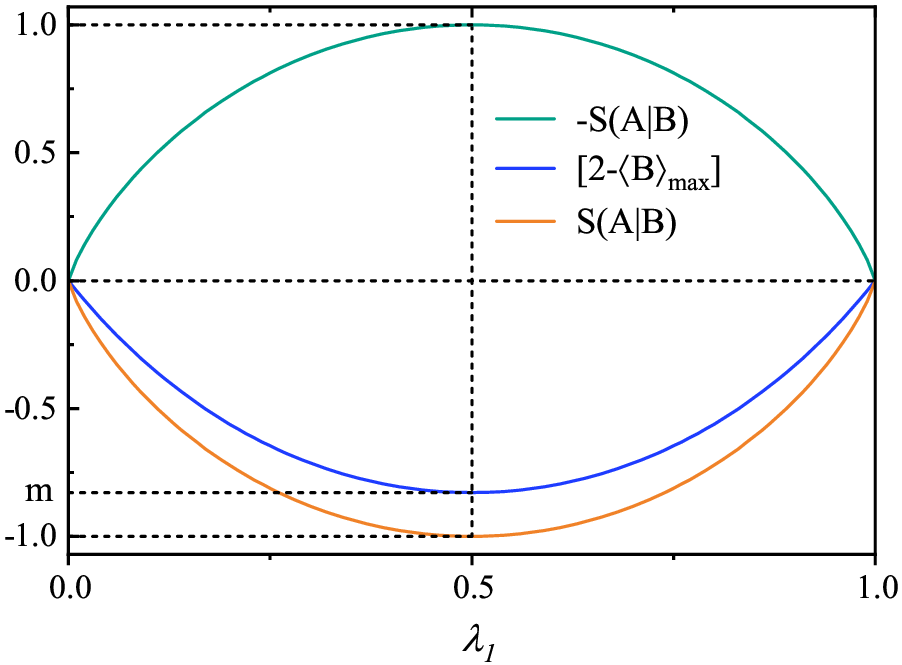,width=0.42\textwidth}
	\caption{The nonlocal correlations $\pm S(A|B)$ and $[2-\langle B\rangle_{\mathrm{max}}]$ as the functions of the parameter $\lambda_1$ for the two-qubit pure state in Eq. \eqref{s74}, where $m=2-2\sqrt{2}$ and we obtain that the relation $2-\langle B\rangle_{\mathrm{max}}\geq S(A|B)$ is satisfied.}
	\label{figs4}
\end{figure}

In Fig. \ref{figs4}, we compare the values of the two nonlocal correlations $S(A|B)$ and $[2-\langle B\rangle_{\mathrm{max}}]$, where both correlations are expressed as the functions of the parameter $\lambda_1$. As shown in the figure, we can obtain the relation
\begin{equation}\label{s79} 
	2-\langle B\rangle_{\mathrm{max}}\geq S(A|B),
\end{equation}
which indicates that the first-order derivatives satisfy
\begin{equation}\label{s80} 
	|\partial_{\lambda_1}[2-\langle B\rangle_{\mathrm{max}}]|\leq |\partial_{\lambda_1}S(A|B)|.
\end{equation}
Furthermore, we can obtain the maximum
\begin{eqnarray}\label{s81} 
	&&\max_{\lambda_1}\{[2-\langle B\rangle_{\mathrm{max}}]-S(A|B)\}\nonumber\\
	&=&\max_{\lambda_1}\{[2-\langle B\rangle_{\mathrm{max}}]+S(A)\}\nonumber\\
	&=&\{[2-\langle B\rangle_{\mathrm{max}}]+S(A)\}|_{\lambda_1=\frac{1}{2}}\nonumber\\
	&=&3-2\sqrt{2},
\end{eqnarray}
where we use the property in Eq. \eqref{s80} and the entropic relation $S(A|B)=-S(B)=-S(A)$ which holds for any two qubit pure state. It should be noted that the results in Eqs. \eqref{s79} and \eqref{s81} are valid for an arbitrary two-qubit pure state, since both the Bell correlation function $\langle B\rangle_{\mathrm{max}}$ and the von Neumann entropy $S(A|B)$ are invariant under local unitary transformations. Therefore, the quantitative trade-off functions between the local preparation contextuality and Bell nonlocality in any two-qubit pure state will be
\begin{eqnarray}\label{s82} 
	H_{QR}(A)+[2-\langle B\rangle_{\mathrm{max}}]&\geq& H_{QR}(A)+S(A|B)\nonumber\\
	&=&1,
\end{eqnarray}
where we use the property in Eq. \eqref{s79} and the relation between local preparation contextuality and bipartite entanglement in Eq. \eqref{s60}.

The above inequality provides the lower bound of the trade-off relation in a shared bipartite pure state as shown in Eq. \eqref{s72} of Theorem 3. Now, we analyze the upper bound which can be written as
\begin{eqnarray}\label{s83} 
	H_{QR}(A)+[2-\langle B\rangle_{\mathrm{max}}]&=&1+S(A)+[2-\langle B\rangle_{\mathrm{max}}]\nonumber\\
	&\leq& 1+3-2\sqrt{2}\nonumber\\
	&=&4-2\sqrt{2},
\end{eqnarray}
where we use the property $H_{QR}(A)=1+S(A)$ in Eq. \eqref{s68} and the maximum of the function $[2-\langle B\rangle_{\mathrm{max}}]+S(A)$ derived in Eq. \eqref{s81}. Combining the lower and upper bounds in Eqs. \eqref{s82} and \eqref{s83}, we complete the proof of Eq. \eqref{s72} for the trade-off relation between local preparation contextuality and Bell nonlocality in the pure state case.

Next, we prove the trade-off relation given in  Eq. \eqref{s71} for an arbitrary two-qubit mixed state. The pure state decomposition of a shared mixed state $\rho_{AB}$ can be written as
\begin{equation}\label{s84} 
	\rho_{AB}=\sum_{i} p_i \proj{\psi_{AB}^i},
\end{equation}
where $p_i$ is the probability of pure state component with the normalization property $\sum_i p_i=1$, and the reduced density matrix of subsystem $A$ is
\begin{equation}\label{s85} 
	\rho_A=\sum_i p_i\rho_A^i=\sum_i p_i\tr_B[\proj{\psi^i_{AB}}],
\end{equation}
where $\rho_A^i$ is the reduced density matrix for the pure state component $\ket{\psi^i_{AB}}$. We now analyze the trade-off functions for the mixed state $\rho_{AB}$ which can be expressed as
\begin{eqnarray}\label{s86} 
	&&	H_{QR}(A)+[2-\langle B\rangle_{\mathrm{max}}]\nonumber\\
	&=&1+S\left(\sum_i p_i\rho_{A}^{i}\right)+[2-\sum_i p_i\langle B\rangle_{\mbox{max}}\left(\ket{\psi^i_{AB}}\right)\nonumber\\
	&\geq& \sum_i p_i\left\{1+S(\rho_A^i)+\left[2-\langle B\rangle_{\mbox{max}}\left(\ket{\psi^i_{AB}}\right)\right]\right\}\nonumber\\
	&=& \sum_i p_i\left\{H_{QR}(A^i)+\left[2-\langle B\rangle_{\mbox{max}}\left(\ket{\psi^i_{AB}}\right)\right]\right\}\nonumber\\
	&\geq& 1,
\end{eqnarray}
where we utilize the properties $H_{QR}(A)=1+S(A)$ and $ \langle B\rangle_{\rho_{AB}}= \sum_{i} p_i \langle B\rangle (\ket{\psi_{AB}^i})=\sum_i p_{i}\bra{\psi^{i}_{AB}}B\ket{\psi^{i}_{AB}}$ \cite{cs17prl-s} in the first equality, the concave property of von Neumann entropy $S(\sum_ip_i \rho_{A}^i)\geq \sum_i p_i S(\rho_{A}^i)$ \cite{nie10book-s} is used in the second inequality, and the last inequality is satisfied due to the result of pure state in Eq. \eqref{s82}. Moreover, the upper bound of the trade-off functions is available for the maximally mixed state $\rho_{AB}=I/4$,
for which we have
\begin{equation}\label{s87} 
	H_{QR}(A)+[2-\langle B\rangle_{\mathrm{max}}]=2+[2-0]=4,
\end{equation}
where both the two terms attain their maximal values. Combining Eqs. \eqref{s86} and \eqref{s87}, we obtain the trade-off relation in Eq. \eqref{s71} for the case of mixed states, and then complete the proof of Theorem 3 in the main text.
\hfill$\blacksquare$

\begin{figure}
	\epsfig{figure=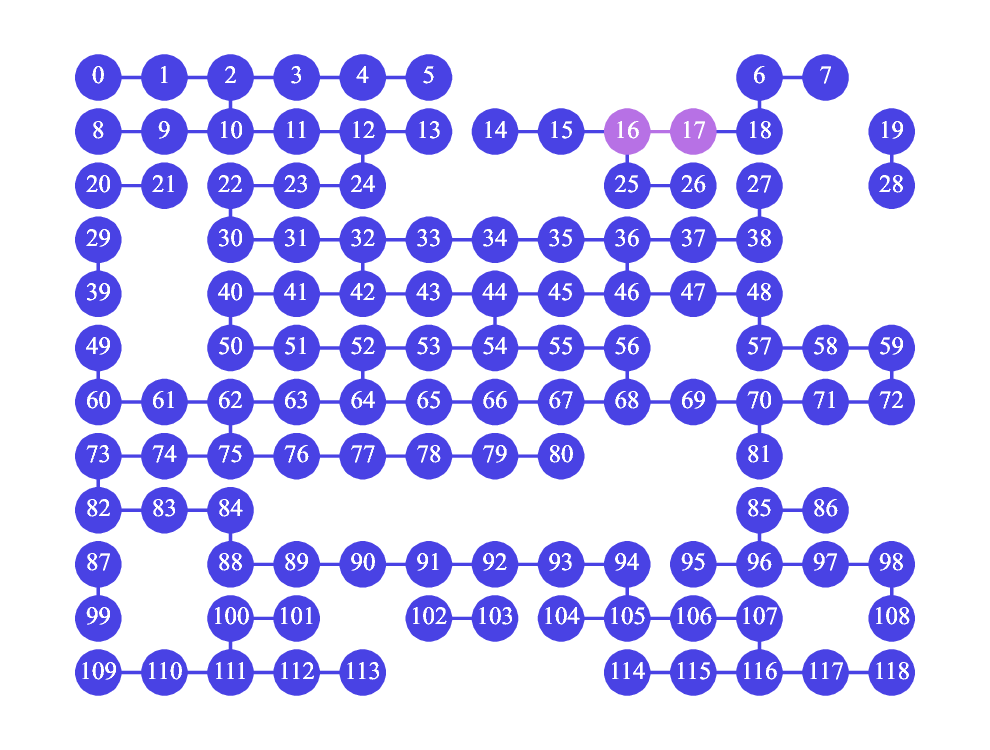,width=0.48\textwidth}
	\caption{The topology graph of \textit{Baihua} device from the Quafu quantum cloud computing cluster. A total of 119 qubits are available, where two purple shaded qubits $Q_{16}$ and $Q_{17}$ are used to perform the demonstration experiments.}
	\label{figs5}
\end{figure}

\section{Device information and readout correction in the experiments}
\label{DEVICE}
In the experimental verification of trade-off relations between local preparation contextuality and nonlocal quantum resources, we utilize a solid-state superconducting system which is operated on \textit{Baihua} device available through the Quafu quantum cloud computing cluster. The platform enables precise control and measurement of quantum states, facilitating the investigation into foundational aspects of quantum mechanics such as quantum contextuality, entanglement, Bell nonlocality, and entropic uncertainty relations.

The topology graph of \textit{Baihua} device is shown in Fig. \ref{figs5}. This device is comprised of 119 qubits, with connections between qubits represented by solid lines indicating their coupling. For our experiments, we utilize qubits $Q_{16}$ and $Q_{17}$, which are highlighted with purple shading in the graph. The basic information of these two physical qubits on \textit{Baihua} device is listed in Table \ref{table-1}, including qubit relaxation times $T_1$ and $T_2$, qubit frequency, anharmonicity, and controlled-$Z$ (CZ) gate fidelity.

Before deploying the \textit{Baihua} device on the cloud platform, a series of calibrations is performed in order to mitigate XY, ZZ, and readout crosstalk, and the energy levels of each qubit are carefully arranged. These operations ensure the negligible influence between qubits in our experiments. Here, it should be noted that our experiment is not designed to perform a loophole-free Bell nonlocality verification \cite{hb15nat-s,gm15prl-s,slk15prl-s}, because the qubits remain co-located rather than being spatially separated \cite{ss23nat-s} due to the physical limitations on \textit{Baihua} chip.

Moreover, due to the existence of readout errors in qubit measurement, it is necessary to perform readout correction to obtain the true qubit probabilities. Let $f_0$ and $f_1$ be the readout probabilities for a qubit prepared in states $\ket{0}$ and $\ket{1}$, respectively. Therefore, the readout fidelity matrix of qubit $Q$ can be defined as
\begin{equation}\label{s88} 
	F = \left( \begin{array}{cc} f_0 & 1 - f_1 \\ 1 - f_0 & f_1  \end{array}  \right).
\end{equation}
According to Bayes' theorem, the uncorrected probabilities of reading out $Q$ in states $\ket{0}$ and $\ket{1}$ can be expressed as
\begin{equation}\label{s89} 
	\mathbf{q} = F \mathbf{p}~,
\end{equation}
where $\mathbf{{q}} = (q_0,q_1)^\top$ and $\mathbf{p} = (p_0,p_1)^\top$ are the uncorrected and corrected probabilities of $Q$, respectively. In our experiments, readout correction must be applied simultaneously to both qubits. Assuming there is no readout crosstalk between the two qubits, the total readout fidelity matrix can be written as the product of the local readout fidelity matrices
\begin{equation}\label{s90} 
	F_{\mathrm{tot}} =  F^{16} \otimes F^{17},
\end{equation}
where the superscripts correspond to the two physical qubits ($Q_{16}$ and $Q_{17}$) used in the experiment. Since experimental techniques can largely eliminate crosstalk, the expression in Eq. \eqref{s90} is effectively valid. Thus, the total corrected probabilities become
\begin{equation}\label{s91} 
	\mathbf{p}_{\mathrm{tot}} = F_{\mathrm{tot}}^{-1} \mathbf{q}_{\mathrm{tot}} ~,
\end{equation}
where
\begin{equation}\label{s92} 
	\mathbf{q}_{\mathrm{tot}} = \left( \begin{array}{cc} \mathbf{q}^{16} \\[4pt] \mathbf{q}^{17} \end{array}  \right), \mathbf{p}_{\mathrm{tot}} = \left( \begin{array}{cc} \mathbf{p}^{16} \\[4pt]  \mathbf{p}^{17} \end{array}  \right).
\end{equation}

In our experiments, all data are obtained using corrected probabilities, with the readout fidelity matrices for the qubits measured contemporaneously with the collection of the corresponding experimental data. This approach ensures that any variations in qubit performance are accurately accounted for, providing reliable and precise results.

\begin{table}
	\centering
	\caption{Basic information of two physical qubits $Q_{16}$ and $Q_{17}$ on \textit{Baihua} device including the qubit relaxation times $T_1$ and $T_2$, qubit frequency, anharmonicity, and controlled-$Z$ (CZ) gate fidelity.}
	\setlength{\tabcolsep}{12pt} 
	\renewcommand{\arraystretch}{1.2} 
	\begin{tabular}{lcc}
		\hline\hline
		Qubit & $Q_{16}$ & $Q_{17}$  \\
		\hline
		$T_1$ ($\mu$s) & 59.2 & 69.0 \\
		$T_2$ ($\mu$s) & 40.1 & 40.3  \\
		Qubit frequency (GHz) & 3.988 & 4.427 \\
		Anharmonicity (GHz) & 0.78 & 0.74 \\
		CZ gate fidelity & \multicolumn{2}{c}{0.9924}\\
		\hline\hline
	\end{tabular}\label{table-1}
\end{table}

\section{Experimental detection of local preparation contextuality and bipartite entanglement}
\label{EXPERIMENTAL 1}

Before the experimental implementation, we first use the Python toolkit PyQuafu \cite{qbac-s} to design quantum circuits locally, specifying both the circuit configuration and the number of shots. These tasks are submitted to a web server, which translates and routes them to the laboratory system. The \textit{Baihua} device executes the circuits, and raw data are preprocessed on-site before being returned via the server. During the experiment, readout fidelity matrices are obtained through additional calibration measurements. The server delivers the measured outcome counts, from which uncorrected probabilities are computed. After applying readout error correction, we calculate the relevant physical observables to obtain the final results.

In the experiment, we prepare a series of two-qubit states
\begin{equation}\label{s93}	
	\ket{\psi(\beta_i)}=\sqrt{\alpha_i}\ket{\phi_{1}}_A\ket{0}_B+\sqrt{\beta_i}\ket{\phi_{2}}_A\ket{1}_B,
\end{equation}
where $\ket{\phi_{1}}=\cos(\pi/8)\ket{0}+\sin(\pi/8)\ket{1}$, $\ket{\phi_{2}}=\sin(\pi/8)\ket{0}-\cos(\pi/8)\ket{1}$ with the parameters $\alpha_i=1-\beta_i$
in which $\beta_i$ changes from $0$ to $0.5$ and $i=1,...,13$. The product state is the case of $\beta_1=0$, and the maximally entangled state corresponds to $\beta_{13}=0.5$. The quantum circuit for state preparation is shown in Fig. 3(a) of the main text, where the corresponding relations between the state parameter $\beta_i$ and three rotation angles $(\theta_1, \theta_2, \theta_3)$ are listed in Table \ref{table-2}.

In Theorem 2 of the main text, it is revealed that there exists a quantitative trade-off relation between local preparation contextuality and bipartite entanglement. According to Eq. (11) of the main text, the optimal measurements of the EUR on subsystem $A$ are $Q=(X+Z)/\sqrt{2}$ and $R=(-X+Z)/\sqrt{2}$. By performing these two measurements, we can obtain the entropic function
\begin{equation}\label{s94}	
	H_{QR}^{(1)}(A)=H_A(Q)+H_A(R).
\end{equation}
Moreover, the measurable upper bound of quantum conditional entropy $S(A|B)$ is available via the EUR with a quantum memory, which can be written as
\begin{equation}\label{s95}	
	S(A|B)_{ub}=H(Q|Z)+H(R|X)-1,
\end{equation}
where the optimal measurements $Z$ and $X$ on Bob's side are derived through the numerical optimization. To specify their optimality, we consider generic forms of Bob's projection measurements:
\begin{equation}\label{s96} 
	\begin{split}
		M_B&=\cos\theta \cos\varphi X+\cos\theta \sin\varphi Y +\sin\theta Z,\\
		M_B'&=\cos\theta' \cos\varphi' X+\cos\theta' \sin\varphi' Y +\sin\theta' Z,
	\end{split}
\end{equation}
where $(X,Y,Z)$ are Pauli matrices in the corresponding directions and $(\theta,\theta',\varphi,\varphi')$ are the optimization parameters.

The optimization procedure of two local measurements $M_B$ and $M'_B$ is derived from the tight upper bound of the conditional entropy, which means that the goal of the numerical optimization is to minimize the function $H(Q|M_B')+H(R|M_B)-1$. Given a value of parameter $\beta$ for the bipartite quantum state $\ket{\psi(\beta)}_{AB}$, we search for the minimum of the function $H(Q|M_B')+H(R|M_B)-1$ by varying the values of the four parameters $(\theta,\theta',\varphi,\varphi')$. Here, the state parameter $\beta$ is constrained within the range of $[0,0.5]$ with a fixed step size of 0.001. Following the aforementioned numerical optimization process, we consistently identify that, irrespective of the value of state parameter $\beta$, the minimum of the function $H(Q|M_B')+H(R|M_B)-1$ corresponds to the optimization parameters $\{\theta=0,\theta'=\pi/2,\varphi=0\}$, which implies the optimal measurements for Bob's subsystem are $X$ and $Z$ as given in Eq. \eqref{s95}.

According to the numerical optimization results of local measurements $Z$ and $X$ on qubit $B$, the negative tight upper bound of  $S(A|B)_{ub}$ in Eq. \eqref{s95} is able to detect quantum entanglement in the prepared two-qubit state. The relevant local measurements $\{Q,R,Z,X\}$ are performed by modulating the rotation angles $\varphi_1$ and $\varphi_2$ in the measurement circuit of Fig. 3(a) of the main text.

For each $\beta_i$, we repeat $30$ sets of experiments, and each set consists of $3000$ single-shot measurements. As shown in Fig. 3(b) of the main text, the solid blue line is the result of theoretical prediction and the orange dashed line represents the noisy numerical simulation. Since two-qubit quantum gate is the primary source of noise, we perform the noisy numerical simulation by introducing depolarizing channels after the controlled-$Z$ gate. The probability of depolarization is set to be 0.0076 according to the CZ gate fidelity of \textit{Baihua} device listed in Table~\ref{table-1}. The detailed data of the trade-off functions for theoretical prediction, noisy numerical simulation and experimental result are listed in Table \ref{table-3}. The experimental results and the noisy simulation exhibit a good agreement, which further verify our theoretical predictions.

\begin{table}
	\centering
	\caption{The corresponding relations between the state parameter $\beta_i$ and the three rotation angles $(\theta_1, \theta_2, \theta_3)$ for the circuit of quantum state preparation in Fig. 3(a) of the main text.}
	\setlength{\tabcolsep}{10pt} 
	\renewcommand{\arraystretch}{1.2} 
	\begin{tabular}{lcccc}
		\hline\hline
		No. &	$\beta_i$  &  $\theta_1/\pi$  &    $\theta_2/\pi$    &    $\theta_3/\pi$ \\
		\hline
		1&	0.00 & 0.2500 & 0.0000  & 0.0000  \\
		2&	0.02 & 0.2625 & -0.1245 & -0.0869 \\
		3&	0.04 & 0.2745 & -0.1726 & -0.1189 \\
		4&	0.06 & 0.2862 & -0.2075 & -0.1411 \\
		5&	0.08 & 0.2976 & -0.2356 & -0.1582 \\
		6&	0.10 & 0.3086 & -0.2594 & -0.1720 \\
		7&	0.12 & 0.3194 & -0.2801 & -0.1834 \\
		8&	0.16 & 0.3403 & -0.3152 & -0.2014 \\
		9&	0.21 & 0.3655 & -0.3515 & -0.2176 \\
		10&	0.26 & 0.3898 & -0.3825 & -0.2292 \\
		11&	0.31 & 0.4134 & -0.4100 & -0.2376 \\
		12&	0.37 & 0.4411 & -0.4401 & -0.2444 \\
		13&	0.50 & 0.5000 & -0.5000 & -0.2500 \\
		\hline\hline
	\end{tabular}\label{table-2}
\end{table}

\section{Experimental detection of local preparation contextuality and Bell nonlocality}
\label{EXPERIMENTAL 2}
\begin{table*}
	\centering
	\caption{The data of theoretical prediction, noisy numerical simulation and experimental result for the functions in the trade-off relation between local preparation contextuality and bipartite entanglement: $H^{th}_{QR}(A), S(A|B)^{th}_{ub}$ are the results of theoretical prediction, $H^{num}_{QR}(A), S(A|B)^{num}_{ub}$ represent the results of noisy numerical simulation, and $H_{QR}^{(1)}(A), S(A|B)_{ub}$ correspond to experimental results.}
	\label{table-3}
	\setlength{\tabcolsep}{5pt}
	\renewcommand{\arraystretch}{1.3} 
	\begin{tabular}{lcccccccc}
		\hline\hline
		$|\psi(\beta_{i})\rangle$  &$\beta_{i}$ & $H^{th}_{QR}(A)$ & $S(A|B)^{th}_{ub}$ & $H^{num}_{QR}(A)$ & $S(A|B)^{num}_{ub}$ & $H_{QR}^{(1)}(A)$ & $S(A|B)_{ub}$\\
		\hline
		$|\psi(\beta_{1})\rangle$  & 0.00 & 1      & 0       & 1.0360 & 0.0360  & 1.0077 $\pm$ 0.0030  & 0.0074 $\pm$ 0.0030  \\
		$|\psi(\beta_{2})\rangle$  & 0.02 & 1.1414 & -0.0573 & 1.1615 & -0.0048 & 1.1529 $\pm$ 0.0040  & -0.0341 $\pm$ 0.0039 \\
		$|\psi(\beta_{3})\rangle$  & 0.04 & 1.2423 & -0.1138 & 1.2581 & -0.0561 & 1.2854 $\pm$ 0.0045  & -0.0417 $\pm$ 0.0063 \\
		$|\psi(\beta_{4})\rangle$  & 0.06 & 1.3274 & -0.1695 & 1.3406 & -0.1079 & 1.3627 $\pm$ 0.0047  & -0.1194 $\pm$ 0.0037 \\
		$|\psi(\beta_{5})\rangle$  & 0.08 & 1.4022 & -0.2243 & 1.4133 & -0.1595 & 1.4321 $\pm$ 0.0052  & -0.1651 $\pm$ 0.0057 \\
		$|\psi(\beta_{6})\rangle$  & 0.10 & 1.4690 & -0.2781 & 1.4786 & -0.2104 & 1.4905 $\pm$ 0.0045  & -0.2140 $\pm$ 0.0052 \\
		$|\psi(\beta_{7})\rangle$  & 0.12 & 1.5294 & -0.3310 & 1.5376 & -0.2604 & 1.5550 $\pm$ 0.0037  & -0.2724 $\pm$ 0.0051 \\
		$|\psi(\beta_{8})\rangle$  & 0.16 & 1.6343 & -0.4333 & 1.6405 & -0.3577 & 1.6449 $\pm$ 0.0033  & -0.3710 $\pm$ 0.0069 \\
		$|\psi(\beta_{9})\rangle$  & 0.21 & 1.7415 & -0.5546 & 1.7457 & -0.4726 & 1.7621 $\pm$ 0.0031  & -0.4968 $\pm$ 0.0063 \\
		$|\psi(\beta_{10})\rangle$ & 0.26 & 1.8268 & -0.6672 & 1.8295 & -0.5788 & 1.8376 $\pm$ 0.0030  & -0.6168 $\pm$ 0.0066 \\
		$|\psi(\beta_{11})\rangle$ & 0.31 & 1.8932 & -0.7693 & 1.8948 & -0.6740 & 1.9008 $\pm$ 0.0024  & -0.7232 $\pm$ 0.0084 \\
		$|\psi(\beta_{12})\rangle$ & 0.37 & 1.9507 & -0.8746 & 1.9514 & -0.7701 & 1.9538 $\pm$ 0.0014  & -0.8385 $\pm$ 0.0063 \\
		$|\psi(\beta_{13})\rangle$ & 0.50 & 2      & -1      & 2.0000 & -0.8716 & 1.9986 $\pm$ 0.0002  & -0.9551 $\pm$ 0.0055 \\
		\hline\hline
	\end{tabular}
\end{table*}

\begin{table*}
	\centering
	\caption{The coefficients for optimal measurements $B_0$ and $B_1$ in the probe of maximal Bell correlation, where $\beta_i$s are the state parameters, and the coefficients ($a_0,b_0,c_0$) and ($a_1,b_1,c_1$) correspond to the optimal local measurements $B_0$ and $B_1$, respectively.}
	\label{table-4}
	\setlength{\tabcolsep}{3.5pt}
	\renewcommand{\arraystretch}{1.3} 
	\begin{tabular}{ccccccccccccccc}
		\hline\hline
		\toprule
		$\beta_{i}$ & 0.00 & 0.02 & 0.04 & 0.06 & 0.08 & 0.10 & 0.12 & 0.16 & 0.21 & 0.26 & 0.31 & 0.37 & 0.50\\
		\hline
		\midrule
		$a_{0}$ & 0 & 0.2696 & 0.3649 & 0.4290 & 0.4769 & 0.5145 & 0.5449 & 0.5931 & 0.6316 & 0.6595 & 0.6790 & 0.6946 & 0.7071 \\
		$b_{0}$ & 0 & 0 & 0 & 0 & 0 & 0 & 0 & 0 & 0 & 0 & 0 & 0 & 0 \\
		$c_{0}$ & 1 & 0.9630 & 0.9310 & 0.9033 & 0.8790 & 0.8575 & 0.8385 & 0.8065 & 0.7753 & 0.7517 & 0.7341 & 0.7194 & 0.7071 \\
		\hline
		$a_{1}$ & 0 & -0.2696 & -0.3649 & -0.4290 & -0.4769 & -0.5145 & -0.5449 & -0.5931 & -0.6316 & -0.6595 & -0.6790 & -0.6946 & -0.7071 \\
		$b_{1}$ & 0 & 0 & 0 & 0 & 0 & 0 & 0 & 0 & 0 & 0 & 0 & 0 & 0 \\
		$c_{1}$ & 1 & 0.9630 & 0.9310 & 0.9033 & 0.8790 & 0.8575 & 0.8385 & 0.8065 & 0.7753 & 0.7517 & 0.7341 & 0.7194 & 0.7071 \\
		\bottomrule
		\hline\hline
	\end{tabular}
\end{table*}

\begin{table*}
	\centering
	\caption{The data of theoretical prediction, noisy numerical simulation and experimental result for the functions in the quantitative trade-off relation between local preparation contextuality and Bell nonlocality: $H^{th}_{QR}(A)$ and $\langle B\rangle^{th}_{\mathrm{max}}$ are the results of theoretical prediction, $H^{num}_{QR}(A)$ and $\langle B\rangle^{num}_{\mathrm{max}}$ represent the results of noisy simulation, and $H_{QR}^{(2)}(A)$ and $\langle B\rangle_{\mathrm{max}}$ correspond to experimental results.}
	\label{table-5}
	\setlength{\tabcolsep}{5pt}
	\renewcommand{\arraystretch}{1.3} 
	\begin{tabular}{lcccccccc}
		\hline\hline
		$|\psi(\beta_{i})\rangle$ &$\beta_{i}$ & $H^{th}_{QR}(A)$ &  $\langle B\rangle^{th}_{\mathrm{max}}$ & $H^{num}_{QR}(A)$ & $\langle B\rangle^{num}_{\mathrm{max}}$ &  $H_{QR}^{(2)}(A)$ & $\langle B\rangle_{\mathrm{max}}$ \\
		\hline
		$|\psi(\beta_{1})\rangle$  & 0.00 & 1      & 2      & 1.0360 &  1.9697 & 1.0196 $\pm$ 0.0037  & 1.9844 $\pm$ 0.0060  \\
		$|\psi(\beta_{2})\rangle$  & 0.02 & 1.1414 & 2.0769 & 1.1615 &  2.0455 & 1.2166 $\pm$ 0.0056  & 2.0100 $\pm$ 0.0080  \\
		$|\psi(\beta_{3})\rangle$  & 0.04 & 1.2423 & 2.1481 & 1.2581 &  2.1156 & 1.3002 $\pm$ 0.0048  & 2.0874 $\pm$ 0.0080  \\
		$|\psi(\beta_{4})\rangle$  & 0.06 & 1.3274 & 2.2141 & 1.3406 &  2.1806 & 1.3783 $\pm$ 0.0055  & 2.1785 $\pm$ 0.0069  \\
		$|\psi(\beta_{5})\rangle$  & 0.08 & 1.4022 & 2.2754 & 1.4133 &  2.2410 & 1.4452 $\pm$ 0.0044  & 2.2118 $\pm$ 0.0097  \\
		$|\psi(\beta_{6})\rangle$  & 0.10 & 1.4690 & 2.3324 & 1.4786 &  2.2971 & 1.5160 $\pm$ 0.0044  & 2.2914 $\pm$ 0.0068  \\
		$|\psi(\beta_{7})\rangle$  & 0.12 & 1.5294 & 2.3853 & 1.5376 &  2.3492 & 1.5568 $\pm$ 0.0043  & 2.3464 $\pm$ 0.0090  \\
		$|\psi(\beta_{8})\rangle$  & 0.16 & 1.6343 & 2.4800 & 1.6405 &  2.4424 & 1.6652 $\pm$ 0.0035  & 2.4637 $\pm$ 0.0082  \\
		$|\psi(\beta_{9})\rangle$  & 0.21 & 1.7415 & 2.5796 & 1.7457 &  2.5406 & 1.7525 $\pm$ 0.0032  & 2.5477 $\pm$ 0.0075  \\
		$|\psi(\beta_{10})\rangle$ & 0.26 & 1.8268 & 2.6605 & 1.8295 &  2.6202 & 1.8383 $\pm$ 0.0022  & 2.6583 $\pm$ 0.0092  \\
		$|\psi(\beta_{11})\rangle$ & 0.31 & 1.8932 & 2.7244 & 1.8948 &  2.6832 & 1.8987 $\pm$ 0.0017  & 2.6474 $\pm$ 0.0101  \\
		$|\psi(\beta_{12})\rangle$ & 0.37 & 1.9507 & 2.7802 & 1.9514 &  2.7381 & 1.9566 $\pm$ 0.0011  & 2.7424 $\pm$ 0.0079  \\
		$|\psi(\beta_{13})\rangle$ & 0.50 & 2      & 2.8284 & 2.0000 &  2.7856 & 1.9985 $\pm$ 0.0002  & 2.8087 $\pm$ 0.0071  \\
		\hline\hline
	\end{tabular}
\end{table*}

In the experiment for probing local preparation contextuality and Bell nonlocality, the two-qubit states are prepared using the quantum circuit shown in Fig. 3(a) of the main text, which have the same forms as those in Eq. \eqref{s93}. The state parameter $\beta_i$ ranges in $[0,0.5]$ with $i=1,2,\dots ,13$, for which the corresponding relationship between $\beta_i$ and rotation angles $(\theta_1, \theta_2, \theta_3)$ are listed in Table \ref{table-2}.

For the subsystem $A$, the witness function for the preparation contextuality in quantum state preparation is
\begin{equation}\label{s97}	
	H_{QR}^{(2)}(A)=H_A(Q)+H_A(R),
\end{equation}
where the optimal measurements are $Q=(X+Z)/\sqrt{2}$ and $R=(-X+Z)/\sqrt{2}$. These two local complementary measurements $Q$ and $R$ are also the local observables for detecting the maximal Bell correlation
\begin{equation}\label{s98}
	\langle B\rangle_{\mathrm{max}}=|\langle A_{0}B_{0}\rangle+\langle A_{0}B_{1}\rangle+\langle A_{1}B_{0}\rangle-\langle A_{1}B_{1}\rangle|,
\end{equation}
where the measurements for subsystem $A$ are
\begin{equation}\label{s99}
	A_0=Q=\frac{X+Z}{\sqrt{2}}, ~~ A_1=R=\frac{-X+Z}{\sqrt{2}},
\end{equation}
and the generic forms of local measurements for subsystem $B$ can be written as
\begin{equation}\label{s100}
	\begin{split}
		B_{0}&=a_{0}X+b_{0}Y+c_{0}Z,\\
		B_{1}&=a_{1}X+b_{1}Y+c_{1}Z,
	\end{split}
\end{equation}
where the coefficients $a_i$ and $b_i$ can be determined via the Horodecki parameter \cite{hr95pla-s}. The relation between $\langle B\rangle_{\mathrm{max}}$ and Horodecki parameter $\mathcal{M}$ is given in Eq. \eqref{s76}. After some calculation, we can derive the coefficients of local measurements $B_i$ which are listed in Table \ref{table-4}, where $\beta_i$s are the state parameters and the optimal local measurements $B_0$ and $B_1$ correspond to the coefficients ($a_0,b_0,c_0$) and ($a_1,b_1,c_1$), respectively.

In Theorem 3, we show that there exists a quantitative trade-off relation between local preparation contextuality and Bell nonlocality in the framework of quantum resource distribution, where the EUR $H_{QR}(A)$ and the deviation function $2-\langle B\rangle_{\mathrm{max}}$ are able to probe the local and nonlocal nonclassical properties. For each state $\ket{\psi(\beta_i)}$, we repeat $30$ sets of experiments, and each set consists of $3000$ single-shot measurements. As shown in Fig. 3(c) of the main text, the solid blue line is the result of theoretical prediction and the orange dashed line represents the noisy numerical simulation, where the simulation strategy is similar to that in Fig. 3(b) of the main text. The detailed data of the trade-off functions for theoretical prediction, noisy numerical simulation and experimental result are listed in Table \ref{table-5}. The experimental results and the noisy simulation exhibit a good agreement, which further verify our theoretical prediction.

Here, it should be noted that, due to the physical limitations on \textit{Baihua} chip, a loophole-free experiment of Bell nonlocality verification for spatially separated superconducting qubits such as the one conducted in Ref. \cite{ss23nat-s} is not performed.

\section{Probe of quantum contextuality via a single optimal measurement}
\label{SINGLE}

\textit{Proposition 1.}---Let $\rho_{A}$ be a single-qubit state with the optimal measurement $Q$ along the direction of its polarization vector. Then the preparation contextuality in the optimal four-state set $\Lambda(\rho_A)$ with the $B_2$ symmetry can be witnessed by the inequality
\begin{equation}\label{s101}
	D_f=\mathrm{max} \left \{|P_0-P_1|-\sqrt{2}/2,0 \right \}>0,
\end{equation}
where $P_0$ and $P_1$ represent the two-outcome probabilities of the measurement $Q$ on the given quantum state.

\begin{table}[b]
	\centering
	\caption{The parameter $\beta_i$ for the prepared states and the experimental data for the indicators  $D_f^{(1)}$ and $D_f^{(2)}$ from two independent experiments.}
	\setlength{\tabcolsep}{8pt} 
	\renewcommand{\arraystretch}{1.2} 
	\begin{tabular}{cccc}
		\hline\hline
		No.  &	 $\beta_i$  &  $D_f^{(1)}$  &    $D_f^{(2)}$  \\
		\hline
		1&	0.00 & 0.2879  $\pm$ 0.0012 & 0.2884  $\pm$ 0.0009\\
		2&	0.02 & 0.2202  $\pm$ 0.0030 & 0.2229  $\pm$ 0.0023\\
		3&	0.04 & 0.1838  $\pm$ 0.0036 & 0.1854  $\pm$ 0.0024\\
		4&	0.06 & 0.1454  $\pm$ 0.0037 & 0.1452  $\pm$ 0.0030\\
		5&	0.08 & 0.1098  $\pm$ 0.0034 & 0.1068  $\pm$ 0.0027\\
		6&	0.10 & 0.0614  $\pm$ 0.0044 & 0.0609  $\pm$ 0.0030\\
		7&	0.12 & 0.0229  $\pm$ 0.0046 & 0.0319  $\pm$ 0.0032\\
		8&	0.16 & -0.0521 $\pm$ 0.0042 & -0.0551 $\pm$ 0.0032\\
		9&	0.21 & -0.1529 $\pm$ 0.0045 & -0.1409 $\pm$ 0.0034\\
		10&	0.26 & -0.2458 $\pm$ 0.0042 & -0.2449 $\pm$ 0.0031\\
		11&	0.31 & -0.3370 $\pm$ 0.0047 & -0.3398 $\pm$ 0.0030\\
		12&	0.37 & -0.4644 $\pm$ 0.0049 & -0.4683 $\pm$ 0.0031\\
		13&	0.50 & -0.6838 $\pm$ 0.0025 & -0.6842 $\pm$ 0.0018\\
		\hline\hline
	\end{tabular}\label{table-6}
\end{table}

\textit{Proof.}---Following an analogous analysis to that presented in Eqs. \eqref{s01}-\eqref{s12} of Sec. I, we can derive that the optimal four-state set $\Lambda(\rho_A)$ is preparation noncontextual when the condition in Eq. \eqref{s13} holds.
In order to give an alternative witness for the  preparation contextuality in the four-state set $\Lambda(\rho_A)$, we define an indicator function
\begin{equation}\label{s102}
	\begin{split}
		D'_f &=(2\langle M_1 \rangle+4c+\gamma+\epsilon+\delta-1)_{\min}\\
		&=2\langle M_1 \rangle+c-1\\
		&=\sqrt{2}|P_0-P_1|+c-1,
	\end{split}
\end{equation}
where $\langle M_1 \rangle$ is the expectation value of the measurement $M_1$ on the single qubit state $\rho_A$, in the second equality we use the ranges of parameters $\epsilon, \gamma, \delta\in{[-c,1-c]}$, and the outcome probabilities of the measurement $Q$ in the third equality are
\begin{equation}\label{s103}
	\begin{split}
		P_0=\tr(Q_+\rho_A)=\frac{1}{2}+\frac{\sqrt{2}\langle M_1 \rangle}{2},\\
		P_1=\tr(Q_-\rho_A)=\frac{1}{2}-\frac{\sqrt{2}\langle M_1 \rangle}{2}.
	\end{split}
\end{equation}
Here, $Q_{\pm}$ are the eigenvectors of the observable $Q$ as shown in Eq. \eqref{s64}, and $\rho_A$ takes the diagonal form in Eq. \eqref{s63} with $\langle M_1 \rangle=\sqrt{2}s_1/2$ being the expectation value.

According to Eq. \eqref{s102}, when the probability difference $|P_0-P_1|\leq \sqrt{2}/2$, the indicator function $D'_f$ can be zero by modulating the parameter $c$ ranging in $[0,1]$, which means that the optimal four-state set $\Lambda(\rho_A)$ admits of a noncontextual ontological explanation. On the other hand, when the probability difference satisfies
\begin{equation}\label{104}
	|P_0-P_1| > \sqrt{2}/2,
\end{equation}
the function $D'_f\neq 0$ which indicates the optimal four-state set $\Lambda(\rho_A)$ being preparation contextual. Therefore, we obtain a convenient criterion based on the probability difference of the two-outcome measurement $Q$. The criterion is stated as
\begin{equation}\label{s105} 
	D_f=\mathrm{max} \left \{ |P_0-P_1|-\sqrt{2}/2,0 \right \}>0,
\end{equation}
which indicates the operational contextuality in the preparation of the optimal four-state set $\Lambda(\rho_A)$ when the probability difference $|P_0-P_1|>\sqrt{2}/2$. Then, the proof of Proposition 1 is completed.
\hfill$\blacksquare$

\begin{figure}
	\centering
	\epsfig{figure=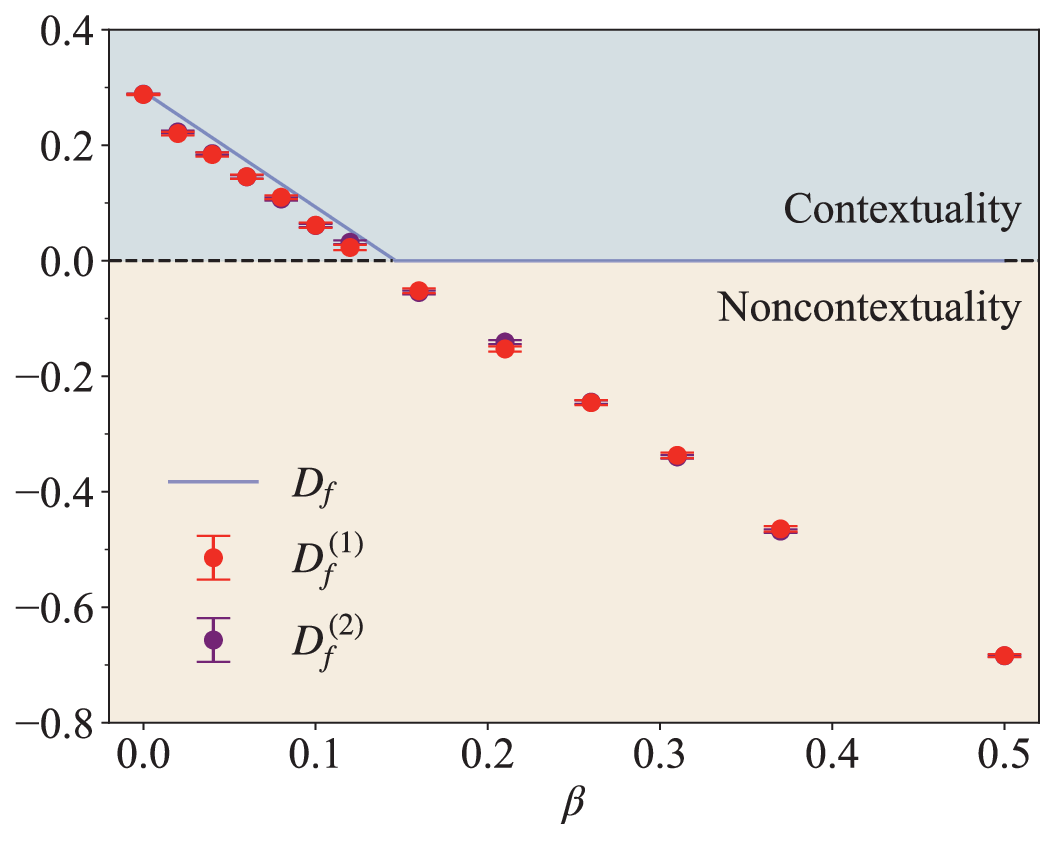,width=0.45\textwidth}
	\caption{The indicator functions $D_f$, $D_f^{(1)}$ and $D_f^{(2)}$ for witnessing the preparation  contextuality. The blue solid line is the theoretical prediction of $D_f$ with the boundary for contextuality and noncontextuality being represented by the black dashed line. The red data points are the experimental result for $D_f^{(1)}(\beta_i)$, and the purple data points are the experimental result for $D_f^{(2)}(\beta_i)$, respectively.}
	\label{figs6}
\end{figure}

The new criterion in Eq. \eqref{s101} is further verified in the superconducting qubit on $Baihua$ device. The experimental data come from the measurement $Q$ on the subsystem $\rho_A$ traced from the prepared two-qubit states $\ket{\psi(\beta_i)}$ as shown in Eq. \eqref{s93} with $i=1, 2, \dots, 13$, which comes from two independent experiments. The first set of experimental data are obtained from the detection of local preparation contextuality and bipartite entanglement, where we use the indicator function
\begin{equation}\label{s106} 
	D_f^{(1)}(\beta_i)= |P_0-P_1|-\sqrt{2}/2.
\end{equation}
The second set of experimental data are achieved from the detection of local preparation contextuality and Bell nonlocality, where we use the indicator function
\begin{equation}\label{s107}
	D_f^{(2)}(\beta_i)= |P_0-P_1|-\sqrt{2}/2.
\end{equation}
The experimental data and the corresponding relation with the prepared states are listed in Table \ref{table-6}. The error bars represent the standard error obtained from $30$ replicates of the experiment and each replicate consists of $3000$ single-shot measurements for the prepared state. In Fig. \ref{figs6}, the theoretical predication for indicator $D_f$, experimental results for the indicators  $D_f^{(1)}$ and $D_f^{(2)}$ are plotted as the function of the state parameter $\beta_i$, which can effectively probe the preparation contextuality. These probing results have a good agreement with those in Figs. 3(b) and 3(c) of the main text, which verifies the validity of our criterion in the proposition.

\section{The faithful criterion for witnessing preparation contextuality in the optimal eight-state set}
\label{IX}
In this section, we provide a faithful criterion for preparation contextuality in the optimal eight-state set $\widetilde{\Lambda}(\vec{s}_1)$ associated with an arbitrary single-qubit state $\vec{s}_1$. First, we construct an eight-state set of $\vec{s}_1$ via the $B_3$-orbit realizability condition and prove the optimality in Sec. \hyperref[IXA]{\text{IXA}}. Then, in Sec. \hyperref[IXB]{\text{IXB}}, we give the proof for this faithful criterion for witnessing preparation contextuality of the optimal eight-state set $\widetilde{\Lambda}(\vec{s}_1)$.

\subsection{The optimal eight-state set with $B_3$ symmetry}
\label{IXA}
\emph{Lemma 2}.---For an arbitrary single-qubit state $\vec{s}_1$, the eight-state set $\widetilde{\Lambda}(\vec{s}_1)$ generated by the $B_3$-orbit realizability condition in the Bloch sphere is the optimal eight-state set.

\emph{Proof.} We start from an arbitrary single-qubit state in Eq. \eqref{s02} which is represented by the polarization vector $\vec{s}_{1}$ in a Bloch sphere. Then we can find the other seven states $\vec{s}_2$, $\vec{s}_3$, $\vec{s}_4$, $\vec{s}_5$, $\vec{s}_6$, $\vec{s}_7$ and $\vec{s}_8$ under the action of the symmetry group of a cube with reflections in a Bloch sphere spanned by $\hat{M}_1$, $\hat{M}_2$ and $\hat{M}_3$ axes (the Coxeter group $B_3$ \cite{hum90book-s}), which satisfies the following equal predictability property
{
\setlength{\abovedisplayskip}{5pt}    
\setlength{\belowdisplayskip}{5pt}
\begin{eqnarray}\label{s131}
\langle M_1\rangle_{\vec{s}_{1}}&=&\langle M_2\rangle_{\vec{s}_{1}}=\langle M_3\rangle_{\vec{s}_{1}},\nonumber\\
\langle M_1\rangle_{\vec{s}_{1}}&=&\langle M_1\rangle_{\vec{s}_{2}}=\langle M_1\rangle_{\vec{s}_{3}}=\langle M_1\rangle_{\vec{s}_{4}}=-\langle M_1\rangle_{\vec{s}_{5}}\nonumber\\
&=&-\langle M_1\rangle_{\vec{s}_{6}}=-\langle M_1\rangle_{\vec{s}_{7}}=-\langle M_1\rangle_{\vec{s}_{8}},\nonumber\\
\langle M_2\rangle_{\vec{s}_{1}}&=&\langle M_2 \rangle_{\vec{s}_{2}}=\langle M_2\rangle_{\vec{s}_{5}}=\langle M_2\rangle_{\vec{s}_{6}}=-\langle M_2\rangle_{\vec{s}_{3}}\nonumber\\	
&=&-\langle M_2 \rangle_{\vec{s}_{4}}=-\langle M_2\rangle_{\vec{s}_{7}}=-\langle M_2\rangle_{\vec{s}_{8}},\nonumber\\
\langle M_3\rangle_{\vec{s}_{1}}&=&\langle M_3 \rangle_{\vec{s}_{3}}=\langle M_3\rangle_{\vec{s}_{5}}=\langle M_3\rangle_{\vec{s}_{7}}=-\langle M_3\rangle_{\vec{s}_{2}}\nonumber\\	
&=&-\langle M_3 \rangle_{\vec{s}_{4}}=-\langle M_3\rangle_{\vec{s}_{6}}=-\langle M_3\rangle_{\vec{s}_{8}}.	\hspace{1cm}
\end{eqnarray}}

\begin{figure}
	\centering
	\epsfig{figure=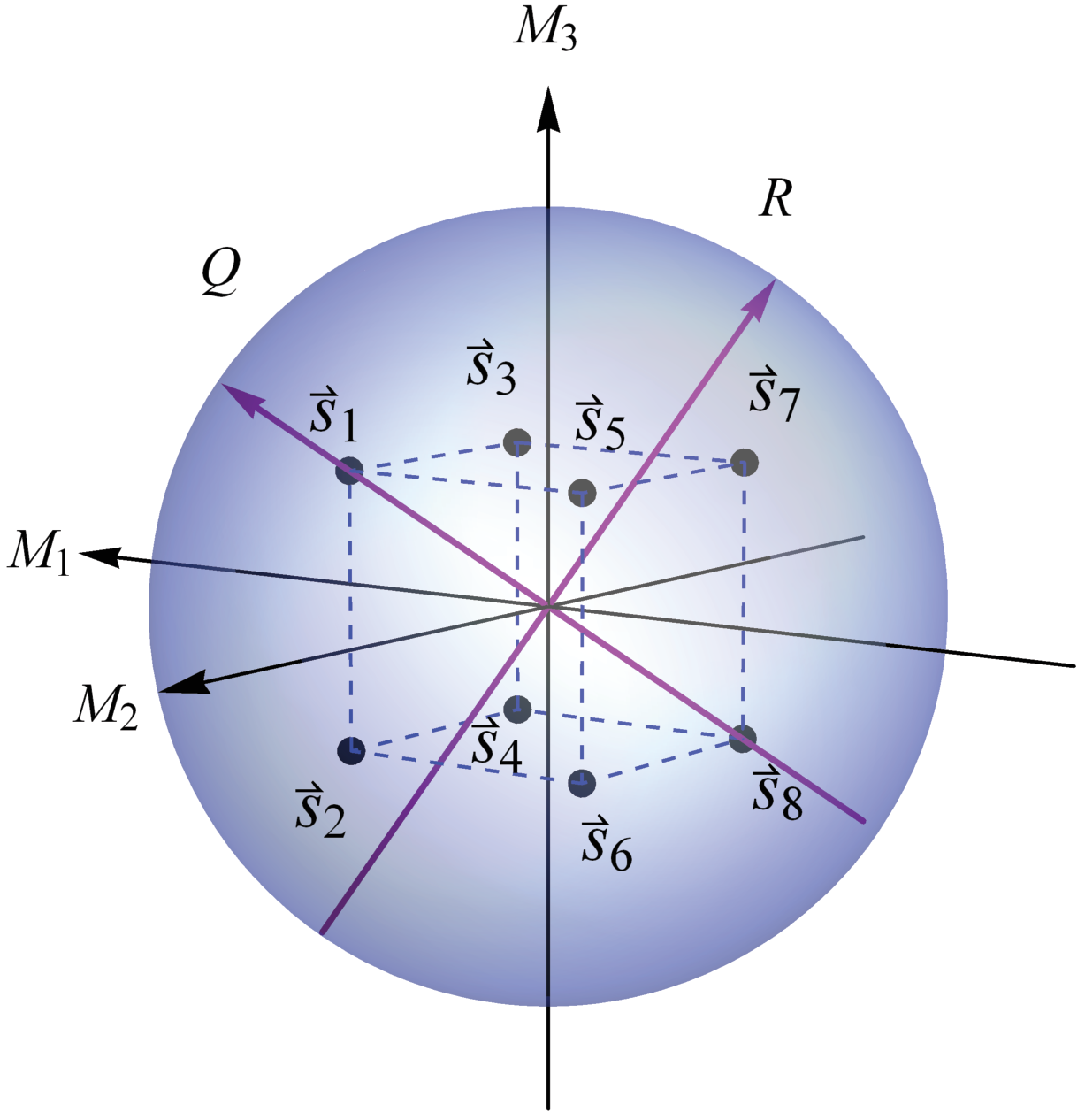,width=0.40\textwidth}
	\caption{The optimal eight-state set $\widetilde{\Lambda}_{\vec{s}_{1}}$ of a single-qubit state $\vec{s}_1$  generated by the $B_3$-orbit realizability with three complementary measurements $M_1$, $M_2$ and $M_3$ in a Bloch sphere, where the pair of measurements $Q$ and $R$ are optimal observables for the EUR in Eq. \eqref{s144} to faithfully witness the preparation contextuality.}
	\label{figs12}
\end{figure}

The eight-state set $\widetilde{\Lambda}_{\vec{s}_{1}}$, generated by the $B_3$ symmetry, satisfies the operational equivalence relation
\begin{eqnarray}\label{s132}
\frac{1}{2}\vec{s}_{1}+\frac{1}{2}\vec{s}_{8}&=&\frac{1}{2}\vec{s}_{2}+\frac{1}{2}\vec{s}_{7}\nonumber\\
&=&\frac{1}{2}\vec{s}_{3}+\frac{1}{2}\vec{s}_{6}\nonumber\\
&=&\frac{1}{2}\vec{s}_{4}+\frac{1}{2}\vec{s}_{5},
\end{eqnarray}
which is related four equivalent preparation procedures. We refer to the relations in Eqs. \eqref{s131} and \eqref{s132} as the $B_3$-orbit realizability condition. As shown in Fig. \ref{figs12}, the eight-state set is illustrated in a Bloch sphere, where the single-qubit state  $\rho(\vec{s}_1)=(I+\vec s_1\cdot \vec{M})/2$, with $\vec{M}=\{M_1,M_2,M_3\}$ being the operator basis, and with the polarization vector $\vec{s}_{1}$ having the form
\begin{eqnarray}\label{s133}
	\vec{s}_{1}=\frac{1}{\sqrt{3}}(M_{1}+M_{2}+M_{3}).
\end{eqnarray}
Then the expectation values of the three mutually orthogonal observables $\{M_1, M_2, M_3\}$ on the single-qubit state $\rho(\vec{s}_1)$ can be expressed as
\begin{equation}\label{s134}
	\begin{split}
		\langle M_1 \rangle_{\vec s_1}&=\tr(M_1\rho)=\frac{1}{\sqrt{3}}s_1,\\
		\langle M_2 \rangle_{\vec s_1}&=\tr(M_2\rho)=\frac{1}{\sqrt{3}}s_1,\\
        \langle M_3 \rangle_{\vec s_1}&=\tr(M_3\rho)=\frac{1}{\sqrt{3}}s_1.
	\end{split}
\end{equation}

After the similar analyses as those in Eqs. \eqref{s06}-\eqref{s14} of the proof for Lemma 1, we can derive the operational property based on the generalized noncontextual ontological model corresponding to the equal predictability in Eq. \eqref{s131} and the operational equivalence in Eq. \eqref{s132}. For the eight-state set $\widetilde{\Lambda}_{\vec{s}_{1}}$ associated with the single-qubit state $\vec{s}_{1}$, preparation noncontextuality implies
\begin{equation}\label{s135}
	\langle M_1 \rangle_{\vec s_1}=\langle M_2 \rangle_{\vec s_1}=\langle M_3 \rangle_{\vec s_1}\leq\frac{1}{3}.
\end{equation}
On the other hand, the eight-state set $\widetilde{\Lambda}_{\vec{s}_1}$ exhibits preparation contextuality whenever the expectation values obey
\begin{equation}\label{s136}
	\langle M_1 \rangle_{\vec s_1}=\langle M_2 \rangle_{\vec s_1}=\langle M_3 \rangle_{\vec s_1}>\frac{1}{3}.
\end{equation}

Next, we prove the optimality of the eight-state set $\widetilde{\Lambda}_{\vec{s}_1}$ with $B_3$ symmetry. The optimal set means that any other eight-state set $\widetilde{\Lambda}'_{\vec{s}_1}$ with the $A_1^3$ symmetry admits of the generalized noncontextual ontological model provided that the eight-state set $\widetilde{\Lambda}_{\vec{s}_1}$  is preparation noncontextual, where it is noted that the Coxeter $A_1^3$ group is a subgroup of $B_3$.

We consider an arbitrary operator basis $\vec M'=\{M_1',M_2',M_3'\}$ in the Bloch sphere. Then, for any given single-qubit state $\vec{s_1}$, the other seven states $\vec{s_i}'$ with $i=2, 3, 4, 5, 6, 7, 8$ can be obtained by the reflection actions under the Coxeter $A_1^3$ group \cite{hum90book}. The eight-state set $\widetilde{\Lambda}'_{\vec{s}_1}$ satisfies the $A_1^3$ equal predictability of a cuboid, which is expressed as
\begin{eqnarray}\label{s137}
\langle M_1'\rangle_{\vec{s}_{1}}&=&\langle M_1'\rangle_{\vec{s}_{2}'}=\langle M_1'\rangle_{\vec{s}_{3}'}=\langle M_1'\rangle_{\vec{s}_{4}'}=-\langle M_1'\rangle_{\vec{s}_{5}'}\nonumber\\
&=&-\langle M_1'\rangle_{\vec{s}_{6}'}=-\langle M_1'\rangle_{\vec{s}_{7}'}=-\langle M_1'\rangle_{\vec{s}_{8}'},\nonumber\\
\langle M_2'\rangle_{\vec{s}_{1}}&=&\langle M_2' \rangle_{\vec{s}_{2}'}=\langle M_2'\rangle_{\vec{s}_{5}'}=\langle M_2'\rangle_{\vec{s}_{6}'}=-\langle M_2'\rangle_{\vec{s}_{3}'}\nonumber\\	
&=&-\langle M_2' \rangle_{\vec{s}_{4}'}=-\langle M_2'\rangle_{\vec{s}_{7}'}=-\langle M_2'\rangle_{\vec{s}_{8}'},\nonumber\\
\langle M_3'\rangle_{\vec{s}_{1}}&=&\langle M_3' \rangle_{\vec{s}_{3}'}=\langle M_3'\rangle_{\vec{s}_{5}'}=\langle M_3'\rangle_{\vec{s}_{7}'}=-\langle M_3'\rangle_{\vec{s}_{2}'}\nonumber\\	
&=&-\langle M_3' \rangle_{\vec{s}_{4}'}=-\langle M_3'\rangle_{\vec{s}_{6}'}=-\langle M_3'\rangle_{\vec{s}_{8}'}.	\hspace{1cm}
\end{eqnarray}
According to the polarization vector representation in Fig. \ref{figs1}, the expectation values of the three mutually orthogonal operators $\{M_1',M_2',M_3'\}$ on the single-qubit state $\vec{s}_1$ have the forms
\begin{equation}\label{s138}
	\begin{split}
		\langle M_1' \rangle_{\vec s_1}&=\tr(M_1'\rho)=s_1\cos\omega_1\cos\omega_2,\\
		\langle M_2' \rangle_{\vec s_1}&=\tr(M_2'\rho)=s_1\sin\omega_1,\\
        \langle M_3' \rangle_{\vec s_1}&=\tr(M_1'\rho)=s_1\cos\omega_1\sin\omega_2.
	\end{split}
\end{equation}
Then, based on the $A_1^3$ equal predictability in Eq. \eqref{s137} and preparation equivalence of eight-state set  $\widetilde{\Lambda}'_{\vec{s}_1}$, we can derive a joint predictability condition via the generalized noncontextual ontological model. More concretely, the condition is that the eight-state set $\widetilde{\Lambda}'_{\vec{s}_1}$ with the $A_1^3$ symmetry is preparation noncontextual when the joint predictability of the three mutually orthogonal observables $\{M_1', M_2', M_3'\}$ on $\vec{s}_1$ satisfies
\begin{eqnarray}\label{s139}
	\langle M_1' \rangle_{\vec{s}_1}+\langle M_2' \rangle_{\vec{s}_1}+\langle M_3' \rangle_{\vec{s}_1}&=&s_1\cos\omega_1\cos\omega_2+s_1\sin\omega_1\nonumber\\
&+&s_1\cos\omega_1\sin\omega_2\nonumber\\
	&\leq& 1.
\end{eqnarray}
Correspondingly, when the joint predictability violates the above relation and obeys the inequality
\begin{equation}\label{s140}
	\langle M_1' \rangle_{\vec{s}_1}+\langle M_2' \rangle_{\vec{s}_1}+\langle M_3' \rangle_{\vec{s}_1} > 1,
\end{equation}
the preparation procedure of the eight-state set  $\widetilde{\Lambda}'_{\vec{s}_1}$ cannot be explained by the generalized noncontextual ontological model and exhibits the operational contextuality.

In this stage, we analyze the maximum of the joint predictability of $M_1'$, $M_2'$ and $M_3'$, which can be written as
\setlength{\abovedisplayskip}{5pt}    
\setlength{\belowdisplayskip}{5pt}
\begin{eqnarray}\label{s141}
	\langle M_1' \rangle_{\vec s_1}+\langle M_2'  \rangle_{\vec s_1}+\langle M_3' \rangle_{\vec{s}_1}
	&=&s_1\cos\omega_1\cos \omega_2+s_1\sin\omega_1\nonumber\\
	&+&s_1\cos\omega_1\sin\omega_2\nonumber\\
	&\leq& \sqrt{2} s_1\cos\omega_1+s_1\sin\omega_1\nonumber\\
	&\leq& \sqrt{3}s_1\nonumber\\
	&=&3\langle M_1 \rangle_{\vec{s}_1},
\end{eqnarray}
where in the first inequality we choose the longitude $\omega_2=\pi/4$, the second inequality holds for the latitude $\omega_1=\arcsin\frac{1}{\sqrt{3}}$, and in the last equality we use the property for the observable $M_1$ in Eq. \eqref{s134}. Furthermore, combining Eq. \eqref{s141} with the noncontextual criteria in Eqs. \eqref{s135} and \eqref{s139} for the eight-state sets with the $B_3$ and $A_1^3$ symmetries respectively, we can obtain the following relation
\begin{eqnarray}\label{s142}
\langle M_1 \rangle_{\vec s_1}=\langle M_2  \rangle_{\vec s_1}&=&\langle M_3  \rangle_{\vec s_1}\nonumber\\
&\leq& 1/3\nonumber\\
&\Rightarrow& \langle M_1' \rangle_{\vec s_1}+\langle M_2'  \rangle_{\vec s_1}+\langle M_3' \rangle_{\vec s_1}\nonumber\\
&\leq& 1.
\end{eqnarray}
The above relation shows the optimality of the eight-state set $\widetilde{\Lambda}_{\vec{s}_1}$ with the $B_3$ symmetry, and indicates that any other eight-state set $\widetilde{\Lambda}'_{\vec{s}_1}$ with the $A_1^3$ symmetry admits of the generalized noncontextual ontological model provided that the optimal set $\widetilde{\Lambda}_{\vec{s}_1}$ is preparation noncontextual. The proof of Lemma 2 is completed. \hfill$\blacksquare$

\subsection{The faithful criterion for the optimal eight-state set}
\label{IXB}
For any given single-qubit state $\rho(\vec{s_1})$ in the Bloch sphere, we can generate the optimal eight-state set $\widetilde{\Lambda}_{\vec{s}_1}$ with the $B_3$ symmetry in Eqs. \eqref{s131} and \eqref{s132}. In this subsection, we first give the faithful criterion for witnessing preparation contextuality in the optimal eight-state set, and then provide an analytical proof for this criterion.

\emph{Theorem 4 \label{Theorem 4}}.---For a single-qubit state $\vec{s}_{1}$, the optimal eight-state set $\widetilde{\Lambda}_{\vec{s}_{1}}$ exhibits the preparation contextuality if and only if the EUR resulting from a pair of optimal complementary observables $Q$ and $R$ on $\vec{s}_1$ satisfies
{
\setlength{\abovedisplayskip}{15pt}    
\setlength{\belowdisplayskip}{15pt}
\begin{equation}\label{s143}
	H(Q)+H(R)< 1+C',
\end{equation}}
where the constant is a binary entropy function $C'=h[(3-\sqrt{3})/6]\simeq 0.7440$, and $H(Q)$ and $H(R)$ are the Shannon entropies of the post-measurement states with the two observables as shown in Fig. \ref{figs12} having the forms
{
\setlength{\abovedisplayskip}{15pt}    
\setlength{\belowdisplayskip}{15pt}
\begin{eqnarray}\label{s144}
	Q&=&\frac{1}{\sqrt{3}}M_1+\frac{1}{\sqrt{3}}M_2+\frac{1}{\sqrt{3}}M_3,\nonumber\\
	R&=&-\frac{1}{\sqrt{6}}M_1-\frac{1}{\sqrt{6}}M_2+\frac{2}{\sqrt{6}}M_3.
\end{eqnarray}}

\textit{Proof.} As shown in Fig. \ref{figs12}, for an arbitrary single-qubit state $\rho(\vec{s}_1)=(I+\vec{s}_1\cdot \vec{M})/2$ in the Bloch sphere, the two observables $Q$ and $R$ can be expressed as Eq. \eqref{s144} in the operator basis $\{M_1, M_2, M_3\}$. We first consider the observable $Q$, which is a two-outcome measurement and can be written as the diagonal representation
\begin{eqnarray}\label{s145}
	Q=Q_+-Q_-,	
\end{eqnarray}
where the two projectors $Q_+$ and $Q_-$ correspond to the eigenvectors with the eigenvalues $+1$ and $-1$ respectively. In the operator basis $\{M_1, M_2, M_3\}$, the two projectors have the forms
\begin{eqnarray}\label{s146}
	Q_+&=&\frac{I+\frac{1}{\sqrt{3}}M_1+\frac{1}{\sqrt{3}}M_2+\frac{1}{\sqrt{3}}M_3}{2},\nonumber\\
	Q_-&=&\frac{I-\frac{1}{\sqrt{3}}M_1-\frac{1}{\sqrt{3}}M_2-\frac{1}{\sqrt{3}}M_3}{2}.
\end{eqnarray}
Therefore, for the given single-qubit state $\rho(\vec{s}_1)$ with the polarization vector $\vec{s}_1$ in Eq. \eqref{s133}, the Shannon entropy of the post-measurement state can be expressed as
\begin{eqnarray}\label{s147}
	H(Q)=h(q)=h\left(\frac{1}{2}-\frac{\sqrt{3}}{2}\langle M_1\rangle_{\vec{s}_1}\right),
\end{eqnarray}
where $h(\cdot)$ is the binary entropy function and $q$ is the probability of the outcome for projector $Q_-$.

Similarly, the two-outcome measurement $R$ in its diagonal representation can be written as
\begin{equation}\label{s148}
	R=R_+-R_-,
\end{equation}
where two projectors corresponding to the eigenvectors with eigenvalues $\pm 1$ have the forms
\begin{eqnarray}\label{s149}
	R_+&=&\frac{I-\frac{1}{\sqrt{6}}M_1-\frac{1}{\sqrt{6}}M_2+\frac{2}{\sqrt{6}}M_3}{2},\nonumber\\
	R_-&=&\frac{I+\frac{1}{\sqrt{6}}M_1+\frac{1}{\sqrt{6}}M_2-\frac{2}{\sqrt{6}}M_3}{2}.
\end{eqnarray}
Then, for the single-qubit state $\rho(\vec{s}_1)$, the Shannon function of the post-measurement state can be expressed as
\begin{eqnarray}\label{s150}
	H(R)=h(r)=h\left(\frac{1}{2}\right)=1,
\end{eqnarray}
where we let $r=1/2$ be the probability of the outcome for projector $R_-$.

\begin{figure}
	\epsfig{figure=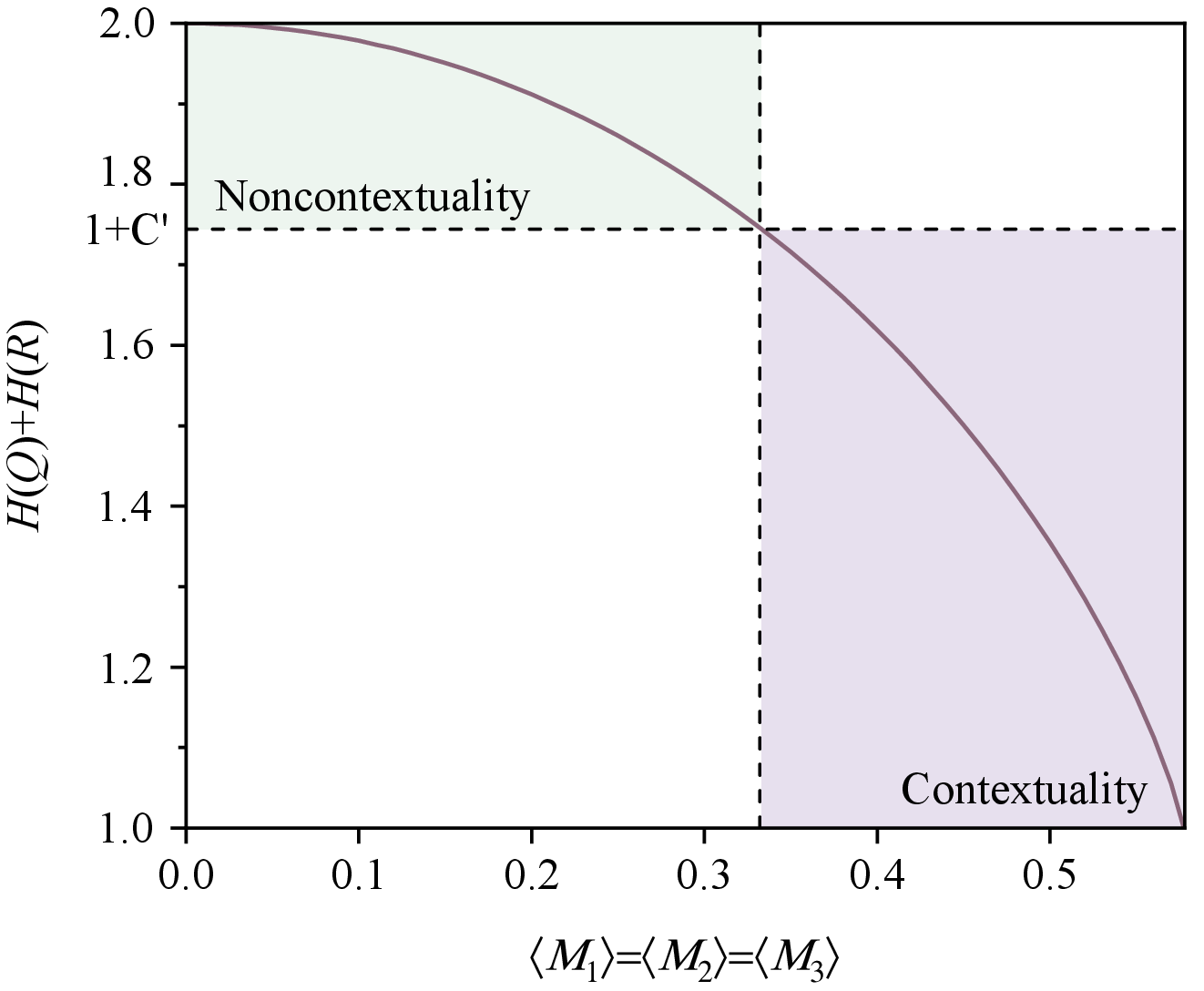,width=0.45\textwidth}
	\caption{The quantitative illustration of the faithful criterion in Theorem 4, where the EUR $H(Q)+H(R)$ (the brown solid line) in the upper left region (the green area) indicates the noncontextuality in the QSP of $\widetilde{\Lambda}_{\vec{s}_1}$ and the EUR in lower right region (the purple area) denotes the preparation contextuality of the set $\widetilde{\Lambda}_{\vec{s}_1}$.}
	\label{figs13}
\end{figure}

Combining Eqs. \eqref{s147} and \eqref{s150}, we can obtain the EUR of the pair of complementary observables $Q$ and $R$
\begin{equation}\label{s151}
	H(Q)+H(R)=1+h\left(\frac{1}{2}-\frac{\sqrt{3}}{2}\langle M_1\rangle_{\vec{s}_1}\right).
\end{equation}
The EUR is a monotonically decreasing function of the parameter $\langle M_1\rangle_{\vec{s}_1}$, where the expectation value ranges in $[0, 1/\sqrt{3}]$ and satisfies the property $\langle M_1\rangle_{\vec{s}_{1}}=\langle M_2\rangle_{\vec{s}_{1}}=\langle M_3\rangle_{\vec{s}_{1}}$ due to the $B_3$ symmetry of the optimal eight-state set $\widetilde{\Lambda}_{\vec{s}_1}$ as shown in Fig. \ref{figs12}.

Next, based on the criterion in Eq. \eqref{s136} and the EUR in Eq. \eqref{s151} as well as its monotonic property, we first prove the sufficient condition of the faithful criterion in Eq. \eqref{s143} of Theorem 4. After some analysis, we have
\begin{eqnarray}\label{s152}
H(Q)+H(R)&<&1+C'=1+h\left(\frac{3-\sqrt{3}}{6}\right)\nonumber\\	
	&\Rightarrow&\frac{1}{2}-\frac{\sqrt{3}}{2}\langle M_1\rangle_{\vec{s}_{1}}<\frac{1}{2}-\frac{\sqrt{3}}{6}\nonumber\\
	&\Rightarrow&\langle M_1\rangle_{\vec{s}_{1}} >\frac{1}{3}\nonumber\\
	&\Rightarrow&\langle M_1\rangle_{\vec{s}_{1}} =\langle M_2\rangle_{\vec{s}_{1}}=\langle M_3\rangle_{\vec{s}_{1}}\nonumber\\
	&>&\frac{1}{3},
\end{eqnarray}
where in the first equality we use the constant $C'=h[(3-\sqrt{3})/6]$, in the second inequality we adopt the expression of EUR in Eq. \eqref{s151} and the monotonic property of binary function $h(\cdot)$, and in the last inequality the $B_3$ symmetry $\langle M_1\rangle_{\vec{s}_{1}}=\langle M_2\rangle_{\vec{s}_{1}}=\langle M_3\rangle_{\vec{s}_{1}}$ is utilized. According to Eq. \eqref{s152}, we can obtain that when the criterion in Eq. \eqref{s143} of Theorem 4 is satisfied the optimal eight-state set $\widetilde{\Lambda}_{\vec{s}_1}$ is preparation contextual and cannot be explained by the generalized noncontextual ontological model. Therefore, the sufficient condition of Theorem 4 is proved.

Next, we prove the necessary condition of Eq. \eqref{s143} in Theorem 4. When the criterion $H(Q)+H(R)<1+C'$ is violated, we can derive the following relations
\begin{eqnarray}\label{s153}
H(Q)+H(R)&\geq& 1+C'=1+h\left(\frac{3-\sqrt{3}}{6}\right)\nonumber\\	
&\Rightarrow& \frac{1}{2}-\frac{\sqrt{3}}{2}\langle M_1\rangle_{\vec{s}_{1}}\geq\frac{1}{2}-\frac{\sqrt{3}}{6}\nonumber\\
&\Rightarrow& \langle M_1\rangle_{\vec{s}_{1}} \leq\frac{1}{3}\nonumber\\
&\Rightarrow& \langle M_1\rangle_{\vec{s}_{1}} =\langle M_2\rangle_{\vec{s}_{1}}=\langle M_3\rangle_{\vec{s}_{1}}\nonumber\\
&\leq&\frac{1}{3},
\end{eqnarray}
where we use Eqs. \eqref{s135} and \eqref{s151} as well as the monotonic property of the EUR function. Thus, according to Eq. \eqref{s153}, we can conclude that, when the EUR criterion in Theorem \hyperref[Theorem 4]{{4}} is violated, the optimal eight-state set $\widetilde{\Lambda}_{\vec{s}_1}$ is noncontextual and admits of a classical ontological model with preparation equivalence.

Combing Eqs. \eqref{s152} and \eqref{s153}, we obtain that the criterion in Eq. \eqref{s143} of Theorem 4 is faithful. The quantitative illustration of this faithful criterion is furhter shown in Fig. \ref{figs13}. The proof of Theorem 4 is completed.
\hfill$\blacksquare$

\section{Trade-off relations for preparation contextuality in the optimal eight-state set}
\label{X}
We consider a shared two-qubit state $\rho_{AB}$  with the reduced state of subsystem A being
\begin{equation}\label{s154}
\rho(A)=(I+\vec s_1\cdot \vec{M})/2,
\end{equation}
where $\vec{s}_{1}=(M_{1}+M_{2}+M_{3})/\sqrt{3}$ is the polarization vector. For the post-measurement state of subsystem A, we can obtain that the optimal measurement $Q$ is along the direction of the polarization vector $\vec{s}_1$ of the reduced quantum state $\rho_A$, and has the form
\begin{equation}\label{s155}
	Q=\frac{1}{\sqrt{3}}(M_{1}+M_{2}+M_{3}).
\end{equation}
Then, for the observable $Q$, the post-measurement state of $\rho_A$ can be written as
\begin{eqnarray}\label{s156}
	\rho_Q^A&=&\frac{1}{2}(1+s_1)Q_++\frac{1}{2}(1-s_1)Q_-\nonumber\\
	&=&\rho_A,
\end{eqnarray}
where $s_1$ is the norm of the polarization vector, $Q_\pm$ denote the eigenvectors corresponding to the eigenvalues $\pm 1$, and satisfy $Q=Q_+-Q_-$ with the expressions given in Eq. \eqref{s146}.

Therefore, the corresponding Shannon entropy for the post-measurement state $\rho_Q^A$ is
\begin{equation}\label{s157}
	H(Q)=h\left[\frac{1+s_1}{2}\right]=S(A),
\end{equation}
where $S(A)=-\mbox{Tr}(\rho_A\mbox{log}\rho_A)$ is the von Neumann entropy of the subsystem A.
For another measurement $R$ with the form
\begin{equation}\label{s158}
	R=-\frac{1}{\sqrt{6}}M_1-\frac{1}{\sqrt{6}}M_2+\frac{2}{\sqrt{6}}M_3,
\end{equation}
we can derive that the Shannon entropy $H_A(R)$ for the post-measurement state is
\begin{eqnarray}\label{s159}
	\rho_R^A=\frac{1}{2}R_+ +\frac{1}{2}R_-=\frac{I}{2},
\end{eqnarray}
where $R_\pm$ with the corresponding eigenvalues $\pm 1$ are the eigenvectors given in Eq. \eqref{s149}, and the complementary property between the measurements $Q$ and $R$ as well as the completeness $R_++R_-=I$ are utilized.

In this case, the Shannon entropy for the post-measurement state $\rho_R^A$ can be expressed as $H(R)=h(1/2)=1$, and we have the EUR
\begin{eqnarray}\label{s160}
	H_{QR}(A)&=&H_A(Q)+H_A(R)\nonumber\\
	&=&1+S(A).
\end{eqnarray}
Then, based on Eq. \eqref{s160} and similar analyses to those in Secs. \hyperref[THEOREM 2]{\text{III}} and \hyperref[THEOREM 3]{\text{IV}} of the SM,
we can obtain that the trade-off relations in Theorems 2 and 3 of the main text are still satisfied.

The quantitative trade-off relation between preparation contextuality of the optimal eight state set $\widetilde{\Lambda}(\rho_A)$ and bipartite entanglement is expressed as
\begin{equation}\label{s161}
	1\leq H_{QR}(A)+S(A|B)\leq 3,
\end{equation}
where $H_{QR}$ is the EUR for subsystem A with the optimal measurements $Q$ and $R$ given in Eqs. \eqref{s155} and \eqref{s158}. When $\rho_{AB}$ is a pure state, we have the trade-off equality $H_{QR}(A)+S(A|B)=1$. According to Eq. \eqref{s143} in Theorem 4, we find that when the two-qubit entanglement is large enough, \textit{i.e.}, $-S(A|B)\geq C'\simeq 0.7440$, preparation procedure of the optimal eight-state set $\widetilde{\Lambda}(\rho_A)$ is necessarily noncontextual due to the property $H_{QR}(A)\geq 1+C'$.

Moreover, the quantitative trade-off relation between preparation contextuality of the optimal eight state set $\widetilde{\Lambda}(\rho_A)$ and bipartite quantum nonlocality has the form
\begin{equation}\label{s162}
	1\leq H_{QR}(A)+[2-\langle B\rangle_{\mathrm{max}}] \leq 4,
\end{equation}
where $\langle B\rangle_{\mathrm{max}}$ is the maximal Bell-CHSH nonlocality exhibited in the shared bipartite state $\rho_{AB}$. When the shared quantum state $\rho_{AB}$ is pure, the upper bound of the above trade-off relation will be $4-2\sqrt{2}$. Then, according to the lower bound of Eq. \eqref{s162}, the local preparation contextuality in the optimal eight-state set $\widetilde{\Lambda}(\rho_A)$ will disappear when the maximal Bell nonlocality in the CHSH scenario satisfies $\langle B\rangle_{\mathrm{max}}\geq 2+C'\simeq 2.7440$.

After a comparison of Eqs. (8)-(9) in the main text and Eqs. (161)-(162) in this SM, we can obtain that the trade-off relations in Theorems 2 and 3 of the main text are still satisfied for preparation contextuality of the optimal eight-state set $\widetilde{\Lambda}(\rho_A)$, and the coexistence regions for local and nonlocal quantum resources are enlarged due to the higher symmetry of the optimal eight-state set.


\begin{thebibliography}{99}
\bibitem{note2}   It should be noted that the optimal set $\Lambda(\rho_A)$ generated by the $B_2$-orbit realizability in Lemma 1 is different from the one for describing the URs in Ref. \cite{lc22prl}, since they serve for the different tasks (see a concrete example in Sec. IIC of the SM \cite{sup25bai}). Moreover, beyond Theorem 1, an alternative criterion can also be derived from the single optimal observable $Q$, and the details of the criterion with its analytical proof are provided in Sec. VIII of the SM \cite{sup25bai}.
\end{thebibliography}
\end{document}